\documentclass[12pt]{article}

\usepackage{axodraw}

\makeatletter

\@addtoreset{equation}{section}

\def\asymp#1%
{\mathrel{\raisebox{-.4em}{$\widetilde{\scriptstyle #1}$}}}
\makeatother

\newcommand\nn         {\nonumber}
\newcommand\plusdist   {`+'-distribution}
\newcommand\xplusdist   {`$x_+$'-distribution}
\newcommand\M[2]       {\ensuremath{|{\cal{M}}^{#1}_{#2}|^2}}
\newcommand\as         {\ensuremath{\alpha_{\mathrm{s}}}}

\newcommand\msbar      {\ensuremath{{\overline {\rm MS}}}}

\newcommand\smfrac[2]  {{\textstyle\frac{#1}{#2}}}
\renewcommand\d        {{\mathrm d}}

\renewcommand\O        {{\mathrm O}}
\newcommand\Oe[1]      {\ensuremath{\mathrm O(\ep^{#1})}}
\newcommand{\bV}       {{\bf V}}
\newcommand{\cV}       {{\cal V}}
\newcommand{\cK}       {{\cal K}}
\newcommand{\cT}       {{\cal T}}

\def\hf{\smfrac{1}{2}}
\def\ep{\epsilon}
\def\ee{$e^+e^-$}
\def\beq{\begin{equation}}
\def\eeq{\end{equation}}
\def\beeq{\begin{eqnarray}}
\def\eeeq{\end{eqnarray}}
\def\ldot{\!\cdot\!}
\def\aand{\!\!\!\!\!\!\!\!&&}
\def\cm{{\cal M}}

\def\tq{{\tilde q}}
\def\tg{{\tilde g}}

\def\bom#1{{\mbox{\boldmath $#1$}}}
\def\to{\rightarrow}

\def\KFS#1{K^{{#1}}_{\scriptscriptstyle\rm F\!.S\!.}}

\def\Kab{\KFS{aa'}}
\def\Kbar{\overline{K}^{aa'}}

\def\cF{{\cal F}}

\newcommand{\la}{\langle}
\newcommand{\ra}{\rangle}

\def\nn{\nonumber}

\def\ID{1 \kern -.45 em 1}

\newcommand{\tpij}{\widetilde p_{ij}}
\newcommand{\tpk}{\widetilde p_k}
\newcommand{\zi}{\tilde z_i}
\newcommand{\zj}{\tilde z_j}
\newcommand{\sijk}{s_{ij,k}}
\newcommand{\yijk}{y_{ij,k}}
\newcommand{\vijk}{v_{ij,k}}
\newcommand{\viji}{v_{ij,i}}                                                    
\newcommand{\tvijk}{\tilde v_{ij,k}}                                            
\newcommand{\ri}{{\mathrm{i}}}
\newcommand{\rd}{{\mathrm{d}}}
\newcommand{\rA}{{\mathrm{A}}}
\newcommand{\rR}{{\mathrm{R}}}
\newcommand{\rV}{{\mathrm{V}}}
\newcommand{\eik}{{\mathrm{eik}}}
\newcommand{\fact}{{\mathrm{C}}}
\newcommand{\coll}{{\mathrm{coll}}}
\newcommand{\reg}{{\mathrm{reg}}}

\newcommand{\eps}{\epsilon}                            

\newcommand{\cD}{{\cal D}}
\newcommand{\alps}{\alpha_{\mathrm{s}}}
\newcommand{\CF}{C_{\mathrm{F}}}
\newcommand{\CA}{C_{\mathrm{A}}}
\newcommand{\TR}{T_{\mathrm{R}}}                                                
\newcommand{\Nc}{N_{\mathrm{c}}}                                                
\def\Li{\mathop{\mathrm{Li}}\nolimits}
\def\Real{\mathop{\mathrm{Re}}\nolimits}
\def\mathswitchr#1{\relax\ifmmode{\mathrm{#1}}\else$\mathrm{#1}$\fi}
\newcommand{\rB}{{\mathswitchr{B}}}
\newcommand{\aLO}{{\mathswitchr{LO}}}
\newcommand{\aNLO}{{\mathswitchr{NLO}}}

\def\draftdate{\relax}
\def\mda{\relax}
\def\mua{\relax}
\def\mla{\relax}
\def\draft{
\def\thtystars{******************************}
\def\sixtystars{\thtystars\thtystars}
\typeout{}
\typeout{\sixtystars**}
\typeout{* Draft mode!
         For final version remove \protect\draft\space in source file *}
\typeout{\sixtystars**}
\typeout{}
\def\draftdate{\today}
\def\mua{\marginpar[\boldmath\hfil$\uparrow$]%
                   {\boldmath$\uparrow$\hfil}%
                    \typeout{marginpar: $\uparrow$}\ignorespaces}
\def\mda{\marginpar[\boldmath\hfil$\downarrow$]%
                   {\boldmath$\downarrow$\hfil}%
                    \typeout{marginpar: $\downarrow$}\ignorespaces}
\def\mla{\marginpar[\boldmath\hfil$\rightarrow$]%
                   {\boldmath$\leftarrow $\hfil}%
                    \typeout{marginpar: $\leftrightarrow$}\ignorespaces}
\overfullrule 5pt
\oddsidemargin -15mm
\marginparwidth 29mm
}

\def\stars{\strut\leaders\hbox{*}\hfill\strut}
\def\starline{\hfil\strut\hfil\hbox to \textwidth {\stars}\hfil}

\begin{document}

\begin{titlepage}
\renewcommand{\thefootnote}{\fnsymbol{footnote}}
\begin{flushright}
CERN--TH/2001-305\\ 
DESY 01--099\\ 
hep-ph/0201036 
     \end{flushright}
\par \vspace{5mm}
\begin{center}
{\Large \bf
The Dipole Formalism  for Next-to-Leading Order\\[.5em]
QCD Calculations with Massive Partons}
\end{center}
\par \vspace{2mm}
\begin{center}
{\bf Stefano Catani}$^1$\footnote{On leave of absence from INFN,
Sezione di Firenze, Florence, Italy.}{\bf, }
{\bf Stefan Dittmaier}$^2$\footnote{Heisenberg fellow 
of the Deutsche Forschungsgemeinschaft DFG.}{\bf,}
{\bf Michael H. Seymour}$^3$
and \\
{\bf Zolt\'an Tr\'ocs\'anyi}$^4$\footnote{Sz\'echenyi fellow of the
Hungarian Ministry of Education.}

\vspace{5mm}

{$^1$ Theory Division, CERN, CH-1211 Geneva 23, Switzerland} \\[.5em]
{$^2$ Deutsches Elektronen-Synchrotron DESY,
D-22603 Hamburg, Germany} \\[.5em]
{$^3$ Theory Group, Department
of Physics \& Astronomy, The University of Manchester,
Manchester, M13 9PL, U.K.} \\[.5em]
{$^4$ University of Debrecen and \\Institute of Nuclear Research of
the Hungarian Academy of Sciences\\ H-4001 Debrecen, PO Box 51, Hungary}

\vspace{5mm}
\end{center}

\par \vspace{2mm}
\begin{center} {\large \bf Abstract} \end{center}
\begin{quote}
\pretolerance 10000
The dipole subtraction method for calculating next-to-leading order
corrections in QCD was originally only formulated for massless partons.
In this paper we extend its definition to include massive partons,
namely quarks, squarks and gluinos.  We pay particular attention to the
quasi-collinear region, which gives rise to terms that are enhanced by
logarithms of the parton masses, $M$.  By ensuring that our subtraction
cross section matches the exact real cross section in all
quasi-collinear regions we achieve uniform convergence both for hard
scales $Q\sim M$ and $Q\gg M$.  Moreover, taking the masses to zero, we
exactly reproduce the previously-calculated massless results.  We give
all the analytical formulae necessary to construct a numerical program
to evaluate the next-to-leading order QCD corrections to arbitrary
observables in an arbitrary process.
\end{quote}

\vspace*{\fill}
\begin{flushleft}
     December 2001
\end{flushleft}
\end{titlepage}

\renewcommand{\thefootnote}{\fnsymbol{footnote}}

\section{Introduction}
\label{intro}

Hard-scattering production of heavy quarks and strongly-interacting heavy
particles (such as squarks, gluinos and so forth) is of topical interest
for Standard Model (SM) and beyond Standard Model physics at high-energy
colliders (see, e.g.\ Refs.~\cite{Altarelli:2000ye,Accomando:1998wt}
and references therein). In order to have reliable theoretical predictions,
it is important to control and explicitly compute QCD radiative
corrections to these production processes beyond the leading order (LO)
approximation in perturbation theory. These computations are very
involved, as has been known since the first next-to-leading order (NLO)
calculations of heavy-quark production in hadron collisions
\cite{HQhtot,HQdist}.

In the case of hard-scattering processes that involve only
{\em massless\/} QCD partons, the practical feasibility of higher-order
computations has been highly simplified by the development of general
algorithms \cite{Giele:1992vf,
Giele:1993dj,Keller:1999tf,Harris:2001sx,Frixione:1996ms,Catani:1996jh,
Catani:1997vz,Nagy:1996bz} to perform NLO calculations.  These
algorithms, based either on the phase-space slicing method
\cite{Fabricius:1981sx} or on the subtraction method
\cite{Ellis:1980wv}, start from the process-dependent QCD matrix
elements and apply process-independent procedures to isolate and cancel
the infrared (soft and collinear) divergences that appear at
intermediate steps of the calculation.  The final output is a
process-independent recipe to construct a modified version of the
original matrix elements. The modified matrix elements can directly be
integrated (usually by numerical Monte Carlo techniques) over the
relevant and process-dependent phase space to compute any infrared and
collinear safe observable at NLO.  

The NLO algorithm based on the dipole formalism \cite{Catani:1996jh}
was fully worked out in Ref.~\cite{Catani:1997vz}. In recent years,
it has been implemented in general purpose Monte Carlo codes for the
calculation of 3-jet \cite{Catani:1996jh} and 4-jet observables
\cite{Nagy:1997yn,Weinzierl:1999yf} in $e^+e^-$ annihilation, 2-jet
\cite{Catani:1997vz,Nagy:2001xb} and 3-jet \cite{Nagy:2001xb} final
states in deep inelastic lepton--hadron scattering, production of three
jets \cite{Nagy:2001fj}, vector-boson pairs \cite{Campbell:1999ah},
vector boson plus massless $b{\bar b}$ \cite{Ellis:1998fv} and
colourless supersymmetric particles \cite{Beenakker:1999xh}
at hadron colliders. It has also been applied to the computation of QCD
corrections to specific processes, such as 4-fermion final states 
\cite{Maina:1996ep} and forward--backward asymmetries \cite{Catani:1999nf}
at high-energy $e^+e^-$ colliders and to radiative quarkonium decays 
\cite{Kramer:1999bf}.

Recent activity has been devoted to extending the dipole formalism by
including the effect of {\em massive\/} partons. A first step was
performed in Ref.~\cite{Dittmaier:2000mb} (see also
Ref.~\cite{Roth:1999kk} for the particular case of small fermion masses)
by developing the dipole formalism to compute NLO QED radiative
corrections to electroweak processes and thus by considering photon
radiation from massive charged fermions in arbitrary helicity configurations.
The QED version of the dipole formalism was applied to study single-photon 
radiation in the scattering processes $\gamma \gamma \to f {\bar f}$, 
$e^-\gamma \to e^-\gamma$ and $\mu^+ \mu^- \to \nu_e {\bar \nu}_e$ 
\cite{Dittmaier:2000mb}, to evaluate NLO electroweak corrections to
$e^+e^- \to WW \to 4f$ \cite{Denner:2000bj} and to $W$-boson production
at hadron colliders \cite{Dittmaier:2001ay} and to compute NLO QCD
corrections to $e^+e^- \to Ht{\bar t}$ (and $Hb{\bar b}$) in the SM and
its minimal supersymmetric extension \cite{Dittmaier:2000tc}.
A QCD extension of the results of Ref.~\cite{Catani:1997vz} to processes 
involving heavy fermions has been presented in Ref.~\cite{Phaf:2001gc}.
In this paper we present our version of the extension of the massless
dipole formalism to QCD heavy partons. A small part of the results were
anticipated in Ref.~\cite{Catani:2001ef}. Full results of our formalism
have already been applied to the computation of the NLO QCD corrections
to the associated production of the SM Higgs boson and a $t{\bar t}$ pair
in hadron collisions \cite{Beenakker:2001rj}. 

In extending the NLO dipole formalism \cite{Catani:1997vz} to arbitrary 
processes with massive partons, we devote particular attention to its
behaviour in the massless limit.  As long as we are interested in
isolating and cancelling the infrared divergences in the cross section,
the extension from massless partons to heavy partons mainly involve
kinematical complications due to the finite value of the parton masses.
QCD radiation from heavy partons, however, can lead to contributions
that, though infrared finite, are proportional to powers of
$\ln Q^2/M^2$, where $M$ is the parton mass and $Q$ is the typical
scale of the hard-scattering process. In kinematical configurations 
where $Q \gg M$, these {\em logarithmically-enhanced\/} contributions
(as well as other types of {\em constant\/} contributions discussed
in Sect.~\ref{se:subnh}) become large and can spoil the numerical
convergence of the calculation. In our formulation of the dipole
method, we are not mainly concerned with the evaluation of these 
terms at NLO (or with their resummation to all orders), but rather in
minimizing the instabilities that these terms can produce.  For
instance, in the case of cross sections that are infrared and collinear
safe in the massless limit, these logarithmic terms cancel in the final
NLO result, but they can still appear at intermediate steps of the
calculation (e.g.\ in the separate evaluation of the real and virtual
contributions) thus leading to reduced convergence in numerical
implementations. These contributions can be singled out in a
process-independent manner, since they are related to the singular
behaviour of the QCD matrix elements in the limit $M \to 0$.  This
singular behaviour is controlled by {\em quasi-collinear\/} factorization
formulae, in the same way that the infrared divergences are controlled
by soft and collinear factorization formulae (see Sect.~\ref{se:limits}).
We exploit these factorization properties not only to cancel the infrared
divergences, but also to stabilize the NLO algorithm with respect to large
numerical contributions that can arise when the hard-scattering kinematics
vary from the regions where $Q \sim M$ to those where $Q \gg M$. In the
case of processes with no initial-state hadrons (see Sect.~\ref{se:subnh}),
we set up the method such that, for those cross sections that are
infrared and collinear safe in the limit of vanishing parton masses,
the NLO algorithm is smooth in the massless limit and, more
importantly, numerically stable for any values of the hard-scattering
scale. In the case of hadron collision processes (see Sect.~\ref{se:subwh}),
similar features are achieved provided the NLO partonic calculation is
properly matched with the definition of the parton distributions of the
colliding hadrons.

In summary, we present a general extension of the dipole formalism that 
includes different species of strongly interacting massive fermions 
(quarks, gluinos) and scalars (squarks) with equal or unequal masses 
(see Sect.~\ref{se:dipoles} and Appendix~\ref{app:SUSYdipoles}). 
We develop the formalism by explicitly carrying out the analytical part
of the NLO calculation for arbitrary infrared and collinear safe
observables in lepton and hadron collisions (see Sect.~\ref{se:xsec}).
The discussion of cross sections that involve fragmentation functions
of massless or light (with respect to the hard-scattering scale)
partons is only sketched, since it does not involve major technical
complications with respect to the results presented here and in
Ref.~\cite{Catani:1997vz}.  An important general feature of our
treatment regards the identification and control of
logarithmically-enhanced contributions of the type $\ln Q^2/M^2$, which
can produce numerical instabilities in kinematical regimes where the
hard-scattering scale $Q$ is much larger than the value~$M$ of the
mass of one or more heavy partons.

This feature, which was also implemented in the case of photon
radiation from massive fermions \cite{Dittmaier:2000mb}, is the main
overall difference with respect to other process-independent treatments
\cite{Dittmaier:2000mb,Roth:1999kk,Phaf:2001gc,Keller:1999tf} of
massive partons in NLO calculations. There are, however, other general
differences. For instance, in Refs.~\cite{Dittmaier:2000mb,Roth:1999kk}
the soft divergences are regularized by introducing an infinitesimal
photon mass (a widespread practice in calculations of electroweak
radiative corrections), while we use dimensional regularization. The
authors of Ref.~\cite{Phaf:2001gc} consider only fermions as heavy
partons and do not consider the cases with different species of massive
fermions of unequal masses. The general treatment of
Ref.~\cite{Keller:1999tf} is based on the phase-space slicing method. 
The authors of Ref.~\cite{Keller:1999tf} pointed out the relevance of
the quasi-collinear limit to control the large terms $\ln Q^2/M^2$ in
the region where $Q \gg M$. They also discussed how these terms can be
incorporated in the definition of parton distributions and
fragmentation functions of heavy quarks but postponed a detailed
treatment to future work.  

In the case of massless partons, the dipole formalism was fully
developed and described in Ref.~\cite{Catani:1997vz}. In the present
extension to include massive partons, we closely follow the
presentation in Ref.~\cite{Catani:1997vz}.  Therefore, we do not repeat
all the steps with the same amount of details as in
Ref.~\cite{Catani:1997vz}. The outline of the paper is as follows.  In
Sect.~\ref{se:sub} we discuss the general features of our
implementation of massive partons in the dipole formalism.  In
Sect.~\ref{notation} we recall the notation and its extension to the 
massive case. In Sect.~\ref{se:limits} we discuss the singular behaviour of 
the real-emission matrix elements in the soft, collinear and quasi-collinear 
limits. In Sect.~\ref{se:dipoles} we give the definition of the dipole
factorization formulae that smoothly interpolate these limiting
regions for any values of the parton masses.  The proof that the
definition consistently matches the singular behaviour of the QCD
matrix elements is not explicitly given: it proceeds along the same
lines as in Ref.~\cite{Catani:1997vz}. In the case of vanishing parton
masses, all the dipole factors reduce to those defined in
Ref.~\cite{Catani:1997vz}.  In Sect.~\ref{se:dipoles} we also derive
the factorization properties of the dipole phase space and perform the
corresponding integration of the dipole factorization formulae. 
Sect.~\ref{se:xsec} contains the explicit expressions necessary for
calculating QCD cross sections in different classes of processes at NLO
accuracy.  In Sect.~\ref{subsec:final}, we write down the final
formulae needed to implement our method, while in
Sects.~\ref{subsec:xsee}--\ref{subsec:twoini} we sketch their
derivation for processes with no (Sect.~\ref{subsec:xsee}), one
(Sect.~\ref{subsec:oneini}) and two (Sect.~\ref{subsec:twoini})
initial-state hadrons. After reading Sect.~\ref{notation},
a reader who is familiar with the method in the massless case
and is interested only in setting up a NLO Monte Carlo
program for a specific process can find all the relevant results
in Sect.~\ref{subsec:final} and in one of the following subsections
(depending on the specific process).  Sect.~\ref{se:summary} contains
our conclusions.

We leave some technical details and discussion of explicit examples to
the appendices. In Appendix~\ref{app:auxint}, we spell out the explicit
expression for the integral of the eikonal term. 
Appendix~\ref{app:Q2fixed} discusses a subtle point in the definition
of the $x$-distributions related to the treatment of initial-state
singularities in NLO calculations. In Appendix~\ref{app:SUSYdipoles},
we give the dipole splitting functions that are relevant for SUSY QCD
calculations. Finally, we present the three simplest examples of our
method in Appendix~\ref{app:examples}.

\section{The general method and its features in the massless limit}
\label{se:sub}

In this section we briefly review\footnote{A concise overview of the
dipole formalism can also be found in Ref.~\cite{Catani:1997fq}.} 
the method of Refs.~\cite{Catani:1996jh,Catani:1997vz} and describe the
main features of our extension to processes with massive partons. 

\subsection{Cross sections without identified or initial-state
hadrons and their smooth massless limit}
\label{se:subnh}

We first consider processes with no initial-state colliding hadrons and
no tagged hadrons in the final state. This is the case, for example, of 
jet production in \ee\ collisions.  Our aim is to calculate a generic
infrared- and collinear-safe cross section $\sigma$ with NLO accuracy.
Assuming that at the LO there are $m$ QCD partons in the final state,
we schematically write
\beq
\label{eq:sigma}
\sigma = \sigma^\aLO + \sigma^\aNLO 
= \int_m\rd\sigma^\rB + \sigma^\aNLO,
\eeq
where the LO contribution $\rd\sigma^\rB$ is the fully differential
Born cross section and the NLO correction $\sigma^\aNLO$ comprises the
real and virtual (one-loop) contributions $\rd\sigma^\rR$ and
$\rd\sigma^\rV$:
\beq
\label{eq:sigmaNLO}
\sigma^\aNLO = \int_{m+1}\rd\sigma^\rR + \int_m\rd\sigma^\rV \;.
\eeq
Here the notation for the integrals indicates that the real contribution 
involves $m+1$ final-state partons (one QCD parton more than in LO), while 
the virtual contribution has the $m$-parton kinematics.

After renormalization of the one-loop matrix element involved in 
$\rd\sigma^\rV$, all the contributions to $\sigma$ are ultraviolet (UV)
finite.  The Born contribution $\rd\sigma^\rB$ is  integrable over the
infrared (IR) region of the phase space, but the real and virtual
contributions to $\sigma^\aNLO$ are separately affected by IR
divergences produced by soft and collinear partons. Although the IR
divergences cancel in the sum on the right-hand side of
Eq.~(\ref{eq:sigmaNLO}), the separate pieces have to be
regularized before any numerical calculation can be attempted.
Analytic continuation of the integrals to $d=4-2\ep$ space-time dimensions
is the only known gauge- and Lorentz-invariant regularization procedure,
but it prevents a straightforward numerical integration. Using dimensional
regularization, the IR divergences are replaced by double and single poles,
$1/\ep^2$ and $1/\ep$, that have to be extracted and cancelled analytically
before performing the limit $\ep \to 0$.

The formalism of Ref.~\cite{Catani:1997vz} deals with the $\ep$ poles by
using the subtraction method. The general idea of the subtraction method is to
introduce an auxiliary cross section $\rd\sigma^\rA$ that has the same 
{\em pointwise\/} singular behaviour (in $d$ dimensions!) as $\rd\sigma^\rR$.
Moreover, $\rd\sigma^\rA$ has to be chosen simple enough, such that it is
{\em analytically\/} integrable in $d$ dimensions over the one-parton
subspaces that cause the soft and collinear divergences. Thus, without
performing any approximations, $\rd\sigma^\rA$ is subtracted from the real
contribution and added back to the virtual contribution. Since 
$\rd\sigma^\rA$ acts as a {\em local\/} counterterm for $\rd\sigma^\rR$,
the difference $[\rd\sigma^\rR-\rd\sigma^\rA]$ is integrable over the entire
$(m+1)$-particle phase space in any number of dimensions and we can safely
take the limit $\ep \to 0$.
Moreover, since the analytical expression of the performed integral 
over the singular subspace,
$\int_1 \rd\sigma^\rA$, explicitly contains all the $\ep$ poles that cancel 
those of the virtual term $\rd\sigma^\rV$, the sum 
$[ \rd\sigma^\rV + \int_1 \rd\sigma^\rA]$ is also (numerically)
integrable in $d=4$ dimensions over the remaining $m$-parton phase space. 
The final structure of the NLO calculation is
\beq
\label{eq:sigmaNLOsub}
\sigma^\aNLO = 
\int_{m+1}\!\left[ \left(\rd\sigma^\rR\right)_{\eps=0} - 
\left(\rd\sigma^\rA\right)_{\eps=0} \;\right]
+ \int_m\!
  \left[\rd\sigma^\rV + \int_1 \rd\sigma^\rA \right]_{\eps=0} \;,
\eeq
and the two terms on the right-hand side are separately integrable.
Usually the integrations are not feasible in analytic form, but they can
always be carried out numerically. For instance, the calculation can be 
implemented in a `partonic Monte Carlo' program, which generates appropriately
weighted partonic events with $m+1$ final-state partons and events with
$m$ final-state partons.

The method described so far is equally applicable to QCD calculations
that involve massless and massive QCD partons. In principle, the extension
from the massless to the massive case can be performed in a straightforward
way. It is sufficient to overcome the (non-trivial)
technical difficulties related to the generalization of $\rd\sigma^\rA$ 
to massive partons and, in particular, to the analytic evaluation
of the integral $\int_1 \rd\sigma^\rA$ over the one-particle phase space with
mass constraints. 

However, in our extension of the formalism of 
Ref.~\cite{Catani:1997vz}, we require additional properties. To explain
the reasons for these additional requirements, it is important to recall
that the finite mass $M$ of the QCD partons plays a different physical role
in different physical processes. In some processes (e.g.\ the total cross
section for heavy-quark hadroproduction \cite{HQhtot}) the finite (and large)
value of $M$ has the essential role of setting the hard scale of the 
cross section. In these cases the massless limit $M \to 0$ is IR {\em unstable}
and the corresponding cross section cannot be computed in QCD perturbation
theory. In other processes (e.g.\ the production of heavy-flavoured 
jets in \ee\ annihilation \cite{HQeejets}), instead, 
the hard scale $Q$ is independent of the mass $M$ and the latter has only the
role of an auxiliary (though important) kinematical scale. These processes
are IR {\em stable\/} in the massless limit, that is, when $M \to 0$ the cross
section is still infrared- and collinear-safe and, thus, perturbatively
computable. 

The processes that are perturbatively stable in the massless limit are often
studied in kinematical regions where the typical hard scale $Q$ is much
larger than the mass $M$ of one (or more) of the heavy partons. In this regime
the integral of the real term $\rd\sigma^\rR(M)$ of the NLO cross section in
Eq.~(\ref{eq:sigmaNLO}) leads to contributions of the type
\beq
\label{eq:log}
\int_{m+1}\!\!\rd\sigma^\rR(M) \quad \rightarrow \quad
\int_0^{Q^2}\!\! \rd{\bf q}_\perp^2 \; \left({\bf q}_\perp^2\right)^{-\ep}
\frac{1}{{\bf q}_\perp^2 + M^2} \asymp{Q \gg M} \ln \frac{Q^2}{M^2} 
+ \O(\ep)\;\;,
\eeq
and
\beq
\label{eq:const}
\int_{m+1}\!\!\rd\sigma^\rR(M) \quad \rightarrow \quad
\int_0^{Q^2}\!\! \rd{\bf q}_\perp^2 \; \left({\bf q}_\perp^2\right)^{-\ep}
\frac{M^2}{\left[{\bf q}_\perp^2 + M^2\right]^2 } \asymp{Q \gg M} 
M^2 \;\frac{1}{M^2} + \O(\ep)\;\;,
\eeq
where ${\bf q}_\perp$ generically denotes the typical transverse momentum
of the heavy parton with mass $M$.
Since these contributions are finite when $\ep \to 0$, naively, they would 
not require any special treatments within the subtraction method. However,
this could lead to serious numerical problems.

The problems are evident in the case of the contribution in
Eq.~(\ref{eq:log}), which is very large when $Q \gg M$.  When computing
the NLO cross section, this large contribution would appear in the
first term (the $(m+1)$-parton integral) on the right-hand side of
Eq.~(\ref{eq:sigmaNLOsub}) and it would be compensated by an equally
large (but with opposite sign) logarithmic contribution arising from
the second term (the $m$-parton integral). Owing to the presence of
several large (although compensating) contributions, a similar naive
procedure would lead to instabilities in {\em any\/} numerical
implementations of the NLO calculation.  The numerical instabilities
would increase by increasing the ratio $Q/M$ and, in particular, they
would prevent from performing the massless limit.

The contribution in Eq.~(\ref{eq:const}) may appear harmless, since
it approaches a constant (finite) value when $M/Q \to 0$. However, 
the constant
behaviour is obtained by combining the factor $M^2$ from the numerator
of the integrand with the factor $1/M^2$ from the integral of the denominator.
Owing to the presence of a linearly divergent (in the limit $M^2/Q^2 \to 0$) 
integral, the contribution in Eq.~(\ref{eq:const}) cannot be evaluated 
numerically by using standard Monte Carlo techniques, since its variance 
increases linearly with $Q^2/M^2$. Finite integrals with infinite variance
(e.g.\ integrands with square root singularities)
often occur in NLO calculations: they can be treated within Monte Carlo
methods by applying importance sampling procedures. In this respect,
the term in Eq.~(\ref{eq:const}) poses no additional conceptual problems.
However, our main point is that the variance of the integral in 
Eq.~(\ref{eq:const}) is not uniform in $M$: it varies from a finite value when
$M ={\rm O}(Q)$ to a divergent value when $M/Q \to 0$.
The presence of these type of contributions 
may thus prevent from straightforwardly constructing a `partonic Monte Carlo'
program that performs equally well in the two kinematical regimes
$M ={\rm O}(Q)$ and $M/Q \ll 1$.

To avoid these numerical problems in the calculation of cross sections
that are IR stable in the massless limit, we set up our massive-parton
formalism by choosing the auxiliary cross section $\rd\sigma^\rA(M)$ in such a
way that the following property is fulfilled:
\beq
\label{eq:smoothlimit}
\lim_{M \to 0} 
\int_{m+1}\!\left[ \left(\rd\sigma^\rR(M)\right)_{\eps=0} \! - 
\left(\rd\sigma^\rA(M)\right)_{\eps=0} \,\right] =
\int_{m+1}\!\left[ \left(\rd\sigma^\rR(M=0)\right)_{\eps=0} \! - 
\left(\rd\sigma^\rA(M=0)\right)_{\eps=0} \,\right] .
\eeq
Note that, to avoid the problems related to the large logarithmic contributions
in Eq.~(\ref{eq:log}), it is sufficient to impose that the integral
of the subtracted cross section on the left-hand side of 
Eq.~(\ref{eq:smoothlimit}) is finite when $M \to 0$. Equation 
(\ref{eq:smoothlimit}) is, instead, a stronger constraint. It implies that,
in the evaluation of the subtracted cross section, the massless limit
(or, more generally, the limit $M/Q \to 0$) commutes with the 
$(m+1)$-parton integral. This guarantees that 
$[ \rd\sigma^\rR(M) - \rd\sigma^\rA(M) ]$ does not contain integrands of the
type in Eq.~(\ref{eq:const}).

Once Eq.~(\ref{eq:smoothlimit}) is satisfied, the $(m+1)$-parton and 
$m$-parton contributions to the NLO cross section in Eq.~(\ref{eq:sigmaNLOsub})
{\em separately\/} have smooth behaviour when $M \to 0$. This behaviour
helps to compute the cross section numerically in the kinematical regimes
where $Q \gg M$. It can also be used to easily check that, in the massless
limit, the massive-parton calculation correctly agrees with the corresponding 
result obtained in the exactly massless case.

The IR divergent contributions to the real cross section $\rd\sigma^\rR$
are process independent. By this we mean that, in the soft and
collinear limits, $\rd\sigma^\rR$ is given by the corresponding
(process-dependent) Born-level cross section $\rd\sigma^\rB$ times universal 
(process-independent) singular factors. Owing to these soft and collinear
factorization properties \cite{factform, Altarelli:1977zs}, it is possible  
to give general prescriptions for constructing the auxiliary cross section
$\rd\sigma^\rA$ in a process-independent manner. 

Within the dipole formalism
\cite{Catani:1997vz}, $\rd\sigma^\rA$ is constructed by a sum over
different contributions, named {\em dipoles}. Each dipole contribution
describes soft and collinear radiation from a pair of ordered partons.
The first parton is called {\it emitter\/} and the second {\it spectator}, 
since only the kinematics of the former leads to the IR singularities.
The dipole configurations can be thought of as being obtained by an effective
two-step process: using the Born-level cross section, an $m$-parton
configuration is first produced and the emitter and spectator are singled out
in all possible ways; then the emitter decays into two partons and
the spectator, which contains information on the colour and spin 
correlations of the real cross section $\rd\sigma^\rR$, is used 
to balance momentum conservation.
The auxiliary cross section $\rd\sigma^\rA$ can symbolically be written as
\beq
\label{eq:genaux}
\rd\sigma^\rA = \sum_{\mathrm{dipoles}} 
\rd\sigma^\rB \otimes \rd V_{\mathrm{dipole}}\:,
\eeq
where the dipole factors $\rd V_{\mathrm{dipole}}$ describe the two-parton 
decays of the emitters. These factors are universal and can be obtained
from the QCD factorization formulae 
(including the associated colour and spin
correlations, as denoted by the symbol $\otimes$)
in the soft and collinear limits.

The product structure in Eq.~(\ref{eq:genaux})
is made possible by the factorization of QCD amplitudes on soft and
collinear poles and by a suitable factorized definition of the phase space.
It permits a factorizable mapping
from the $(m+1)$-parton phase space to an $m$-parton subspace, 
identified by the partonic variables in $\rd\sigma^\rB$, times
a single-parton phase space, identified by the partonic variables
in $\rd V_{\mathrm{dipole}}$. The single-parton phase space is 
process-independent: it describes the two-parton decay of the dipole and 
embodies the kinematical dependence on the degrees of freedom that lead
to the IR singularities. This mapping makes $\rd V_{\mathrm{dipole}}$
fully integrable analytically and in a process-independent manner.
We can symbolically write:
\beq
\label{dsA1}
\int_{m+1} \rd\sigma^\rA = \sum_{\mathrm{dipoles}} \;\int_m 
\;\rd\sigma^\rB \otimes
\int_1 \;\rd V_{\mathrm{dipole}} = \int_m \left[ \rd\sigma^\rB \otimes {\bom I}
\right] \;\;,
\eeq
where the universal factor ${\bom I}$ is defined by
\beq
\label{Ifac}
{\bom I} = \sum_{\mathrm{dipoles}} \;\int_1 \;\rd V_{\mathrm{dipole}} \;\;,
\eeq
and contains all the $\ep$ poles that are necessary to cancel the (equal and
with opposite sign) poles in $\rd\sigma^\rV$. As a byproduct of this
cancellation mechanism, Eq.~(\ref{dsA1}) can also be used to indirectly
derive explicit information on the $\ep$ singularities in the virtual 
contribution.
Since $\rd\sigma^\rV$ and $\rd\sigma^\rB$ are respectively obtained
from one-loop and tree-level QCD amplitudes,
the factorization structure on the right-hand side of Eq.~(\ref{dsA1})
implies that the IR divergences of the one-loop amplitudes can be obtained
from the corresponding tree amplitudes in terms of a universal factorization
formula \cite{Giele:1992vf, Kunszt:1994mc, Catani:1997vz}.

The structure of the final NLO result is given as follows in terms of two
contributions, $\sigma^{\aNLO\,\{m+1\}}$ and $\sigma^{\aNLO\,\{m\}}$, 
with $(m+1)$-parton and $m$-parton kinematics, respectively, 
which are separately finite
and integrable in four space-time dimensions:
\beeq
\label{sNLO3}
\sigma^\aNLO &=& \sigma^{\aNLO\,\{m+1\}} + \sigma^{\aNLO\,\{m\}}  \\
&=&\int_{m+1} \left[ \left( \rd\sigma^\rR \right)_{\ep=0} -
\left( \sum_{\mathrm{dipoles}} \;\rd\sigma^\rB \otimes 
\;\rd V_{\mathrm{dipole}} \right)_{\ep=0} \;\right]
+  \int_m \left[ \rd\sigma^\rV + \rd\sigma^\rB \otimes {\bom I}
\right]_{\ep=0} \;\;.\nonumber
\eeeq
Equation (\ref{sNLO3}) represents the dipole formalism implementation of 
the general subtraction formula (\ref{eq:sigmaNLOsub}).

The explicit expressions of the dipole factors $\rd V_{\mathrm{dipole}}$ 
and ${\bom I}$ for all the cross sections with massless QCD partons were
obtained in Ref.~\cite{Catani:1997vz}. In this paper we present their 
generalization to the case of massive QCD partons in the final state. As in 
Ref.~\cite{Catani:1997vz}, the dipole factors $\rd V_{\mathrm{dipole}}$
are constructed by starting from the QCD factorization formulae in the
soft and collinear limits. Moreover, to implement the smoothness condition in 
Eq.~(\ref{eq:smoothlimit}), the dipole factors have to match the behaviour
of the QCD matrix elements in the so-called {\em quasi-collinear\/} limit 
\cite{Catani:2001ef}. The collinear limit describes the splitting process
of one massless parton in two massless partons when the relative transverse
momentum ${\bf q}_\perp$ of the latter vanishes.
The quasi-collinear limit (see Sect.~\ref{se:cqclim}) is a generalization of the
collinear limit to the splitting processes of massive partons.
It is obtained by letting ${\bf q}_\perp$ and the parton masses  
vanish {\em uniformly}, i.e.\ we consider the limit ${\bf q}_\perp , M \to 0$
at fixed ratio ${\bf q}_\perp/M$. In the quasi-collinear 
limit, 
the integrands of the type in Eqs.~(\ref{eq:log}) {\em and\/} (\ref{eq:const})
are formally regarded as being as singular as those that are proportional to 
$1/{\bf q}_\perp^{2}$ and
lead to the collinear divergences. Therefore,
constructing the dipole factors in Eq.~(\ref{eq:genaux}) in such a way that 
they have the same quasi-collinear limit as the real cross section 
$\rd\sigma^\rR$, our (subtracted) $(m+1)$-parton contribution, 
$\sigma^{\aNLO\,\{m+1\}}$, to Eq.~(\ref{sNLO3}) fulfils the smoothness
condition in Eq.~(\ref{eq:smoothlimit}). In addition, the dipole factors
of the present paper exactly coincide with those of Ref.~\cite{Catani:1997vz}
in the massless limit. We do that for the practical purpose of facilitating
a direct comparison with the results in Ref.~\cite{Catani:1997vz}. 
 
Analogously to the massless-parton case, the cancellation of the $\ep$ poles
in the $m$-parton contribution $\sigma^{\aNLO\,\{m\}}$ to Eq.~(\ref{sNLO3})
can be exploited to derive a universal factorization formula that controls
the IR divergences of any one-loop amplitude with massless and massive 
partons. Having fulfilled the smoothness condition (\ref{eq:smoothlimit})
and performing the analytical integration over the one-parton phase
space in Eq.~(\ref{Ifac}) uniformly in the parton masses,
our $\sigma^{\aNLO\,\{m\}}$ can be used to obtain such a factorization formula
in a form that explicitly exhibits not only the $\ep$ poles, but also
(i) the logarithmic terms (of the type $\ln^2 M$ and $\ln M$) that become 
singular when $M \to 0$ and (ii) the constant (finite) terms that originate
in the calculation of the one-loop amplitude from the non-commutativity of the 
limits $M \to 0$ and $\ep \to 0$. This result was anticipated in 
Ref.~\cite{Catani:2001ef}.

\subsection{Cross sections in hadron collisions}
\label{se:subwh}

The general procedure we have outlined so far applies
to all processes with no initial-state hadrons (for instance, jets and
heavy-quark production in \ee\ annihilation).
In lepton--hadron and hadron--hadron collision processes,
the hadronic cross section is computed by convoluting the corresponding 
partonic cross section $\sigma(xp_{\mathrm{had}})$ with the 
(process-independent) parton distributions $f(x)$ of the incoming hadrons. 
Here, $p_{\mathrm{had}}$ and $x$ generically denote the momentum of the 
colliding hadron and the momentum fraction carried by the colliding parton,
respectively. The presence of initial-state partons, carrying a well
defined momentum, makes the partonic subprocess collinear unsafe, so the
partonic cross section cannot naively be computed in QCD perturbation theory.
However, {\em provided\/} the incoming parton is {\em massless},
the universal factorization theorem \cite{Collins:1989gx}
of collinear singularities holds, and perturbative QCD can still be applied.
The initial-state collinear divergences of the partonic cross section
can be factorized and reabsorbed in the definition of the 
non-perturbative parton distributions. This procedure can consistently be
carried out at any perturbative order. In particular, at NLO the 
perturbatively-computable partonic cross section is given by
\beq
\label{eq:sigmaNLOin}
\sigma^\aNLO(p) = \int_{m+1}\rd\sigma^\rR(p) + \int_m\rd\sigma^\rV(p) +
\int_m \rd\sigma^\fact(p) \:;
\eeq
where
\beq
\label{eq:sigmacc}
\int_m \rd\sigma^\fact(p) = \int_0^1 \rd x \int_m 
\rd\sigma^{\rB}(xp) \;\Gamma(x) \;.
\eeq
Comparing Eq.~(\ref{eq:sigmaNLO}) with Eq.~(\ref{eq:sigmaNLOin}), we see that
the latter contains the additional collinear counterterm $\rd\sigma^\fact(p)$,
which arises from the redefinition of the
parton distributions. As symbolically written in Eq.~(\ref{eq:sigmacc}),
the collinear counterterm is given by the convolution of the Born cross section
with a process-independent factor $\Gamma(x)$, 
which is divergent when $\ep \to 0$.

We have emphasized the fact that the partons that initiate the
hard-scattering subprocess have to be massless. The reason is that
in the presence of two or more incoming massive partons the 
Bloch--Nordsieck mechanism \cite{Bloch:1937pw}
of cancellation of {\em soft\/} divergences is violated in QCD \cite{BNviol}.
The violation is due to non-cancelled soft divergences that are proportional 
to the mass $M$ of the incoming partons and are process dependent.
Although the violation occurs starting from next-to-next-to-leading order
(NNLO), its process dependence
spoils the general\footnote{By general we mean to all perturbative
orders and for arbitrary non-vanishing values of the mass of the incoming
partons.} validity of the factorization theorem 
\cite{Collins:1989gx}. Since the factorization theorem cannot formally
be used to perturbatively define and compute partonic cross sections with 
incoming massive partons, we do not consider these cross sections.

Note, however, that our distinction between {\em massless\/} and 
{\em massive\/} incoming partons has only a formal meaning. By `mass'
of the incoming parton we do not mean the actual physical mass of the
parton.  We only mean that the perturbative calculation is performed by
setting this mass equal to zero. The practical calculational procedure
is as follows.  If the physical mass of a parton is much smaller than
the typical value $Q$ of the hard-scattering scale in the process, this
parton has to be treated as a massless parton and included both in the
initial and in the final state. If the case is not so, the parton has
to be treated as massive and not considered as an incoming parton.  We
discuss this point further at the end of this subsection.

The subtraction method and the dipole formalism can be used to compute
the NLO partonic cross section in Eq.~(\ref{eq:sigmaNLOin}). Using the
dipole formalism, the auxiliary cross section $\rd\sigma^\rA$ still has the
form in Eq.~(\ref{eq:genaux}) and the sum over the dipoles is extended 
\cite{Catani:1997vz} to include initial-state dipoles that describe the
collinear decays of any initial-state {\em massless\/} parton (the
emitter) into two {\em massless\/} partons.  Furthermore, in
considering the final-state emitters, the sum over spectators has to be
extended to include initial-state massless partons, leading to dipole
kinematics that differ from those with a final-state spectator.
In the case with only massless partons in the final state, the initial-state
dipoles were worked out in  Ref.~\cite{Catani:1997vz}.
In this paper we extend the results of Ref.~\cite{Catani:1997vz} to include
the case with massive partons in the final state.

Independently of the fact that the final-state partons are massless or massive,
the initial-state dipoles can be integrated over the collinear region.
Their integral cancels the divergences of the collinear counterterm 
$\rd\sigma^\fact(xp)$, and the final result for NLO cross section in 
Eq.~(\ref{eq:sigmaNLOin}) can be written symbolically as
\beeq\label{sNLO4}
\sigma^\aNLO(p) &=&
\sigma^{\aNLO\,\{m+1\}}(p) + \sigma^{\aNLO\,\{m\}}(p) + \int_0^1 dx \;
{\hat \sigma}^{\aNLO\,\{m\}}(x;xp) \nonumber \\
&=&
\int_{m+1} \left[ \left( \rd\sigma^\rR(p) \right)_{\ep=0} -
\left( \sum_{\mathrm{dipoles}} \;\rd\sigma^\rB(p) \otimes 
\;\rd V_{\mathrm{dipole}}
\right)_{\ep=0} \;\right]  \\
&+&  \int_m \left[ \rd\sigma^\rV(p) +
\rd\sigma^\rB(p) \otimes {\bom I}
\right]_{\ep=0} + \int_0^1 dx \;
\int_m \left[ \rd\sigma^\rB(xp) \otimes \left( {\bom P}
+ {\bom K} \right)(x) \right]_{\ep=0} \;\;. \nonumber
\eeeq
The contributions
$\sigma^{\aNLO\,\{m+1\}}(p)$ and $\sigma^{\aNLO\,\{m\}}(p)$ (with $(m+1)$-parton
and $m$-parton kinematics, respectively) are analogous to those
in Eq.~(\ref{sNLO3}).
The last term on the right-hand side of Eq.~(\ref{sNLO4}) is a finite
(in four dimensions) remainder that is left after cancellation of the $\ep$
poles of the collinear counterterm in Eq.~(\ref{eq:sigmacc}).
This term
involves a cross section ${\hat \sigma}^{\aNLO\,\{m\}}(x;xp)$ with $m$-parton
kinematics and an additional one-dimensional integration with respect to the
longitudinal momentum fraction $x$. This integration arises from the 
convolution of the Born-level cross section $\rd\sigma^\rB(xp)$ with 
$x$-dependent functions ${\bom P}$ and ${\bom K}$ that are universal and finite
for $\ep \to 0$. The explicit expressions of ${\bom P}$ and ${\bom K}$
in the massless case were given in Ref.~\cite{Catani:1997vz}. In this paper
we obtain their generalization to the case of partonic cross sections with
massive partons in the final state.

In Ref.~\cite{Catani:1997vz} an extensive discussion was devoted to a fully
detailed treatment of cross sections with identified hadrons in the 
final state, i.e.\ cross sections that involve the introduction of parton
fragmentation functions of the outgoing hadrons. For simplicity,
in this paper we do not present
an analogous discussion of these cross sections, since the extension 
to include the associated production of final-state heavy partons is
straightforward. More precisely, we can consider two main kinematical
configurations according to whether the momentum $Q$ of the final-state
massive parton is $a)$ not observed or comparable to its mass $M$ and $b)$ 
much larger than its mass. In the case $a)$, the heavy parton does not require
additional (with respect to other cross sections)
special treatment, apart from it being included as spectator in the final-state
dipoles that are introduced \cite{Catani:1997vz}
to describe the collinear decay of the massless emitter that undergoes
final-state fragmentation. In the case $b)$, a proper treatment of the large
contributions $\ln Q^2/M^2$ is required, which can be simply achieved by
considering the massive parton as effectively massless and by introducing
its (process-independent) perturbative fragmentation function \cite{perff}.

Considering the case of cross section calculations in processes with no 
initial-state hadrons, in Sect.~\ref{se:subnh} we have discussed how we set up 
the dipole formalism with massive partons
to guarantee a smooth and numerically stable interpolation
throughout the kinematical region where the hard-scattering scale $Q$
varies from $Q \sim M$ to $Q \gg M$. In hadron collisions, a QCD parton with a
non-vanishing physical mass $M$ has to be regarded as effectively massive or
massless depending on whether $Q \sim M$ or $Q \gg M$. Moreover,
the transition between the former to the latter region needs the
introduction and {\em scheme-dependent\/} definition
of the parton distribution of the effectively massless parton.
Evidently, a smooth interpolation between the two regions requires
a proper and careful matching between the definition of the partonic
cross section and that of the parton distribution. This general theoretical
issue has been the subject of many investigations in the recent
literature (see e.g.\ the list of references in
Ref.~\cite{Demina:1999ze}), and
several practical implementations are available,
mostly in the context of deep-inelastic heavy-quark production. Once the
matching conditions and the parton distributions
within a certain scheme are precisely specified, smoothness conditions
similar to Eq.~(\ref{eq:smoothlimit}) can be imposed also on 
NLO calculations of cross sections that involve parton distributions
and fragmentation functions. The procedure outlined in 
Sect.~\ref{se:subnh} to guarantee the validity of Eq.~(\ref{eq:smoothlimit})
can thus be applied to set up the dipole formalism such that
it performs in a manner that is smooth and numerically
stable with respect to large variations of the ratio $Q/M$. 
This is left to future studies.

\subsection{Summary of the general method}
\label{subsec:summary}

The starting points of the calculation of QCD radiative corrections
at NLO are the tree-level and one-loop matrix elements that
respectively enter in the expression of the cross section contributions
 $\rd\sigma^\rR$ and $\rd\sigma^\rV$ in Eq.~(\ref{eq:sigmaNLO})
(or Eq.~(\ref{eq:sigmaNLOin})). The
separate integration of these matrix elements is not trivial because of
their infrared divergences. The goal of the dipole formalism is to
construct `effective matrix elements' that can straightforwardly be
integrated in four space-time dimensions. The final output of the
formalism is summarized in Eq.~(\ref{sNLO3}) (or Eq.~(\ref{sNLO4})),
which give the expression of the NLO cross sections in terms of the
`effective matrix elements'. The formalism provides explicit
expressions for the universal factors $\rd V_{\mathrm{dipole}}$
and~${\bom I}$ (and ${\bom P}$, ${\bom K}$), which we present 
in Sect.~\ref{se:xsec}.
%
Having these factors
at our disposal, the only other ingredients necessary for 
the full NLO calculation are simply related to the evaluation of the original
matrix elements. We need: 
\begin{itemize}
\item
a set of independent colour projections of the matrix element squared
at the Born level, summed over parton polarizations, in $d$ dimensions;
\item
the one-loop contribution $\rd\sigma^\rV$ in $d$ dimensions;
\item
an additional projection of the Born-level matrix element over the
helicity of each external gluon in four dimensions;
\item
the real emission  contribution $\rd\sigma^\rR$ in four dimensions.
\end{itemize}
Since the comment presented 
in the `Note added' section of Ref.~\cite{Catani:1997vz} is valid
independently of the value of the parton masses, this list can be simplified
if one uses the {\em dimensional-reduction\/} scheme
for regularizing the one-loop matrix elements. In this case the Born-level
calculation in the first item above
can directly be carried out in {\em four\/} space-time dimensions.

\section{Notation}
\label{notation}

\subsection{Matrix elements
}

We consider processes with coloured particles (partons) in the initial 
and final states.  Any number of additional non-coloured particles is
allowed, too, but they will be suppressed in the notation.  Partons in
the initial state are labelled by $a,b,\dots$, partons in the final
state by $i,j,k,\dots$; generically we write $\{i;a\}$ for all partons.
When we do not make a distinction between initial- and final-state partons,
we use the common labels $I,J,\dots$ and we shortly write $\{I\}=\{i;a\}$.

The colour indices of the partons are denoted by $c_a$, $c_i$, which
range over $1,\dots,N_c^2-1$ for gluons (or any other partons, such as gluinos,
in the adjoint representation of the gauge group) and over $1,\dots,N_c$ for
quarks and antiquarks (or any other partons, such as squarks, in the 
fundamental representation).  Spin indices are generically denoted by $s_a$,
$s_i$. As in Ref.~\cite{Catani:1997vz}, 
we formally introduce an orthogonal basis of unit vectors 
$|\{c_i;c_a\}\ra\otimes|\{s_i;s_a\}\ra$ in the space of colour and spin, 
in such a way that an amplitude ${\cal M}^{\{c_i,s_i;c_a,s_a\}}(\{p_i;p_a\})$
with definite colour, spin and momenta $\{p_i,p_a\}$ can be written as 
\beq
{\cal M}^{\{c_i,s_i;c_a,s_a\}}(\{p_i;p_a\}) \equiv
\left(\prod_b \sqrt{n_c(b)}\right) \:
\Big(\la \{c_i;c_a\}| \otimes \la \{s_i;s_a\}|\Big)|\{i;a\}\ra\:.
\eeq
Thus $|\{i;a\}\ra$ is an abstract vector in colour and spin space, and
its normalization is fixed by including a factor of $1/\sqrt{n_c(b)}$
for each initial-state parton $b$ carrying $n_c(b)$ colour degrees of 
freedom.  Then the squared amplitude summed over colours and spins and
averaged over initial-state colours is
\beq
\label{eq:M2}
\frac{1}{\prod_b n_c(b)}\,
|{\cal M}(\{p_i;p_a\})|^2 = \la \{i;a\} | \{i;a\}\ra.
\eeq

Colour interactions at the QCD vertices are represented by associating
colour charges ${\bom T}_i$ or ${\bom T}_a$ with the emission of a gluon
from each parton $i$ or $a$. The colour charge ${\bom T}_i= \{T_i^n \} $
is a vector with respect to the colour indices $n$ of the emitted gluon
and an $SU(N_c)$ matrix with respect to the colour indices of the parton 
$i$; analogously ${\bom T}_a$ describes gluon emission from the initial-state
parton $a$.
More precisely, for a final-state parton $i$ the action onto the colour
space is defined by
\beq
\label{eq:colmat}
\la c_1,\dots,c_i,\dots,c_m| T_i^n | b_1,\dots,b_i,\dots,b_m \ra 
= \delta_{c_1b_1} \dots T_{c_ib_i}^n \dots \delta_{c_m b_m} \;,
\eeq
where $T_{c b}^n$ is the colour-charge matrix in the representation of
the final-state particle $i$, i.e.\ $T_{c b}^n=\ri f_{cnb}$ if $i$ is a
gluon or a gluino, $T_{\alpha\beta}^n=t_{\alpha\beta}^n$ if $i$ is a (s)quark
and $T_{\alpha\beta}^n=-t_{\beta\alpha}^n$ if $i$ is an anti(s)quark. 
The colour-charge operator of an initial-state parton $a$, for which 
a relation analogous to Eq.~(\ref{eq:colmat}) holds,
is defined by crossing symmetry, that is by
$T^n_{\alpha \beta} = - t^n_{\beta \alpha }$ if $a$ is a (s)quark  and
$T^n_{\alpha \beta} = t^n_{\alpha \beta}$ if $a$ is an anti(s)quark.
Using this notation, we also define the square of colour-correlated
tree-amplitudes with $m$ final-state partons, $|{\cal{M}}_{m}^{i,k}|^2$,
as follows 
\beeq
\label{eq:colam}
|{\cal{M}}_{m}^{j,k}|^2 \aand \equiv
\left(\prod_b n_c(b)\right)\,
{}_m\la \{i, a\}| \,{\bom T}_j \cdot {\bom T}_k \,|\{i, a\}\ra_m
\nn \\ \aand =
\left[
{\cal M}_m^{a_1 \ldots b_j  \ldots b_k  \ldots a_m} (\{p_i;p_a\})
\right]^*
\, T_{b_ja_j}^n \, T_{b_ka_k}^n
\, {\cal M}_m^{a_1 \ldots a_j  \ldots a_k  \ldots a_m} (\{p_i;p_a\}) 
\:,
\eeeq
and analogously for the cases in which $j$ and/or $k$ are replaced by
initial-state partons.

In our notation, each vector $|\{i;a\}\ra$ is a colour-singlet state, so
colour conservation is simply
\beq
\biggl(\sum_j {\bom T}_j + \sum_b {\bom T}_b \biggr) \;| \{i;a\}\ra = 0,
\eeq
where the sum over $j$ and $b$ extends over all 
the external partons of the state vector $|\{i;a\}\ra$
in the final and initial state, respectively.

The colour-charge algebra for the product 
$({\bom T}_i)^n ({\bom T}_j)^n \equiv {\bom T}_i \cdot {\bom T}_j$ is:
\beq
{\bom T}_i \cdot {\bom T}_j ={\bom T}_j \cdot {\bom T}_i \;\;\;\;{\rm if}
\;\;i \neq j; \;\;\;\;\;\;{\bom T}_i^2= C_i ,
\eeq
and analogously for initial-state partons.  Here $C_i$ is the quadratic
Casimir operator in the representation of particle $i$ and we have
$\CF= \TR(\Nc^2-1)/\Nc= (\Nc^2-1)/(2\Nc)$ in the fundamental and
$\CA=2\,\TR \Nc=\Nc$ in the adjoint representation, i.e.~we are using
the customary normalization $\TR=1/2$.

\subsection{Dimensional regularization, one-loop amplitudes and renormalization}

We employ dimensional regularization in $d=4-2\ep$ space-time
dimensions to regulate both the IR and UV
divergences.  More precisely, we use conventional dimensional
regularization (CDR), where quarks (spin-$\hf$ Dirac fermions)
possess 2 spin polarizations, gluons have 
$d-2$ helicity states and all particle momenta are taken as $d$-dimensional.
The $d$-dimensional phase space for $m$ outgoing particles with momenta
$p_1,\dots,p_m$, masses $m_1,\dots,m_m$ and total momentum $P$ is denoted by
\beq
\label{eq:psf}
\rd\phi_m(p_1,\dots,p_m;P) = 
\left[ \,\prod_{i=1}^m \frac{\rd^{d}p_i}{(2\pi)^{d-1}}
\,\delta_+(p_i^2-m_i^2) \right] 
(2\pi)^d \delta^{(d)}(p_1+\dots+p_m-P).
\eeq

The dimensional-regularization scale, which appears in the calculation
of the matrix elements, is denoted by $\mu$. The dependence on $\mu$ cancels
after having combined the matrix elements in the NLO calculation
of physical cross sections, although the latter eventually depend
on the renormalization and factorization scales $\mu_R$ and $\mu_F$. 
To avoid a cumbersome notation,
we therefore set $\mu_R=\mu$, while $\mu$ and $\mu_F$ will differ in general.

The virtual contribution $\rd\sigma^\rV$ to the NLO cross section is
proportional to the real part of the interference between the tree-level
and one-loop matrix elements. Although this interference is not positive
definite, we denote it by $| \cm(\{p_i,p_a\})|^2_{(\mathrm{1-loop})}$,
since it represents the one-loop correction to the square of the tree-level 
matrix element. Actually, $\rd\sigma^\rV$
is proportional to the renormalized one-loop correction
$| \cm(\{p_i,p_a\})|^2_{(\mathrm{1-loop})}$, which is
obtained from the corresponding bare quantity by adding ultraviolet
counterterms that implement charge (coupling) and mass renormalization.
The renormalized QCD coupling at the renormalization scale $\mu$ is
shortly denoted by $\as$. It can be defined either in the customary \msbar\ 
renormalization scheme or in the renormalization scheme of Ref.~\cite{asren},
which 
differs from the \msbar\ scheme in the treatment of heavy partons and
guarantees their decoupling 
in the infinite mass limit.
The renormalized mass parameters (i.e.\ parton masses and related parameters
such as those which appear in Yukawa couplings) are denoted by $m_i$ and can
refer either to the \msbar\ renormalization scheme or to the pole-mass
definition. The results of this paper are independent of these different
renormalization prescriptions. In particular, since these prescriptions 
coincide in the limit of vanishing parton masses, they do not affect the
massless limit of IR stable cross sections.

Once renormalization has been performed, the one-loop contribution
$| \cm |^2_{(\mathrm{1-loop})}$ no longer contains $1/\ep$
poles of UV origin, but it still contains $1/\ep$
poles and {\em finite\/} terms of IR origin. The latter depend on the 
dimensional
regularization procedure used for evaluating the loop integral.
The regularization-scheme dependence of the virtual corrections has to be 
consistently matched to that of the real ones, to guarantee the
regularization-scheme independence of the NLO cross section.
Since in this paper we deal with the real corrections by using the CDR scheme,
we need the result for $| \cm |^2_{(\mathrm{1-loop})}$ within this regularization scheme.
One-loop matrix elements are sometimes evaluated by using
dimensional-regularization prescriptions that differ from CDR (the difference
can be due to the dimensionality of the momenta of the external particles 
and/or to the number of polarizations of both external and internal particles).
If $| \cm |^2_{(\mathrm{1-loop})}$ is known in a scheme that is different from CDR,
its expression in the CDR scheme can be obtained by introducing a 
correction proportional to the corresponding tree-level amplitude.
The correction term for several different regularization schemes can be found
in Refs.~\cite{Kunszt:1994sd, Catani:1997pk} for the massless-parton case
and in Ref.~\cite{Catani:2001ef} for the general case with massless and massive
partons.


\section{Factorization in the soft and (quasi-)collinear limits}
\label{se:limits}

We consider a generic tree-level matrix element ${\cal M}_{m+1}$ with
$m+1$ massive or massless QCD partons in the final state and up to two
massless initial-state partons. At NLO, the dependence of the squared matrix 
element $|{\cal M}_{m+1}|^2$ on the momenta of the final-state partons 
is singular in two different situations. The first
situation occurs in the {\it soft\/} region, where the momentum of a
final-state gluon tends to zero in any fixed direction. The second
situation corresponds to the {\it (quasi-)collinear\/} region, where a
final-state parton becomes collinear to another (initial- or final-state)
parton.

\subsection{Soft limit}

In the soft region the momentum $p_j$ of a final-state gluon $j$ tends
to zero in any fixed direction $q$, i.e.\ $p_j=\lambda q$ with
$\lambda\to 0$ for fixed $q$. In this limit, the squared matrix element  
diverges as $1/\lambda^2$ and its divergent part can be computed in terms of
the eikonal current of the soft gluon \cite{factform}.
As described in Ref.~\cite{Catani:1997vz} in more
detail, using partial fractioning and colour conservation,
the singular behaviour of $|{\cal M}_{m+1}|^2$ can
be written as a sum over emitter--spectator ($I$--$K$) pairs,
\beeq
\label{eq:slim}
\lefteqn{
{}_{m+1,a\dots}\langle \dots,j,\dots;a,\dots|
|\dots,j,\dots;a,\dots\rangle_{m+1,a\dots} \;\asymp{
 \lambda\to 0}
} 
\\*
&&-\frac{8\pi\mu^{2\eps}\alps}{\lambda^2}
\sum_{I \ne j} \frac{1}{p_I q} \sum_{K\ne j,I}
{}_{m,a\dots}\langle \dots;a,\dots|
{\bom T}_I\cdot{\bom T}_K \left[ \frac{p_I p_K}{(p_I+p_K)q}-\frac{m_I^2}{2p_I q} \right]
|\dots;a,\dots\rangle_{m,a\dots} \;, \nn
\eeeq
where $|\dots\rangle_{m,a\dots}$ corresponds to the $m$-parton matrix element
that results from the original $(m+1)$-parton matrix element upon omitting the
gluon $j$, and the indices $I,K$ label both initial- and final-state 
partons.
This formula is valid in the general case of massive or massless partons.

\subsection{Collinear and quasi-collinear limits}
\label{se:cqclim} 

In the massless case, the squared matrix element 
$|{\cal M}_{m+1}|^2$ diverges when two external partons become collinear
\cite{Altarelli:1977zs}.
If at least one of the external partons is massive, the collinear divergence
is screened by the finite value of the parton mass. Nonetheless, the cross
section contribution of the matrix element from this phase-space region is 
strongly (logarithmically) enhanced when the parton mass becomes small.
As discussed in Sect.~\ref{se:subnh}, to control these enhanced contributions
we have to study the behaviour of the matrix element in the 
quasi-collinear limit \cite{Catani:2001ef, Keller:1999tf}.
Since the quasi-collinear limit is a generalization of the customary
collinear limit, the latter can in turn be obtained as a special case
from the former.

We consider two final-state partons $i$ and $j$ that can be produced
by a QCD vertex through the splitting process $\widetilde{ij}\to i+j$.
The partons $i$ and $j$ have momenta $p_i$ and $p_j$ and masses $m_i$ and
$m_j$ ($p_i^2=m_i^2, p_j^2=m_j^2$). The mass of the parton $\widetilde{ij}$
is denoted by $m_{ij}$ and is constrained by  $m_{ij} \leq m_i + m_j$.
This constraint forbids the on-shell decay of the parton $\widetilde{ij}$. 
We introduce the following Sudakov parametrization of the parton momenta:
\beq
p_i^\mu = z p^\mu + k_\perp^\mu 
- \frac{k_\perp^2+ z^2 m_{ij}^2- m_i^2}{z}\frac{n^\mu}{2pn}\:, \quad
p_j^\mu = (1-z) p^\mu - k_\perp^\mu 
- \frac{k_\perp^2+(1-z)^2m_{ij}^2 - m_j^2}{1-z}\frac{n^\mu}{2pn}\:,
\label{eq:qclimgen}
\eeq
where $p^\mu$ is a time-like momentum (with $p^2=m_{ij}^2$) that 
points towards the collinear (forward) direction, 
$n^\mu$ is an auxiliary light-like vector ($n^2=0$)
and $k_\perp$ is the momentum component that is 
orthogonal to both $p$ and $n$ ($pk_\perp=nk_\perp=0$). The invariant mass
of the final-state partons is
\beq
\label{eq:scpro}
(p_i + p_j)^2 = -\frac{k_\perp^2}{z(1-z)}
+\frac{m_i^2}{z} +\frac{m_j^2}{1-z}\:.
\eeq

The quasi-collinear region is reached when $k_\perp$ becomes of ${\rm O}(m)$ 
and small. This region can be identified by performing the {\em uniform\/}
rescaling
\beq
k_\perp \to \lambda k_\perp, \quad m_i\to \lambda m_i, 
\quad m_j\to \lambda m_j, \quad m_{ij}\to \lambda m_{ij}
\eeq
and studying the limit $\lambda \to 0$. Neglecting terms that are less singular
than $1/\lambda^2$ the squared matrix element $|{\cal M}_{m+1}|^2$ behaves 
as~\cite{Catani:2001ef}
\beeq
\lefteqn{
{}_{m+1,a\dots}\langle \dots,i,j,\dots;a,\dots|
|\dots,i,j,\dots;a,\dots\rangle_{m+1,a\dots}
\;\;   \asymp{
\lambda\to 0} 
}
\nn\\
\label{eq:colllim}
&&\!\!
\frac{1}{\lambda^2} \frac{8\pi\mu^{2\eps}\alps}{(p_i+p_j)^2-m_{ij}^2} \:
{}_{m,a\dots}\langle \dots,\widetilde{ij},\dots;a,\dots|
\hat P_{\widetilde{ij},i}(z,k_\perp,\{m\};\eps)
|\dots,\widetilde{ij},\dots;a,\dots\rangle_{m,a\dots},
\hspace{2em}
\eeeq
where $|\dots\rangle_{m,a\ldots}$ corresponds to the $m$-parton matrix
element that
is obtained from the $(m+1)$-parton matrix element by replacing the parton
pair $i,j$ by the single parton $\widetilde{ij}$.

The kernel $\hat P_{\widetilde{ij},i}(z,k_\perp,\{m\};\eps)$ in
Eq.~(\ref{eq:colllim}) is the generalization of the $d$-dimensional
Altarelli--Parisi splitting function \cite{Altarelli:1977zs}
from the collinear to the quasi-collinear limits. As in the collinear case,
it depends on the momentum fraction $z$, on the transverse momentum
$k_\perp$ and on the helicity of the parton $\widetilde{ij}$ in the $m$-parton
matrix element. In the quasi-collinear case, it also depends on the masses
$m_i, m_j, m_{ij}$ of the partons involved in the splitting process.
The mass dependence is shortly denoted by $\{m\}$.

The generalized Altarelli--Parisi kernels for the QCD splitting processes 
$Q \to Q+g$, $Q \to g +Q$ and $g \to Q +{\bar Q}$,
are
\beeq
\langle s| 
\hat P_{QQ}(z,k_\perp,m_Q;\eps) |s'\rangle
\aand = \delta_{ss'} \;\CF
\left[ \frac{1+z^2}{1-z} - \eps(1-z) - \frac{m_Q^2}{p_Q p_g} \right],
\\ 
\langle s| 
\hat P_{Qg}(z,k_\perp,m_Q;\eps) |s'\rangle
\aand = 
\delta_{ss'} \;\CF
\left[ \frac{1+(1-z)^2}{z} - \eps z - \frac{m_Q^2}{p_g p_Q} \right],
\\ 
\langle\mu| 
\hat P_{gQ}(z,k_\perp,m_Q;\eps) |\nu\rangle
\aand = \TR \left[ -g^{\mu\nu} - 
4 \frac{k_\perp^\mu k_\perp^\nu}{(p_Q+p_{\bar Q})^2} \right],
\eeeq
where $s,s'$ and $\mu,\nu$ denote the indices of the spin correlations
for the parton $\widetilde{ij}$, and the $k_\perp$ and mass dependence
of the scalar products $p_ip_j$ is given in Eq.~(\ref{eq:scpro}).
Of course, the $g\to gg$ splitting function
remains unaffected by the inclusion of mass terms,
\beq
\langle\mu| \hat P_{gg}(z,k_\perp;\eps) |\nu\rangle
= 2\CA \left[ -g^{\mu\nu}\left(\frac{z}{1-z}+\frac{1-z}{z}\right)
-2(1-\eps)z(1-z)\frac{k_\perp^\mu k_\perp^\nu}{k_\perp^2} \right].
\eeq
We also need the spin-averaged form of the splitting functions.
For quarks this is obtained by contraction with $\delta_{ss'}/2$ and
for a gluon with on-shell momentum $p$, by contraction with
\beq
\frac{1}{d-2}d_{\mu\nu}(p)= \frac{1}{2(1-\eps)}
\left[-g_{\mu\nu}+(\mbox{gauge terms})\right],
\eeq
where the gauge terms are proportional either to $p^\mu$ or to $p^\nu$ and
\beq
-g^{\mu\nu}d_{\mu\nu}(p)=d-2, \qquad
p^\mu d_{\mu\nu}(p)=0.
\eeq
The spin-averaged QCD splitting functions only depend on $z$, $\eps$ and 
the ratio
\beq
\mu_{ij}^2 = \frac{m_i^2+m_j^2}{(p_i+p_j)^2-m_{ij}^2} \;, 
\eeq
and are denoted by 
$\langle\hat P_{\widetilde{ij},i}(z;\eps;\mu_{ij}^2)\rangle$. 
They explicitly read
\beeq
\label{avhpqq}
\langle\hat P_{QQ}(z;\eps;\mu_{Qg}^2)\rangle
\aand = \CF
\left[ \frac{1+z^2}{1-z} - \eps(1-z) - 2\mu_{Qg}^2  \right],
\\ 
\langle\hat P_{Qg}(z;\eps;\mu_{gQ}^2)\rangle
\aand = 
\CF
\left[ \frac{1+(1-z)^2}{z} - \eps z - 2\mu_{gQ}^2  \right],
\\ 
\langle\hat P_{gQ}(z;\eps)\rangle
\aand = \TR \left[ 1 - \frac{2z(1-z)-\mu_{Q{\bar Q}}^2}{1-\eps} \right],
\\
\label{avhpgg}
\langle\hat P_{gg}(z;\eps)\rangle
\aand = 2\CA \left[ \frac{z}{1-z} + \frac{1-z}{z} + z(1-z) \right].
\eeeq
These generalized, spin-averaged splitting functions and some of their
counterparts in SUSY QCD have already been given in
Ref.~\cite{Catani:2001ef}. The four-dimensional QED analogue of these
functions describes the splitting processes $e^\pm \to e^\pm +\gamma$, 
$\gamma \to e^+ + e^-$ and is well known \cite{Baier:1973ms,Keller:1999tf,
Dittmaier:2000mb}.

Setting $m_i=m_j=m_{ij}=0$ in all the expressions of this subsection,
we recover the known expressions for the collinear limit (see e.g.\ 
Ref.~\cite{Catani:1997vz}).

The  matrix element ${\cal M}_{m+1,a\dots}$ diverges also when a massless
final-state parton $i$ becomes collinear to a massless initial-state
parton $a$. The singularity is related to the splitting $a\to
\widetilde{ai}+i$, where $\widetilde{ai}$ is the incoming parton of the
related $m$-parton process. Since in this paper we restrict ourselves 
to massless incoming partons, none of the partons $a$, $i$,
$\widetilde{ai}$ is massive and we can completely take over the
factorization for massless partons described in Ref.~\cite{Catani:1997vz}.
There, the collinear limit is defined by $k_\perp\to 0$ in
\beq
p_i^\mu = (1-x)p_a^\mu+k_\perp^\mu
-\frac{k_\perp^2}{1-x}\frac{n^\mu}{2p_an}, \qquad
p_a p_i = -\frac{k_\perp^2}{2(1-x)}.
\eeq
In this limit $|{\cal M}_{m+1}|^2$ behaves as
\beeq
\lefteqn{
{}_{m+1,a\dots}\langle \dots,i,\dots;a,\dots|
|\dots,i,\dots;a,\dots\rangle_{m+1,a\dots}
}
\nn\\
\label{eq:colis}
& \asymp{
k_\perp \to 0} \; &
\frac{4\pi\mu^{2\eps}\alps}{x(p_a p_i)} \;
{}_{m,\widetilde{ai}\dots}\langle \dots;\widetilde{ai},\dots|
\hat P_{a,\widetilde{ai}}(x,k_\perp;\eps)
|\dots;\widetilde{ai},\dots\rangle_{m,\widetilde{ai}\dots},
\eeeq
where $|\dots\rangle_{m,\widetilde{ai}\ldots}$ corresponds to the
$m$-parton matrix element that
is obtained from the $(m+1)$-parton matrix element by replacing the parton
pair $a,i$ by the single incoming parton $\widetilde{ai}$, which has the
reduced momentum $x p_a$ with $0<x<1$.

%
%
%

\section{Dipole factorization formulae}
\label{se:dipoles}

In the actual calculation of cross sections, Eqs.~(\ref{eq:slim}),
(\ref{eq:colllim}) and (\ref{eq:colis}) cannot straightforwardly be used
as `true' factorization formulae, because the momenta of the partons in
the matrix elements on the right-hand sides are unambiguously defined
only in the strict soft and (quasi-)collinear limits.  In this section
we define dipole factorization formulae that, besides having the
correct limiting behaviour, implement momentum conservation away from
the limits. Momentum conservation is implemented not only exactly but
also in a factorizable way.  The dipole factorization formulae
can thus be used to introduce the dipole factors $\rd V_{\mathrm{dipole}}$
that are needed to construct (see Eq.~(\ref{eq:genaux})) the auxiliary
cross section $\rd\sigma^\rA$ and its related integrated counterpart in 
Eq.~(\ref{eq:sigmaNLOsub}). 

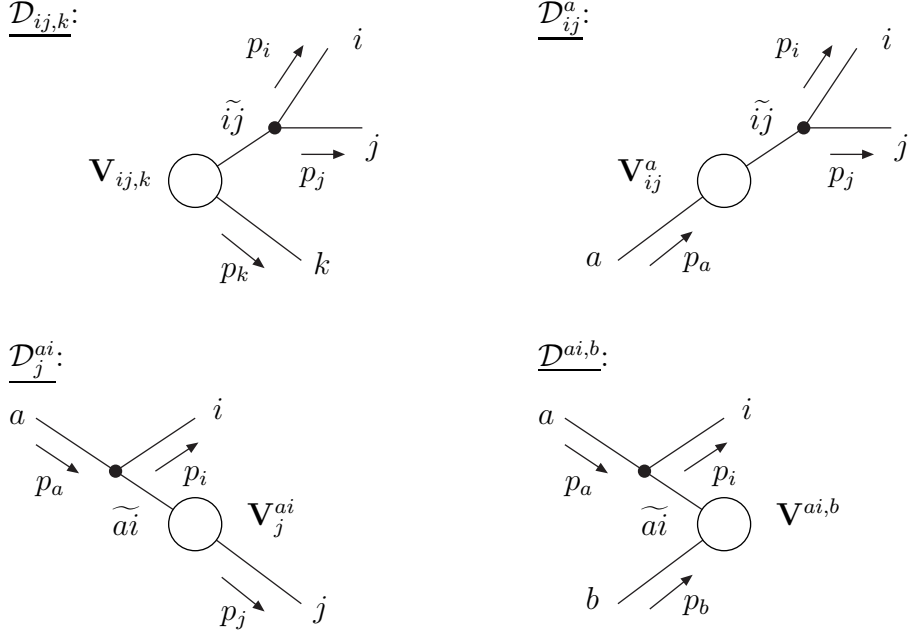
\begin{figure}
\centerline{
\begin{picture}(360,230)(0,0)
\put(0,120){
  \begin{picture}(160,120)(0,0)
  \Line(80,50)(110, 70)
  \Line(110, 70)(130,100)
  \Line(80,50)(120, 20)
  \LongArrow( 90,30)(105, 18)
  \LongArrow(110,85)(120, 100)
  \LongArrow(120,60)(135, 60)
  \Line(110,70)(143,70)
  \Vertex(110,70){2.5}
  \GCirc(80,50){10}{1}
  \put( 90, 70){$\widetilde{ij}$}
  \put(140, 100){$i$}
  \put(145, 60){$j$}
  \put(125, 15){$k$}
  \put( 10,110){\underline{$\cD_{ij,k}$}:}
  \put( 40, 50){$\bV_{ij,k}$}
  \put( 90, 14){$p_k$}
  \put(100, 99){$p_i$}
  \put(120, 50){$p_j$}
  \end{picture} }
\put(200,120){
  \begin{picture}(160,120)(0,0)
  \Line(80,50)(110,70)
  \Line(110,70)(130,100)
  \Line(80,50)( 40, 20)
  \LongArrow( 52,18)( 67, 30)
  \LongArrow(110,85)(120, 100)
  \LongArrow(120,60)(135, 60)
  \Line(110,70)(143,70)
  \Vertex(110,70){2.5}
  \GCirc(80,50){10}{1}
  \put( 90, 70){$\widetilde{ij}$}
  \put(140, 100){$i$}
  \put(145, 60){$j$}  
  \put( 28, 18){$a$}
  \put( 10,110){\underline{$\cD_{ij}^a$}:}
  \put( 40, 50){$\bV_{ij}^a$}
  \put( 65, 18){$p_a$}
  \put(100, 99){$p_i$}
  \put(120, 50){$p_j$}
  \end{picture} }
\put(0,-10){
  \begin{picture}(160,120)(0,0)
  \Line(80,50)( 20, 90)
  \Line(80,50)(120, 20)
  \LongArrow( 90,30)(105, 18)
  \LongArrow( 65,70)( 80, 80)
  \LongArrow( 20,80)( 35, 70)
  \Line( 50,70)( 80,90)
  \Vertex( 50,70){2.5}
  \GCirc(80,50){10}{1}
  \put( 49, 46){$\widetilde{ai}$}
  \put( 10, 88){$a$}
  \put( 87, 90){$i$}
  \put(125, 15){$j$}
  \put( 10,110){\underline{$\cD_j^{ai}$}:}
  \put(100, 50){$\bV_j^{ai}$}
  \put( 90, 14){$p_j$}
  \put( 20, 63){$p_a$}
  \put( 76, 67){$p_i$}
  \end{picture} }
\put(200,-10){
  \begin{picture}(160,120)(0,0)
  \Line(80,50)( 20, 90)
  \Line(80,50)( 40, 20)
  \LongArrow( 52,18)( 67, 30)
  \LongArrow( 65,70)( 80, 80)
  \LongArrow( 20,80)( 35, 70)
  \Line( 50,70)( 80,90)
  \Vertex( 50,70){2.5}
  \GCirc(80,50){10}{1}
  \put( 49, 46){$\widetilde{ai}$}
  \put( 10, 88){$a$}
  \put( 87, 90){$i$}
  \put( 28, 18){$b$}
  \put( 10,110){\underline{$\cD^{ai,b}$}:}
  \put(100, 50){$\bV^{ai,b}$}
  \put( 65, 18){$p_b$}
  \put( 20, 63){$p_a$}
  \put( 76, 67){$p_i$}
  \end{picture} }
\end{picture} } 
\caption{Effective diagrams for the different emitter--spectator cases.}
\label{fig:effdiags}
\end{figure}

We present the dipole factorization formulae as in Ref.~\cite{Catani:1997vz}:
\beq
\label{eq:dff}
|{\cal M}_{m+1}|^2 =
\sum_{i,j}\sum_{k\ne i,j} \cD_{ij,k} +
\sum_{i,j}\sum_a \cD_{ij}^a +
\sum_{a,i}\sum_{j\ne i} \cD_j^{ai} +
\sum_{a,i}\sum_{b\ne a} \cD^{ai,b} + \dots \:,
\eeq
where the terms on the right-hand side approximate the matrix element
in different singular regions and the dots stand for terms that lead to
finite integrals.  Four different situations can be distinguished for
the emitter--spectator pairs (see Fig.~\ref{fig:effdiags}),
since either one can belong to the final or initial states.
The first two terms on the right-hand side control the singularities of
the $(m+1)$-parton matrix element when two final-state partons $i$ and
$j$ become \mbox{(quasi-)}collinear: the emitter is a final-state
parton and the spectator can be either in the final ($\cD_{ij,k}$) or
in the initial ($\cD_{ij}^a$) states.
The third and fourth terms on the right-hand side
control the singularities of the $(m+1)$-parton
matrix element when a final-state parton $i$ and an initial-state parton $a$
become collinear: the emitter is an initial-state parton and the spectator
can be either in the final ($\cD_j^{ai}$) or in the initial ($\cD^{ai,b}$)
states. When the parton $i$ is soft, all the four dipole functions $\cD$
become singular.  These four types of dipoles are considered in the
following subsections.

The dipole factors introduced in Ref.~\cite{Phaf:2001gc} differ from ours
in several respects. For instance, they do not deal with the case of
massive partons with unequal masses and, typically, they treat the
kinematic recoil differently.  The main overall differences arise from
the fact that they are not aimed to control the quasi-collinear region.

\subsection{Final-state emitter and final-state spectator}

The dipole contribution $\cD_{ij,k}$ (see Fig.~\ref{fig:effdiags}) to the 
factorization formula (\ref{eq:dff}) is
\beeq
\label{eq:Dijk}
&&
\cD_{ij,k}(p_1,\dots,p_{m+1}) =
\nn\\ && \qquad
-\frac{1}{(p_i+p_j)^2-m_{ij}^2} \;
{}_m\langle\dots,\widetilde{ij},\dots,\tilde k,\dots|
\frac{{\bom T}_k\cdot{\bom T}_{ij}}{{\bom T}_{ij}^2} \bV_{ij,k}
|\dots,\widetilde{ij},\dots,\tilde k,\dots\rangle_m\:.\qquad
\eeeq   
The final-state parton momenta $p_i$, $p_j$ and $p_k$ have arbitrary
masses and their total outgoing momentum is denoted by $Q$,
\beq
\label{eq:Qfff}
p_i^2 = m_i^2\:, \quad
p_j^2 = m_j^2\:, \quad
p_k^2 = m_k^2\:, \qquad\qquad
Q=p_i+p_j+p_k\:.
\eeq
The $m$-parton matrix elements in Eq.~(\ref{eq:Dijk}) are obtained from the
original $(m+1)$-parton matrix element by replacing $i$ and $j$ with
the parent parton $\widetilde{ij}$ in the splitting process 
$\widetilde{ij}\to i+j$ and by changing the momentum of the spectator
parton $k$. The on-shell mass of the emitter $\widetilde{ij}$ is $m_{ij}$
and we only consider situations with $m_{ij}\le m_i+m_j$. 
In Eq.~(\ref{eq:Dijk}) we have split off the colour structure from the spin
functions $\bV_{ij,k}$, which are given in Sect.~\ref{subsec:ffdipoles}.

\subsubsection{Kinematics and phase-space factorization}
\label{se:ffkin}

The auxiliary momenta $\tpij$ and $\tpk$ of the emitter and spectator
are defined in such a way that
$\tpij$ plays the role of $p$ in the quasi-collinear limit
(\ref{eq:qclimgen}), but at the same time they obey their mass-shell
conditions and total momentum conservation,
\beq
\tpij^2 = m_{ij}^2\:, \quad 
\tpk^2 = m_k^2\:,     \qquad\qquad
Q^\mu = \tpij^\mu+\tpk^\mu\:. 
\eeq
 
For later use, we introduce the rescaled parton masses $\mu_n$ and the
relative velocities $v_{p,q}$ between two massive momenta
$p^\mu$ and $q^\mu$,
\beq
\mu_n = \frac{m_n}{\sqrt{Q^2}} \quad (n=i,j,k,ij), \qquad\qquad
v_{p,q} = \sqrt{1-\frac{p^2 q^2}{(pq)^2}}.
\eeq
The relative velocities can also be written as
\beq
v_{p,q} = \frac{\sqrt{\lambda((p+q)^2,p^2,q^2)}}{(p+q)^2-p^2-q^2}
\eeq
in terms of the customary triangular function $\lambda(x,y,z)$,
\beq
\lambda(x,y,z) = x^2+y^2+z^2-2xy-2xz-2yz\:.
\eeq
For instance, the velocity $\tvijk$ between $\tpij$ and $\tpk$ is given by
\beq
\tvijk = \frac{\sqrt{\lambda(1,\mu_{ij}^2,\mu_k^2)}}{1-\mu_{ij}^2-\mu_k^2}\:.
\label{eq:tvijk}
\eeq

Our definition of the momenta $\tpij$ and $\tpk$ in terms of the original
momenta $p_i$, $p_j$ and $p_k$ (or $Q=p_i+p_j+p_k$) is:
\beeq
\tpk^\mu \aand =
\frac{\sqrt{\lambda(Q^2,m_{ij}^2,m_k^2)}}
{\sqrt{\lambda(Q^2,(p_i+p_j)^2,m_k^2)}}
\left( p_k^\mu-\frac{Q p_k}{Q^2}Q^\mu \right)
+\frac{Q^2+m_k^2-m_{ij}^2}{2Q^2} Q^\mu\:,
\nn\\[.5em]
\label{eq:fsm}
\tpij^\mu \aand = Q^\mu-\tpk^\mu\:,
\eeeq
and it coincides with the one given in Ref.~\cite{Dittmaier:2000mb}
for the specific case with $m_i=m_{ij}$ considered there.
The definition in Eq.~(\ref{eq:fsm}) is symmetric with respect to $i
\leftrightarrow j$ and is uniformly applicable to any configuration of the 
parton masses $m_n$ $(n=i,j,k,ij)$. 
Different definitions of the momenta $\tpij$ and $\tpk$
can be introduced. The authors of Ref.~\cite{Phaf:2001gc} consider only
the specific cases with $\{m_i=m_j=m_{ij}=0, m_k\neq 0\}$,
$\{m_i=m_{ij}, m_k=m_j=0\}$ and $\{m_i=m_{ij}=m_k, m_j=0\}$: in the last two
cases their definition coincides with ours.

Next we exactly factorize the three-particle phase space 
$\rd\phi(p_i,p_j,p_k;Q)$ of Eq.~(\ref{eq:psf})
in terms of the two-particle phase space $\rd\phi(\tpij,\tpk;Q)$ and a
single-particle phase-space factor $[\rd p_i(\tpij,\tpk)]$,
\beq
\label{eq:psfact}
\rd\phi(p_i,p_j,p_k;Q) =
\rd\phi(\tpij,\tpk;Q) \;
[\rd p_i(\tpij,\tpk)] \;\Theta(1-\mu_i-\mu_j-\mu_k)\:.
\eeq
In App.~B of Ref.~\cite{Dittmaier:2000mb} the derivation of this phase-space
splitting was outlined for the case with $m_i=m_{ij}$ in four space-time dimensions.
The generalization to $m_i\ne m_{ij}$ in $d$ dimensions is
straightforward; we present only the results.  We obtain
\beeq
\int [\rd p_i(\tpij,\tpk)] \aand =
\frac{1}{4} (2\pi)^{-3+2\eps} (Q^2)^{1-\eps}
(1-\mu_i^2-\mu_j^2-\mu_k^2)^{2-2\eps}
\left[ \lambda(1,\mu_{ij}^2,\mu_k^2) \right]^{\frac{-1+2\eps}{2}}
\int\rd^{d-3}\Omega
\nn\\[.5em]
&& {}\times
\int_{y_-}^{y_+}\rd\yijk\, (1-\yijk)^{1-2\eps}
\left[\mu_i^2+\mu_j^2+(1-\mu_i^2-\mu_j^2-\mu_k^2)\yijk\right]^{-\eps}
\nn\\[.5em]
&& {}\times
\int_{z_-(\yijk)}^{z_+(\yijk)}\rd\zi\,
\left[z_+(\yijk)-\zi\right]^{-\eps} \left[\zi-z_-(\yijk)\right]^{-\eps}\:,
\label{eq:psff}
\eeeq
where the variables $\zi$ and $\yijk$ 
are defined as in Refs.~\cite{Catani:1997vz,Dittmaier:2000mb}:
\beq
\zi = 1-\zj = \frac{p_i p_k}{p_i p_k + p_j p_k}\:, \qquad
\yijk = \frac{p_i p_j}{p_i p_j + p_i p_k + p_j p_k} \;.
\label{eq:ziyijk}
\eeq
Their integration boundary is given by
\beeq
y_- \aand = \frac{2\mu_i\mu_j}{1-\mu_i^2-\mu_j^2-\mu_k^2}\:, \qquad
y_+ = 1-\frac{2\mu_k(1-\mu_k)}{1-\mu_i^2-\mu_j^2-\mu_k^2}\:,
\nn\\[.5em]
z_\pm(\yijk) \aand = \frac{2\mu_i^2+(1-\mu_i^2-\mu_j^2-\mu_k^2)\yijk}
{2[\mu_i^2+\mu_j^2+(1-\mu_i^2-\mu_j^2-\mu_k^2)\yijk]} (1\pm\viji\vijk) \;,
\label{eq:yzbound}
\eeeq
where $\viji$ ($\vijk$) is the relative velocity between 
$p_i+p_j$ and $p_i$ ($p_k$).
The relative velocities are functions of $\yijk$ and explicitly read
\beeq
\vijk \aand = 
\frac{\sqrt{[2\mu_k^2+(1-\mu_i^2-\mu_j^2-\mu_k^2)(1-\yijk)]^2-4\mu_k^2}}
{(1-\mu_i^2-\mu_j^2-\mu_k^2)(1-\yijk)},
\nn\\[.5em]
\viji \aand = 
\frac{\sqrt{(1-\mu_i^2-\mu_j^2-\mu_k^2)^2\yijk^2-4\mu_i^2\mu_j^2}}
{(1-\mu_i^2-\mu_j^2-\mu_k^2)\yijk+2\mu_i^2}.
\label{eq:vijk}
\eeeq
The $\rd^{d-3}\Omega$ integration extends over the solid angle
perpendicular to $\tpij$ and $\tpk$ and thus
\beq
\int\rd^{d-3}\Omega = \frac{2\pi}{\pi^\eps\Gamma(1-\eps)}.
\label{eq:dOmega}
\eeq                                              

\subsubsection{The dipole splitting functions}
\label{subsec:ffdipoles}

We give the functions $\bV_{ij,k}$ in Eq.~(\ref{eq:Dijk}) for the
three QCD splitting processes $\widetilde{ij}\to i+j$:
\begin{itemize}
\item
$Q\to g(p_i) + Q(p_j)$: \quad $m_i=0$ and $m_j=m_{ij}=m_Q$,
\item
$g\to Q(p_i) + \bar Q(p_j)$: \quad $m_i=m_j=m_Q$ and $m_{ij}=0$,
\item
$g\to g(p_i) +g(p_j)$: \quad $m_i=m_j=m_{ij}=0$. 
\end{itemize}
The case $\bar Q\to g\bar Q$ is formally identical to $Q\to gQ$.  The
spectator mass $m_k$ may be zero or non-zero in all cases.  Note that
in general $\bV_{ij,k}$ is non-diagonal in the helicity space of
the parton $\widetilde{ij}$. For the analytical integration over the
singular degrees of freedom (see Sect.~\ref{subsub:FFinteg})
we also need the spin-averaged functions
$\langle\bV_{ij,k}\rangle$.  Denoting $s,s'$ or $\mu,\nu$ the
helicities of $\widetilde{ij}=Q$ or $\widetilde{ij}=g$ in
$\langle\dots,\widetilde{ij},\dots||\dots,\widetilde{ij},\dots\rangle$,
we define the dipole functions $\bV_{ij,k}$ as follows:
\beeq
\label{eq:V_gQk}
\langle s|\bV_{g Q,k}|s'\rangle \aand =
8\pi\mu^{2\eps}\alps\CF \left\{
\frac{2}{1-\zj(1-\yijk)}
-\frac{\tvijk}{\vijk}\left[1+\zj+\frac{m_Q^2}{p_i p_j}+\eps(1-\zj)\right]
\right\} \delta_{ss'}
\nn\\[.5em]
\aand = \langle\bV_{gQ,k}\rangle \delta_{ss'}\:,
\\[.5em]
\langle\mu|\bV_{Q\bar Q,k}|\nu\rangle \aand =
8\pi\mu^{2\eps}\alps\TR \frac{1}{\vijk}
\left\{ -g^{\mu\nu}\left[1-\frac{2\kappa}{1-\eps}
  \left(z_+z_- -\frac{m_Q^2}{(p_i+p_j)^2}\right) \right] \right.
\nn\\*
&& \qquad\qquad\qquad\qquad \left. {}
-\frac{4}{(p_i+p_j)^2}
\left[\zi^{(m)}p_i^\mu-\zj^{(m)}p_j^\mu\right]
\left[\zi^{(m)}p_i^\nu-\zj^{(m)}p_j^\nu\right]
\right\}\:,
\label{eq:V_QQk}
\\[.5em]
\langle\bV_{Q\bar Q,k}\rangle \aand =
8\pi\mu^{2\eps}\alps\TR \frac{1}{\vijk}
\Biggl\{ 1-\frac{2}{1-\eps}
\Biggl[ \zi(1-\zi)-(1-\kappa)z_+z_-
\nn\\*
&& \qquad\qquad\qquad\qquad {}
-\frac{\kappa\mu_Q^2}{2\mu_Q^2+(1-2\mu_Q^2-\mu_k^2)\yijk} \Biggr]
\Biggr\}\:,
\\[.5em]    
\langle\mu|\bV_{gg,k}|\nu\rangle \aand =
16\pi\mu^{2\eps}\alps\CA
\left\{ -g^{\mu\nu}\left[\frac{1}{1-\zi(1-\yijk)}
+\frac{1}{1-\zj(1-\yijk)}-\frac{2-\kappa z_+z_-}{\vijk} \right]
\right.
\nn\\
&& \qquad\qquad\qquad\quad \left. {}
+\frac{1}{\vijk}\frac{1-\eps}{p_i p_j}
\Big[\zi^{(m)}p_i^\mu-\zj^{(m)}p_j^\mu\Big]
\Big[\zi^{(m)}p_i^\nu-\zj^{(m)}p_j^\nu\Big]
\right\}\:,
\label{eq:V_ggk}
\\[.5em]      
\langle\bV_{gg,k}\rangle \aand =
16\pi\mu^{2\eps}\alps\CA
\left\{ \frac{1}{1-\zi(1-\yijk)}+\frac{1}{1-\zj(1-\yijk)}
\right.
\nn\\
&& \qquad\qquad\qquad\quad \left. {}
+\frac{\zi(1-\zi)-(1-\kappa)z_+z_- -2}{\vijk}
\right\}\:,
\eeeq
where $z_\pm$ is given in Eq.~(\ref{eq:yzbound}) and the new variables 
$\zi^{(m)}$ and $\zj^{(m)}$ are
\beq
\zi^{(m)} = \zi - \frac{1}{2}(1-\vijk)\:, \quad
\zj^{(m)} = \zj - \frac{1}{2}(1-\vijk)\:.
\eeq
The constant $\kappa$ is a free parameter, which only redistributes
non-singular contributions between the different terms within square
brackets in Eq.~(\ref{eq:sigmaNLOsub}).  Choosing different values of
$\kappa$, we can simplify the expressions of either the subtraction
term ($\kappa = 0$) or its integral ($\kappa = 2/3$).
Of course, the final result for $\sigma^\aNLO$ must not depend on
$\kappa$, i.e.\ its $\kappa$-independence can serve as a consistency
check of the calculation. The velocity factors ${\tvijk}/{\vijk}$ and 
$1/{\vijk}$ are also harmless in the singular limits; they are introduced
in the definition of the functions $\bV_{ij,k}$ only to simplify their 
integration in analytic form. The definition of the four-dimensional 
$\bV_{gQ,k}$ in Ref.~\cite{Dittmaier:2000mb} differs from ours by some
harmless velocity factors. 

In App.~\ref{app:SUSYdipoles}, we give the dipole splitting functions 
$\bV_{ij,k}$ that are relevant for SUSY QCD calculations.

\subsubsection{The integrated dipole functions}
\label{subsub:FFinteg}

The spin-correlation terms of the splitting functions in Eqs.~(\ref{eq:V_QQk})
and (\ref{eq:V_ggk}) have been defined such that 
$\left[\zi^{(m)}p_i^\mu-\zj^{(m)}p_j^\mu\right]\tilde p_{ij,\mu}=0$,
when $m_i=m_j, m_{ij}=0$. This property ensures \cite{Catani:1997vz} that
the spin correlations vanishes after azimuthal integration. In the evaluation
of the integrals of the dipole factors over the phase space
$[\rd p_i(\tpij,\tpk)]$, we can thus replace the spin matrices
$\bV_{ij,k}$ with their spin averages $\langle\bV_{ij,k}\rangle$. We define
\beq
\label{eq:Iijk_def}
\int [\rd p_i(\tpij,\tpk)] \,
\frac{1}{(p_i+p_j)^2-m_{ij}^2} \, \langle\bV_{ij,k}\rangle
\,\equiv\, \frac{\alps}{2\pi}\frac{1}{\Gamma(1-\eps)}
\biggl(\frac{4\pi\mu^2}{Q^2}\biggr)^\eps I_{ij,k}(\eps)\:,
\eeq  
where $I_{ij,k}(\eps)$ depends also on the parton masses and
$\tpij \cdot \tpk$.  In the massless case, the integrals
$I_{ij,k}(\eps)$ do not depend on $\tpij \cdot \tpk$ and become the
functions ${\cal V}_{ij}(\ep)$ of Ref.~\cite{Catani:1997vz}.
For non-vanishing values of the parton masses, these integrals cannot be
exactly performed in $d$ dimensions in terms of simple functions. 
We evaluate them by neglecting corrections of O$(\ep)$. Note, however,
that we want to compute them for arbitrary (finite or vanishing)
masses, so we do not want to spoil the commutativity of the massless
limit with the limit $\ep \to 0$.  We thus perform the $\ep$ expansion
{\em uniformly\/} in the parton masses, i.e.\  the coefficients of the
O$(\ep)$ corrections that we neglect are not singular when one or more
parton masses vanish. For instance, we do not expand in $\ep$ terms
of the form $\mu_j^{-2\eps}$, whose limits $\mu_j\to 0$ and $\eps\to 0$
do not commute.

We decompose the integrals $I_{ij,k}(\eps)$ into an {\it eikonal\/}
part, $I^{\eik}(\eps)$, which contains the soft integrals, and a
remaining {\it collinear\/} part, $I^{\coll}_{ij,k}(\eps)$:
\beeq
\label{eq:I_gQk}
I_{gQ,k}(\mu_Q,\mu_k;\eps) \aand =
\CF\left[ 2I^{\eik}(\mu_Q,\mu_k;\eps)
        + I^{\coll}_{gQ,k}(\mu_Q,\mu_k;\eps) \right]\:,
\\
I_{Q\bar Q,k}(\mu_Q,\mu_k;\eps) \aand =
\TR\,I^{\coll}_{Q\bar Q,k}(\mu_Q,\mu_k;\eps)\:,
\\
I_{gg,k}(\mu_k;\eps) \aand =
2\CA\left[ 2I^{\eik}(0,\mu_k;\eps)
         + I^{\coll}_{gg,k}(\mu_k;\eps) \right]\:.
\eeeq

The eikonal integral is defined by
\beq
\label{eq:Ieik}
\frac{\alps}{2\pi}\frac{1}{\Gamma(1-\eps)}
\biggl(\frac{4\pi\mu^2}{Q^2}\biggr)^\eps I^{\eik}(\mu_j,\mu_k;\eps) =
\int [\rd p_i(\tpij,\tpk)] \, \frac{1}{2p_i p_j} \,
\frac{8\pi\mu^{2\eps}\alps}{1-\zj(1-\yijk)},
\eeq
where $m_i=0$ and $m_{ij}=m_j$ in the kinematics defined in 
Sect.~\ref{subsec:ffdipoles}.  This integral can be obtained as a complicated
expression of dilogarithmic functions (see App.~\ref{app:auxint}).
However, for the calculation of
physical cross sections (see Sect.~\ref{se:xsec}),
we explicitly need only the symmetric part of $I^{\eik}$
and that is simpler. We define
\beq
I^{\eik}_{\pm}(\mu_j,\mu_k;\eps) = 
\frac{1}{2} \left[ I^{\eik}(\mu_j,\mu_k;\eps)
               \pm I^{\eik}(\mu_k,\mu_j;\eps) \right]\:,
\eeq
so that
\beq
I^{\eik}(\mu_j,\mu_k;\eps) =
I^{\eik}_+(\mu_j,\mu_k;\eps) + I^{\eik}_-(\mu_j,\mu_k;\eps)\:.
\eeq
The symmetric part of the eikonal integral is
given by 
\beeq
I^{\eik}_+(\mu_j,\mu_k;\eps)
\aand = \frac{1}{\tvijk} \left[
\left(1-(\mu_j+\mu_k)^2\right)^{-2\eps}\frac{1}{2\eps^2}
\left(1-\frac{1}{2}\rho_j^{-2\eps}-\frac{1}{2}\rho_k^{-2\eps}\right)
\right.
\nn\\
&& \quad {}
+\frac{\pi^2}{24}(6-\mu_j^{-2\eps}-\mu_k^{-2\eps}) 
+2\Li_2(-\rho)-2\Li_2(1-\rho)
\nn\\
&& \quad \left. {}
-\frac{1}{2}\Li_2\left(1-\rho_j^2\right)
-\frac{1}{2}\Li_2\left(1-\rho_k^2\right)
\right]
+\Oe{}\:,
\eeeq
where
\beeq
\label{eq:rhon}
\rho_n(\mu_j,\mu_k) \aand = \sqrt{
\frac{1-\tvijk+2\mu_n^2/(1-\mu_j^2-\mu_k^2)}
     {1+\tvijk+2\mu_n^2/(1-\mu_j^2-\mu_k^2)}} \qquad (n=j,k) ,
\nn\\
\rho \aand = \sqrt{\frac{1-\tvijk}{1+\tvijk}}\:.
\eeeq

For the special case $\mu_j=0$, which is needed for the $g\to gg$
splitting, the result for $I^{\eik}_+(0,\mu_k;\eps)$ can be easily read
off.  Using $\rho_j(0,\mu_k) = 0$, $\rho_k(0,\mu_k) = \mu_k$ and
$\rho = 0$, we obtain
\beeq
\label{eq:Ieiksym2}
I^{\eik}_+(0, \mu_k;\eps) =
(1 - \mu_k^2)^{-2\ep}\frac{1}{2\ep^2}
\left(1 - \frac12 \mu_k^{-2\ep} \right)
- \frac{\pi^2}{6} - \frac{\pi^2}{24}\mu_k^{-2\ep}
- \frac12\Li_2(1 - \mu_k^2)
+\Oe{}\:.
\eeeq
If both masses vanish, then $I^{\eik}_-(0, 0)=0$, and we recover the 
massless result \cite{Catani:1997vz}: 
\beq
I^{\eik}(0, 0;\eps)=I^{\eik}_+(0, 0;\eps)
= \frac{1}{2\ep^2} - \frac{\pi^2}{4} +\Oe{} \;.
\eeq
For the case of non-vanishing masses, we have 
$\rho_j \rho_k = \rho$, therefore, we can use the relation
\beq
1 - \frac12 \rho_j^{-2\ep} - \frac12 \rho_k^{-2\ep}
= \ep\ln \rho - \ep^2 (\ln^2 \rho_j + \ln^2 \rho_k)
+\Oe{3}\:,
\eeq
to obtain the following expansion in $\ep$:
\beeq
\nn
&&
\label{eq:Ieiksym2exp}
I^{\eik}_+(\mu_j, \mu_k;\eps) =
\nn \\ && \qquad
\frac{1}{\tilde{v}_{ij,k}}
\Bigg[
\frac{1}{2\ep} \ln \rho
- \ln \rho\,\ln\Big(1 - (\mu_j + \mu_k)^2\Big)
- \frac12 \ln^2\rho_j - \frac12 \ln^2\rho_k
\nn \\ && \qquad\qquad
+ \frac{\pi^2}{6}
+ 2 \Li_2(-\rho) - 2 \Li_2(1 - \rho)
- \frac12 \Li_2(1 - \rho_j^2) - \frac12 \Li_2(1 - \rho_k^2)\Bigg]
\nn \\  && \qquad\qquad
+\Oe{} \:.
\eeeq

The integrals of the collinear parts are simpler. The explicit results
are
\beeq
\hspace*{-1em}
I^{\coll}_{gQ,k}(\mu_Q,\mu_k;\eps) \aand =
\frac{3}{2\eps}-\frac{\mu_Q^{-2\eps}}{2\eps}
-2\mu_Q^{-2\eps}
-2\ln\left[(1-\mu_k)^2-\mu_Q^2\right]+\ln(1-\mu_k)
\nn\\
&& \quad {}
-\frac{2\mu_Q^2}{1-\mu_Q^2-\mu_k^2}\ln\left(\frac{\mu_Q}{1-\mu_k}\right)
+5-\frac{\mu_k}{1-\mu_k}-\frac{2\mu_k(1-2\mu_k)}{1-\mu_Q^2-\mu_k^2}
\nn\\ 
&& \quad {}
+\Oe{}\:,
\\[.5em]
\hspace*{-1em}
I^{\coll}_{Q\bar Q,k}(\mu_Q,\mu_k;\eps) \aand =
-\frac{2}{3}\left[\frac{1}{\eps}\left(1 - \mu_Q^{-2\eps}\right)
-2\ln(1-\mu_k)-2\ln\left(\frac{1+\rho_1}{2}\right)
+\frac{2}{3}\rho_1(3+\rho_1^2)
-\frac{2\rho_1^3\mu_k}{1+\mu_k}
\right]
\hspace*{-1em}
\nn\\
&& \quad {}
+\left(\kappa-\frac{2}{3}\right) \frac{2\mu_k^2}{1-\mu_k^2} \left[
\rho_2^3\ln\left(\frac{\rho_2-\rho_1}{\rho_2+\rho_1}\right)
-\ln\left(\frac{1-\rho_1}{1+\rho_1}\right)
-\frac{8\rho_1\mu_Q^2}{1-\mu_k^2} \right]
\nn\\ 
&& \quad {}
+\Oe{}\:,
\label{IcollQQ}
\\[.5em]
I^{\coll}_{gg,k}(\mu_k;\eps) \aand =
\frac{11}{6\eps}+\frac{50}{9}
-\frac{11}{3} \left[ \frac{\mu_k}{1+\mu_k} + \ln(1-\mu_k) \right]
+\frac{(2-3\kappa)\mu_k^2}{3(1-\mu_k^2)}
\ln\left(\frac{2\mu_k}{1+\mu_k}\right) \nn \\
&& \quad {} +\Oe{}\:,
\hspace{2em}
\eeeq
with
\beq
\label{eq:rho_idef}
\rho_1 = \sqrt{1-\frac{4\mu_Q^2}{(1-\mu_k)^2}}\:, \qquad
\rho_2 = \sqrt{1-\frac{4\mu_Q^2}{1-\mu_k^2}}\:.
\eeq

Finally, we inspect the special case of parametrically-small masses
which is relevant if one considers massless fermions and collinear
singularities regularized by small fermion masses, as is often done in
electroweak physics. The corresponding asymptotic forms of $I^{\eik}_+$
and $I^{\coll}_{ij,k}$ in this limit read
\beeq
I^{\eik}_+ & \;\asymp{m_j,m_k\to 0}\; &
 \frac{1}{4\ep} \ln\frac{m_j^2}{\sijk}
+\frac{1}{4\ep} \ln\frac{m_k^2}{\sijk}
- \frac18 \ln^2\frac{m_j^2}{\sijk}
- \frac18 \ln^2\frac{m_k^2}{\sijk}
- \frac{\pi^2}{3}
\:,
\nn\\
\label{eq:massreg}
I_{gQ,k}^{\coll}  & \;\asymp{m_Q,m_k\to 0}\; &
\frac{1}{\eps} + \frac12\ln\frac{m_Q^2}{\sijk} +3 \:,
\nn\\
I_{Q\bar Q,k}^{\coll} & \;\asymp{m_Q,m_k\to 0}\; &
-\frac{2}{3}\ln\frac{m_Q^2}{\sijk} -\frac{16}{9} \:,
\nn\\
I_{gg,k}^{\coll} & \;\asymp{m_k\to 0}\; &
\frac{11}{6\eps} +\frac{50}{9} \:, 
\eeeq
where $\sijk = 2 \tpij \tpk$. Note that in these expressions the fermion mass
is only regarded as collinear regulator. Therefore,
on the right-hand side we have neglected
not only contributions of $\Oe{}$, but also all the contributions of 
$\O(m^2/\sijk)$, even in coefficients of
$\ep$ poles.
The collinear logarithmic contributions in the $Q\to gQ$ splitting 
are related to those in the fermion--photon splitting considered in 
Ref.~\cite{Dittmaier:2000mb}, where soft divergences were regularized
by an infinitesimal photon mass.

\subsection{Final-state emitter and initial-state spectator}
\label{se:finalinitial}

\newcommand{\xija}{x_{ij,a}}
\newcommand{\tpa}{\tilde p_a}

The dipole contribution $\cD_{ij}^a$ (see Fig.~\ref{fig:effdiags}) to the 
factorization formula (\ref{eq:dff}) is defined in close analogy to that in
Eq.~(\ref{eq:Dijk}):
\beeq
\label{eq:fidipole}
\lefteqn{ \cD_{ij}^a(p_1,\dots,p_{m+1};p_a,\dots) =}
\nn \\
&& 
-\frac{1}{(p_i+p_j)^2-m_{ij}^2} \; \frac{1}{\xija} \;
{}_{m,a}\langle\dots,\widetilde{ij},\dots;\tilde{a},\dots|
\frac{{\bom T}_a\cdot{\bom T}_{ij}}{{\bom T}_{ij}^2} \bV_{ij}^a
|\dots,\widetilde{ij},\dots;\tilde{a},\dots\rangle_{m,a}\:.
\hspace*{3em}
\eeeq  
The $m$-parton matrix elements on the right-hand side are obtained from
the original $(m+1)$-parton matrix element by replacing the final-state
partons $i$ and $j$ with the emitter parton $\widetilde{ij}$ and the
initial-state parton $a$ with the spectator parton $\tilde{a}$. 

The final-state parton momenta $p_i$ and $p_j$ have arbitrary masses,
the initial-state momentum $p_a$ is massless and their total
transferred momentum is denoted by $Q$ (not to be confused with $Q$
in Eq.~(\ref{eq:Qfff})),
\beq
p_i^2 = m_i^2, \qquad p_j^2 = m_j^2, \qquad p_a^2 = 0,
\qquad \qquad
Q = p_i + p_j - p_a.
\eeq
The on-shell mass of the emitter is $m_{ij}$,
with $m_{ij}\le m_i+m_j$.

As discussed in Sect.~\ref{se:subwh}, we do not consider massive partons 
in the initial state in this paper.  A treatment of massive
initial-state particles is described in Ref.~\cite{Dittmaier:2000mb}
for photon emission off fermions.

\subsubsection{Kinematics and phase-space factorization}
\label{se:fikin}

We introduce the auxiliary variables
\beeq
\xija \aand = \frac{p_a p_i+p_a p_j-p_i p_j+\frac{1}{2}(m_{ij}^2-m_i^2-m_j^2)}
{p_a p_i + p_a p_j}, 
\nn\\
\zi \aand = \frac{p_a p_i}{p_a p_i + p_a p_j}, \qquad
\zj = \frac{p_a p_j}{p_a p_i + p_a p_j},
\eeeq
and we define the momenta of the emitter and spectator as
\beq
\label{eq:mfeis}
\tpa^\mu = \xija p_a^\mu, \qquad
\tpij^\mu = p_i^\mu + p_j^\mu - (1-\xija)p_a^\mu .
\eeq
These momenta obey their mass-shell relations and momentum conservation,
\beq
\tpij^2 = m_{ij}^2, \qquad\tpa^2 = 0, \qquad \qquad
Q = \tpij - \tpa.
\eeq
We also define the rescaled parton masses $\mu_n$ as
\beq
\mu_n = \frac{m_n}{\sqrt{2\tpij p_a}}, \quad n=i,j,ij.
\eeq
The momentum definition in Eq.~(\ref{eq:mfeis}) coincides with that of
Ref.~\cite{Phaf:2001gc} for the specific case (with $m_{ij}=m_j$ and $m_i=0$) 
considered there and with Ref.~\cite{Dittmaier:2000mb} for $m_{ij}=m_i$
and $m_j=0$, but differs for arbitrary combinations of masses.

Since the introduction of the rescaled momentum $\tpa=\xija p_a$
defines a new (boosted) frame, the factorization of the two-particle
phase space $\rd\phi(\tpij;Q+\tpa)$ from the three-particle
phase space $\rd\phi(p_i,p_j;Q+p_a)$ involves a convolution over a
boost parameter $x$,
\beq
\label{eq:psconv}
\rd\phi(p_i,p_j;Q+p_a) =
\int_0^1\!\rd x
\; \rd\phi\Big(\tpij;Q+x p_a\Big)
\; [\rd p_i(\tpij;p_a,x)] \;\Theta(x_+ - x) \;,
\eeq
with the upper limit
\beq
x_+ = 1+\mu_{ij}^2-(\mu_i+\mu_j)^2.
\eeq
The single-parton phase space $[\rd p_i(\tpij;p_a,x)]$ is
\beeq
\int [\rd p_i(\tpij;p_a,x)] \aand =
\frac{1}{4} (2\pi)^{-3+2\eps} 
(2\tpij p_a)^{1-\eps}
\int_0^{x_+} \rd \xija\, \delta(x-\xija) \,
(1-x+\mu_{ij}^2)^{-\eps} 
\nn\\[.5em]
&& {}\times 
\int\rd^{d-3}\Omega
\int_{z_-(x)}^{z_+(x)}\rd\zi\, 
\left[z_+(x)-\zi\right]^{-\eps}\left[\zi-z_-(x)\right]^{-\eps}
\label{eq:psfi}
\eeeq
with the limits
\beq
z_\pm(x) = \frac{1-x+\mu_{ij}^2+\mu_i^2-\mu_j^2 \pm
\sqrt{(1-x+\mu_{ij}^2-\mu_i^2-\mu_j^2)^2
-4\mu_i^2 \mu_j^2}}
{2(1-x+\mu_{ij}^2)}.
\label{eq:zlimits}
\eeq
The $\rd^{d-3}\Omega$ integration is over the solid angle $\Omega$ 
perpendicular to $\tpij$ and $p_a$.

\subsubsection{The dipole splitting functions}
\label{se:fidipoles}

We give the functions $\bV_{ij}^a$ in Eq.~(\ref{eq:fidipole}) for the 
two different QCD splitting processes $\widetilde{ij}\to i+j$:
\begin{itemize}
\item
$Q\to g(p_i) + Q(p_j)$: \quad $m_i=0$ and $m_j=m_{ij}=m_Q$,
\item
$g\to Q(p_i) + {\bar Q}(p_j)$: \quad $m_i=m_j=m_Q$ and $m_{ij}=0$.
\end{itemize}
The case $\bar Q\to g\bar Q$ is formally identical to $Q\to gQ$.
The splitting $g\to gg$ is already given in Eq.~(5.40) of
Ref.~\cite{Catani:1997vz}, 
since no massive parton is involved in this case.
Specifically, we define the dipole functions 
\beeq
\label{eq:V_gQa}
\langle s|\bV_{gQ}^a|s'\rangle \aand =
8\pi\mu^{2\eps}\alps\CF \left\{
\frac{2}{2-\xija-\zj}-1-\zj-\frac{m_Q^2}{p_i p_j}
-\eps(1-\zj) \right\} \delta_{ss'}
\nn\\[.5em]
\aand = \langle\bV_{gQ}^a\rangle \delta_{ss'},
\\[.5em]
\langle\mu|\bV_{Q\bar Q}^a|\nu\rangle \aand =
8\pi\mu^{2\eps}\alps\TR
\left\{ -g^{\mu\nu}
-\frac{4}{(p_i+p_j)^2} 
\Big[\zi p_i^\mu-\zj p_j^\mu\Big]
\Big[\zi p_i^\nu-\zj p_j^\nu\Big]
\right\},
\hspace*{2em}
\label{eq:V_QQbara}\\[.5em]
\langle\bV_{Q\bar Q}^a\rangle \aand =
8\pi\mu^{2\eps}\alps\TR 
\Biggl\{ 1-\frac{2}{1-\eps}  (z_+ -\zi)(\zi-z_-)
\Biggr\} \;,
\eeeq
with $z_\pm$ given in Eq.~(\ref{eq:zlimits}).

In App.~\ref{app:SUSYdipoles}, we give the dipole splitting functions 
$\bV_{ij}^a$ that are relevant for SUSY QCD calculations.

\subsubsection{The integrated dipole functions}
\label{subsub:FIinteg}

The integral of the spin-averaged dipole functions
$\langle\bV_{ij}^a\rangle$ over the singular phase space
$[\rd p_i(\tpij;p_a,x)]$ is defined by
\beq
\label{eq:Iija_def}
\int [\rd p_i(\tpij;p_a,x)] \,
\frac{1}{(p_i+p_j)^2-m_{ij}^2} \, \langle\bV_{ij}^a\rangle
\equiv \frac{\alps}{2\pi}\frac{1}{\Gamma(1-\eps)}
\biggl(\frac{4\pi\mu^2}{2\tpij p_a}\biggr)^\eps 
I_{ij}^a(x;\eps) \;,
\eeq
where $I_{ij}^a(x; \eps)$ depends also on $\tpij \cdot p_a$ and the
parton masses.  In the massless case, $I_{ij}^a$ does not depend on
$\tpij \cdot p_a$ and becomes the function ${\cal V}_{ij}(x;\ep)$ of
Ref.~\cite{Catani:1997vz}.

For $\ep = 0$ the function $I_{gQ}^a(x;\ep)$ is well-defined
everywhere, except at the endpoint $x=1$ where the
integral is singular as $\frac{\ln(1-x)}{1-x}$. To avoid this
singularity, the limit $\ep \to 0$ has to be performed uniformly in $x$,
which is achieved by decomposing $I_{gQ}^a(x;\ep)$ into a continuum and
an endpoint part using the usual \plusdist, which is defined by
\beq
\label{eq:pdist}
\int_0^1\rd x\, \Big(f(x)\Big)_+ g(x) \equiv
\int_0^1\rd x\, f(x) \left[g(x)-g(1)\right]\:.
\eeq
In this manner, $I_{gQ}^a(x;\ep)$ defines an $x$-distribution whose
coefficients contain poles in $\ep$.  

In the case of the  splitting process $g\to Q\bar Q$, the endpoint is
located at $x_+ = 1 - 4\mu_Q^2$, where the function $I_{Q\bar
Q}^a(x;\ep)$ is not singular. However, it becomes singular as
$\frac{1}{1-x}$ for vanishing quark masses. To ensure a smooth massless
behaviour, the limit $\ep \to 0$ has to be performed uniformly both in
$x$ and in the quark mass. To achieve this, we have to
use a modified version of the \plusdist, which we call \xplusdist,
\beq
\label{eq:xpdist}
\int_0^1\rd x\, \Big(f(x)\Big)_{x_+} g(x) \equiv
\int_0^1\rd x\, f(x) \Theta(x_+ - x) \left[g(x)-g(x_+)\right]\:.
\eeq
The \xplusdist\ reduces to the usual \plusdist\ by setting $x_+ =1$ and,
in particular, it tends to the \plusdist\ smoothly when $\mu_Q \to 0$ in
$I_{Q\bar Q}^a(x;\ep)$.

After decomposing $I_{ij}^a$ into a continuum and an endpoint part, we
further decompose the endpoint part into a singular ($J_{ij}^{a;\rm S}$)
and a finite ($J_{ij}^{a;\rm NS}$) piece, such that the latter vanishes
for vanishing masses:  
\beeq
\label{eq:I_gQa}
I_{gQ}^a(x;\ep) \aand = \CF \left\{
[J_{gQ}^a(x, \mu_Q)]_{+} + \delta(1-x)\,
\Big[J_{gQ}^{a;\rm S}(\mu_Q;\ep) + J_{gQ}^{a;\rm NS}(\mu_Q)\Big]
\right\}
+\Oe{}\:,
\\
\label{eq:I_QQa}
I_{Q\bar Q}^a(x;\ep) \aand = \TR \left\{
[J_{Q\bar Q}^a(x, \mu_Q)]_{x_+} + \delta(x_+-x)\,
\Big[J_{Q\bar Q}^{a;\rm S}(\mu_Q;\ep) + J_{Q\bar Q}^{a;\rm NS}(\mu_Q)\Big]
\right\}
+\Oe{}\:,
\eeeq
where $x_+ = 1 - 4 \mu_Q^2$ and
$J_{ij}^{a;\rm NS}(\mu_Q=0) = 0\:.$

The three contributions to Eq.~(\ref{eq:I_gQa}) are 
\beeq
[J_{gQ}^a(x, \mu_Q)]_+ \aand = 
\left( \frac{1-x}{2(1-x+\mu_Q^2)^2} 
-\frac{2}{1-x}\left[1+\ln(1-x+\mu_Q^2)\right] \right)_+
\nn\\ && \quad 
+ \left(\frac{2}{1-x}\right)_+ \ln(2+\mu_Q^2-x)\:,
\label{eq:JfigQacont}
\eeeq
\beq
J_{gQ}^{a;\rm S}(\mu_Q;\ep) =
\frac{1}{\eps^2} 
-\mu_Q^{-2\eps}
\left(\frac{1}{\eps^2}+\frac{1}{2\eps}+\frac{\pi^2}{6}+2\right)
-\frac{1}{\eps}\ln(1+\mu_Q^2)
+  \frac{3}{2\eps} + \frac{7}{2} - \frac{\pi^2}{2}\:,
\label{eq:JfigQaS}
\eeq
\beeq
\label{eq:fi_intdip_gQ}
J_{gQ}^{a;\rm NS}(\mu_Q) \aand = 
\frac{1}{2}\ln^2(1+\mu_Q^2) -2 \ln \mu_Q^2 \ln(1+\mu_Q^2) -4\Li_2(-\mu_Q^2)
\nn \\ && 
+\frac{1}{2}\ln(1+\mu_Q^2)
+\frac{\mu_Q^2}{2(1+\mu_Q^2)}\;.
\eeeq
Note that the notation in Eq.~(\ref{eq:JfigQacont}) is symbolic:
$[J_{gQ}^a(x, \mu_Q)]_+$ {\em contains\/} all the plus distribution
contributions, but is not itself a plus distribution.  The same is true
of $[J_{gg}^a(x)]_+$ defined in Eq.~(\ref{eq:Jfiggacont}).

Setting the mass of the quark to zero, we immediately obtain
the corresponding expression in the massless theory
(see Eq.~(5.57) in Ref.~\cite{Catani:1997vz}).
If the quark mass is non-vanishing, we can expand
the singular part of the endpoint contribution as 
\beq
J_{gQ}^{a;\rm S}(\mu_Q>0) = 
  \frac{1}{\eps}\ln \frac{\mu_Q^2}{1+\mu_Q^2}
- \frac{1}{2}\ln^2\!\mu_Q^2
+ \frac{1}{\eps}
+ \frac{1}{2}\ln\mu_Q^2
+ \frac{3}{2} - \frac{2}{3}\pi^2\:.
\eeq

The three contributions to Eq.~(\ref{eq:I_QQa}) are 
\beq
[J_{Q\bar Q}^a(x, \mu_Q)]_{x_+} = 
\frac{2}{3} 
\left(
\frac{1-x+2\mu_Q^2}{(1-x)^2} 
\sqrt{1-\frac{4\mu_Q^2}{1-x}}
\right)_{x_+}\:,
\label{eq:JfiQQacont}
\eeq
\beq
J_{Q\bar Q}^{a;\rm S}(\mu_Q;\ep) = 
-\frac{2}{3\eps}\left(1 - \mu_Q^{-2\eps}\right) -\frac{10}{9}\:,
\label{eq:JfiQQaS}
\eeq
\beq
J_{Q\bar Q}^{a;\rm NS}(\mu_Q) = 
\frac{10}{9}\Big(1 - \sqrt{1-4\mu_Q^2}\Big)
-\frac{8}{9}\mu_Q^2\sqrt{1-4\mu_Q^2}
+\frac{4}{3}\ln\left(\frac{1+\sqrt{1-4\mu_Q^2}}{2}\right)
\:.
\label{eq:fi_intdip_QQ}
\eeq
The massless result of Eq.~(5.58) in Ref.~\cite{Catani:1997vz} is
again recovered by simply setting $m_Q=0$.
If $m_Q \neq 0$, the collinear pole $1/\eps$ is
replaced by the logarithm of the scaled quark mass,
\beq
J_{Q\bar Q}^{a;\rm S}(\mu_Q>0) = 
-\frac{2}{3}\ln\mu_Q^2 -\frac{10}{9}\:.
\eeq

As mentioned above, the case of the splitting process $g \to gg$ is already 
discussed in Ref.~\cite{Catani:1997vz}. We recall the result by using the
present notation: 
\beq
\label{eq:I_gga}
I_{gg}^a(x;\ep) = 2\CA \left\{
[J_{gg}^a(x)]_{+} + \delta(1-x)\,J_{gg}^{a;\rm S}(\ep)
\right\}+\Oe{} \:,
\eeq
where
\beeq
\label{eq:Jfiggacont}
[J_{gg}^a(x)]_+ \aand = 
\left(\frac{2}{1-x}\ln\frac{1}{1-x}-\frac{11}{6}\frac{1}{1-x} \right)_+
+\frac{2}{1-x}\ln(2-x)\:,
\\
J_{gg}^{a;\rm S}(x) \aand = 
\frac{1}{\eps^2} + \frac{11}{6\eps}
+ \frac{67}{18}-\frac{\pi^2}{2} \:.
\eeeq

Finally, we consider the asymptotics in the case of a parametrically-small
quark mass $m_Q = \mu_Q\,\sqrt{2\tpij p_a}$, used as collinear regulator:
\beeq
I_{gQ}^a(x, \mu_Q) \aand\;\asymp{\mu_Q \to 0}\;
\CF\,\Bigg[
- 2\left(\frac{\ln(1-x)}{1-x}\right)_+
- \frac{3}{2}\left(\frac{1}{1-x}\right)_+ 
+ \frac{2}{1-x} \ln(2-x)
\nn\\* && \qquad\qquad\;\;
+\delta(1-x) \left(
\frac{1}{\eps}\left(\ln\mu_Q^2+1\right)
-\frac12 \ln^2\mu_Q^2
+\frac12 \ln\mu_Q^2
-\frac{2}{3}\pi^2
+\frac{3}{2}
\right)\Bigg] \:,
\nn\\* 
\\[.5em]
I_{Q\bar Q}^a(x, \mu_Q) \aand\;\asymp{\mu_Q \to 0}\; 
\TR\,\Bigg[\frac{2}{3} \left(\frac{1}{1-x}\right)_+
-\delta(1-x)\left(\frac{2}{3}\ln\mu_Q^2 +\frac{10}{9} \right)\Bigg] \:.
\eeeq
Analogously to Eq.~(\ref{eq:massreg}), on the right-hand side we have
neglected all the contributions of $\O(\mu_Q)$, independently of their
$\eps$ dependence.  Comparing these results with Eqs.~(5.57) and (5.58)
in Ref.~\cite{Catani:1997vz}, we notice that the continuum parts reduce
to those of the exactly massless case, as expected since all the
singularities are related to the final-state emitter.  The $1/\eps$
collinear poles of the massless theory in the endpoint parts are
replaced by the mass logarithms, which in the $Q\to gQ$ splitting are
related to those of the fermion-photon splitting \cite{Dittmaier:2000mb}.

We conclude this section with a comment on a subtle point related to the
definition of the $x$-distributions or, more precisely, to the definition
of the space of test functions onto which the $x$-distributions act.
Indeed, the integrals $I_{ij}^a$ depend on two variables, $x$ and
$\tpij$, which are both integration variables in the calculation of the
cross section.  Our definition in Eqs.~(\ref{eq:I_gQa}) and
(\ref{eq:I_QQa}), in terms of continuum and endpoint parts, considers
$I_{ij}^a$ as an $x$-distribution that acts on functions of the
variables $x$ and $\tpij$, and $\tpij$ is regarded as an additional and
independent integration variable that has to be kept fixed in the
subtraction prescriptions of Eqs.~(\ref{eq:pdist}) and (\ref{eq:xpdist}).
However, the $x$-distributions can also be defined in a different way, 
for instance, by regarding $Q=\tpij-x p_a$ (instead of $\tpij$) as the
independent integration variable. This alternative definition is
explicitly worked out in App.~\ref{app:Q2fixed}, where we show that the
replacement $\tpij \to Q=\tpij-x p_a$, being $x$ dependent, has to be
handled with care in the definition of the $x$-distributions. 

\subsection{Initial-state emitter and final-state spectator}

We define the dipole contribution $\cD_j^{ai}$ (see
Fig.~\ref{fig:effdiags}) to the factorization formula (\ref{eq:dff}) in
close analogy to that in Eq.~(\ref{eq:Dijk}):
\beeq
\lefteqn{ \cD_j^{ai}(p_1,\dots,p_{m+1};p_a,\dots) =}
\nn\\
&& 
-\frac{1}{2p_a p_i} \; \frac{1}{\xija} \;
{}_{m,\widetilde{ai}}\langle\dots,\tilde{j},\dots;\widetilde{ai},\dots|
\frac{{\bom T}_j\cdot{\bom T}_{ai}}{{\bom T}_{ai}^2} \bV_j^{ai}
|\dots,\tilde{j},\dots;\widetilde{ai},\dots\rangle_{m,\widetilde{ai}}\:.
\hspace*{2em}
\eeeq
The $m$-parton matrix elements on the right-hand side are obtained
from the original $(m+1)$-parton matrix elements by
replacing the initial-state parton $a$ with the  emitter parton
$\widetilde{ai}$, the final-state parton $j$ with $\tilde{j}$ and
dropping the final-state parton $i$.  

The final-state parton momentum $p_i$ and the initial-state momentum
$p_a$ are massless and the mass of spectator momentum $p_j$ is
arbitrary. The total transferred momentum is denoted by $Q$ (not to be
confused with $Q$ in Eq.~(\ref{eq:Qfff})),
\beq
p_i^2 = 0, \qquad p_j^2 = m_j^2, \qquad p_a^2 = 0,
\qquad \qquad
Q = p_i + p_j - p_a.
\eeq
The on-shell mass of the emitter is $m_{ai} = 0$.

\subsubsection{Kinematics and phase-space factorization}
\label{se:ifkin}

\newcommand{\tpai}{\tilde p_{ai}}
\newcommand{\tpj}{\tilde p_j}
We can completely take over the kinematics of Sect.~\ref{se:fikin}
after replacing $\tpa\to\tpai$, $\tpij\to\tpj$, $m_i\to 0$ and
$m_{ij} \to m_j$ there. Thus, the new momenta read 
\beq
\tpai^\mu = \xija p_a^\mu, \qquad
\tpj^\mu = p_i^\mu + p_j^\mu - (1-\xija)p_a^\mu,
\eeq
where
\beq
\xija = \frac{p_a p_i+p_a p_j-p_i p_j}{p_a p_i + p_a p_j}, \qquad
\zi = \frac{p_a p_i}{p_a p_i + p_a p_j}, \qquad
\zj = \frac{p_a p_j}{p_a p_i + p_a p_j}.
\eeq
They obey the mass-shell relations 
\beq
p_a^2 = p_i^2 = \tpai^2 = 0, \qquad
p_j^2 = \tpj^2 = m_j^2
\eeq
and momentum conservation
\beq
Q = p_i + p_j - p_a = \tpj - \tpai.
\eeq
The rescaled mass $\mu_j$ of the parton $j$ is defined by
\beq
\mu_j = \frac{m_j}{\sqrt{2\tpj p_a}}\:.
\eeq

For the sake of completeness we also write down the phase-space factorization,
\beq
\label{eq:psconv2}
\int\rd\phi(p_i,p_j;Q+p_a) =
\int_0^1\rd x\,
\int\rd\phi\Big(\tpj;Q+x p_a\Big)
\int [\rd p_i(\tpj;p_a,x)]
\eeq
with
\beeq
\int [\rd p_i(\tpj;p_a,x)] \aand =
\frac{1}{4} (2\pi)^{-3+2\eps} 
(2\tpj p_a)^{1-\eps}
\int_0^1 \rd \xija\, \delta(x-\xija) \,
(1-x+\mu_j^2)^{-\eps} 
\nn\\[.5em]
&& {}\times 
\int\rd^{d-3}\Omega
\int_0^{z_+(x)}\rd\zi\, 
\left[z_+(x)-\zi\right]^{-\eps}\zi^{-\eps}\:,
\eeeq
where the upper limit of the $z$-integration is
\beq
z_+(x) = \frac{1-x} {1-x+\mu_j^2}.
\eeq

\subsubsection{The dipole functions}

We consider the splittings $q\to qg$ or $g q$, $g\to q\bar q$ or $\bar qq$,
$g\to gg$ of massless partons in the presence of the massive spectator
parton $j$. The corresponding dipole functions read as follows:
\beeq
\langle s|\bV^{qg}_j|s'\rangle \aand =
8\pi\mu^{2\eps}\alps\CF \left\{
\frac{2}{2-\xija-\zj}-1-\xija-\eps(1-\xija) \right\} \delta_{ss'}
\nn\\*[.5em]
\aand = \langle\bV^{qg}_j\rangle \delta_{ss'}\:,
\\[.5em]
\langle s|\bV^{g\bar q}_j|s'\rangle \aand =
8\pi\mu^{2\eps}\alps\TR \left\{
1-\eps-2\xija(1-\xija) \right\} \delta_{ss'}
\nn\\[.5em]
\aand = \langle\bV^{g\bar q}_j\rangle \delta_{ss'}\:,
\\[.5em]
\langle\mu|\bV^{qq}_j|\nu\rangle \aand =
8\pi\mu^{2\eps}\alps\CF \left\{
-g^{\mu\nu}\xija+\frac{1-\xija}{\xija}\frac{2\zi\zj}{p_i p_j}
\biggl(\frac{p_i^\mu}{\zi}-\frac{p_j^\mu}{\zj}\biggr)
\biggl(\frac{p_i^\nu}{\zi}-\frac{p_j^\nu}{\zj}\biggr) \right\}\:,
\label{eq:V_qqIF}\\[.5em]
\langle\bV^{qq}_j\rangle \aand =
8\pi\mu^{2\eps}\alps\CF \left\{
\xija+\frac{2(1-\xija)}{\xija(1-\eps)}
-\frac{2\mu_j^2}{\xija(1-\eps)} \frac{\zi}{\zj} \right\}\:,
\\[.5em]
\langle\mu|\bV^{gg}_j|\nu\rangle \aand =
16\pi\mu^{2\eps}\alps\CA \left\{
-g^{\mu\nu}\biggl[\frac{1}{2-\xija-\zj}-1+\xija(1-\xija)\biggr]
\right.
\nn\\[.5em]
&& \qquad\qquad\qquad
\left. {}
+(1-\eps)\frac{1-\xija}{\xija}\frac{\zi\zj}{p_i p_j}
\biggl(\frac{p_i^\mu}{\zi}-\frac{p_j^\mu}{\zj}\biggr)
\biggl(\frac{p_i^\nu}{\zi}-\frac{p_j^\nu}{\zj}\biggr) \right\}\:,
\label{eq:V_ggIF}\\[.5em]
\langle\bV^{gg}_j\rangle \aand =
16\pi\mu^{2\eps}\alps\CA \left\{
\frac{1}{2-\xija-\zj}-1+\xija(1-\xija)
+\frac{1-\xija}{\xija}
-\frac{\mu_j^2}{\xija}\frac{\zi}{\zj} 
\right\}\:.
\nn\\
\eeeq

\subsubsection{The integrated dipole functions}
\label{subsub:IFinteg}

We define the integral of $\langle\bV^{ai}_j\rangle$
over the singular phase space $[\rd p_i(\tpj;p_a,x)]$ by
\beq
\int [\rd p_i(\tpj;p_a,x)] \,
\frac{1}{2p_a p_i} \, 
\frac{n_{\mathrm{s}}(\widetilde{ai})}{n_{\mathrm{s}}(a)} \,
\langle\bV^{ai}_j\rangle
\equiv \frac{\alps}{2\pi}\frac{1}{\Gamma(1-\eps)}
\biggl(\frac{4\pi\mu^2}{2\tpj p_a}\biggr)^\eps 
I_j^{a,\widetilde{ai}}(x,\mu_j;\ep),
\eeq
where the factor $n_{\mathrm{s}}(\widetilde{ai})/n_{\mathrm{s}}(a)$
turns the spin-average over $\widetilde{ai}$ into the one over $a$.
Splitting off the massless part $I_j^{ab}(x,0;\ep)$, the results for 
$I_j^{ab}(x,\mu_j;\ep)$ can be written in the compact form
\beeq
\label{eq:Ijab}
I_j^{ab}(x,\mu_j;\ep) \aand = I_j^{ab}(x,0;\ep)
+ P^{ab}_\reg(x) \ln\left(\frac{1-x}{1-x+\mu_j^2}\right)
\nn\\ \nn
&& {} + \delta_{gb}\,{\bom T}_a^2\,\frac{2\mu_j^2}{x}
\ln\left(\frac{\mu_j^2}{1-x+\mu_j^2}\right)
+ 2\delta^{ab}\,{\bom T}_a^2\,\left(\frac{1}{1-x}\right)_+
\ln\left(\frac{2-x}{2-x+\mu_j^2}\right)
\\ \nn
&& {} + \delta^{ab}\,\delta(1-x)\,{\bom T}_a^2
\left[ \frac{1}{\eps}\ln(1+\mu_j^2)+\frac{1}{2}\ln^2(1+\mu_j^2)
+2\Li_2\left(\frac{1}{1+\mu_j^2}\right)-\frac{\pi^2}{3} \right]
\\ 
&& {}
+\Oe{}\:,
\eeeq
where the auxiliary functions $P^{ab}_\reg(x)$ are given by
\beeq
P^{qq}_\reg(x) \aand = -\CF (1+x)\:,\qquad\qquad
P^{gq}_\reg(x) = \TR \left[ x^2+(1-x)^2 \right]\:,
\nn \\[.5em]
P^{qg}_\reg(x) \aand = \CF \frac{1+(1-x)^2}{x}\:,\qquad\;
P^{gg}_\reg(x) = 2\CA \left[ \frac{1-x}{x}-1+x(1-x) \right]\:.
\label{eq:Preg}
\eeeq
Since the kinematics of the spectator parton $j$ is not connected with
any singularity, the limit $m_j\to 0$ is smooth in all cases.
The massless part can be written in the following compact form,
\beeq
\label{eq:Iab}
I^{ab}_j(x,0;\ep) \aand = 
-\frac{1}{\eps} P^{ab}(x) + P^{ab}_\reg(x)\ln(1-x)
+2\,\delta^{ab}\,{\bom T}_a^2 \left[
2\,\left( \frac{\ln(1-x)}{1-x} \right)_+
-\frac{\ln(2-x)}{1-x}\right]
\nn\\[.5em]
&&
+{\hat P}^{\prime\, ab}(x)
+ \delta^{ab} \, \delta(1-x)
\left[ {\bom T}_a^2 \left(\frac{1}{\eps^2}+\frac{\pi^2}{6}
\right)+\gamma_a\frac{1}{\eps}\right]
+\Oe{}\:,
\eeeq
where $\gamma_a$ are the following flavour constants:
\beq
\gamma_q = \frac{3}{2} \CF \:, \qquad\qquad
\gamma_g = \frac{11}{6} \CA - \frac{2}{3} \TR N_f \:,
\label{eq:gamma_g}
\eeq
and
${\hat P}^{\prime\, ab}(x)$ is the $\eps$-dependent part of
the $d$-dimensional Altarelli-Parisi splitting functions in
Eqs.~(\ref{avhpqq}--\ref{avhpgg}):
\beq
\label{avhpep}
\la {\hat P}_{ab}(x;\ep) \ra \;= \;\la {\hat P}_{ab}(x;\ep=0) \ra
- \,\ep \;{\hat P}^{\prime\, ab}(x) + {\rm O}(\ep^2) \:.
\eeq
Explicitly,
\beeq
\label{eq:Pprime}
{\hat P}^{\prime\, qq}(x) \aand = 
{\hat P}^{\prime\, qg}(1-x) = \CF\,(1-x)\:,\quad
{\hat P}^{\prime\, gq}(x) = 2 \TR\,x(1-x)\:,\quad
{\hat P}^{\prime\, gg}(x) = 0\:.\quad~
\eeeq
The functions $P^{ab}(x)$ in Eq.~(\ref{eq:Iab}) are the usual
regularized Altarelli--Parisi functions that can be written as
\cite{Catani:1997vz}
\beeq
\label{eq:Pab}
P^{ab}(x) =
P^{ab}_{{\rm reg}}(x) 
+ \delta^{ab} \left[
2 \,{\bom T}_a^2 \left( \frac{1}{1-x} \right)_+ + \gamma_a \,\delta(1-x)
\right]\:.
\eeeq

\subsection{Initial-state emitter and initial-state spectator}

Since we allow for massive partons only in the final state in this
paper, the case of emitter and spectator both from the initial state is
already completely covered in Ref.~\cite{Catani:1997vz}, where all cases for
massless QCD partons are presented. Results for the $Q\to Qg$ splitting
with a massive quark in the initial state can be deduced from the 
treatment of photon emission off fermions described in Ref.~\cite{Dittmaier:2000mb}.

\section{QCD cross sections at NLO}
\label{se:xsec}

In this section we describe our method for evaluating QCD cross
sections of massive and/or massless partons. We start with the precise
definitions of the three terms on the right-hand side of
Eq.~(\ref{sNLO4}), which can be considered the final expressions, ready
for implementing in a partonic Monte Carlo program.
Then we derive the explicit forms of the insertion operators ${\bom I}$,
${\bom P}$ and ${\bom K}$ relevant to the three different cases of
initial state: elementary particle reactions involving zero, one or two
initial-state hadrons.
More details on our notation can be found in Sects.~6-11 of
Ref.~\cite{Catani:1997vz}.

\subsection{Final formulae}
\label{subsec:final}

In order to present the final formulae of our algorithm, we start with
the most general case of two incident hadrons and then point out the
simplifications if one or both of the incoming particles are leptons.

In the case of processes with two initial-state hadrons carrying momenta
$p_A^\mu$ and $p_B^\mu$, the calculation of the QCD cross section
requires the introduction of parton distributions. If we denote by
$f_{a/A}(\eta, \mu_F^2)$ the density of partons of type $a$ in the
incoming hadron $A$, the hadronic cross section is given by
\beq
\label{eq:2hxs}
\sigma(p_A,p_B) = \sum_{a,b} \int_0^1\!\rd\eta_a \, f_{a/A}(\eta_a, \mu_F^2)
\,\int_0^1 \rd\eta_b \, f_{b/B}(\eta_b, \mu_F^2)
\,\sigma_{ab}(\eta_a p_A,\eta_b p_B)\:,
\eeq
where $\sigma_{ab}$ is the partonic cross section, which in the first
two orders of the perturbative expansion has the schematic form given in
Eq.~(\ref{eq:sigma}),
\beq
\sigma_{ab}(p_a,p_b) =
\sigma_{ab}^\aLO(p_a,p_b)
+ \sigma_{ab}^\aNLO(p_a,p_b;\mu_F^2)\:.
\eeq
If one or two of the colliding particles is pointlike, i.e.~a lepton in
the Standard Model, then the integration becomes trivial by
replacing the corresponding parton density by
\mbox{$\delta(1-\eta)\delta_{aA}$}.
The LO parton-level cross section is given by
\beeq
\label{eq:LOppfin}
\sigma^{\rm LO}_{ab}(p_a,p_b) \aand =
\int_m \! \rd\sigma^\rB_{ab}(p_a,p_b) 
\nn \\
\aand = \int \! \rd\Phi^{(m)}(p_a,p_b)
\, \frac{1}{n_c(a) n_c(b)} \,| \cm_{m,ab}(\{p_i,m_i\};p_a,p_b)|^2
\,F_J^{(m)}(\{p_i\};p_a,p_b) \:,\quad
\eeeq
where $a$ and $b$ denote the flavours of the incoming partons, $p_a$ and
$p_b$ are their momenta and $n_c(a), n_c(b)$ are their number of colours
(for an incoming lepton $n_c$ is set to 1).
The matrix element $| \cm_{m,ab}|^2$ is the square of the tree-level
amplitude to produce $m$ final-state partons.  The factor 
\beq
\label{eq:dPhi}
\rd\Phi^{(m)}(p_a,p_b) =
{\cal N}_{in} \frac{1}{n_s(a) n_s(b) \cF(p_a \cdot p_b)}
\sum_{\{ m \} } \rd\phi_m(p_1, \dots ,p_m;p_a+p_b)
\frac{1}{S_{\{ m \} }}
\eeq
contains
(i) the averaging over the number of initial-state polarizations
($n_s(i)$ number of states for parton $i$, $i=a$ or $b$),
(ii) the flux factor $\cF(p_a\cdot p_b)$,
(iii) a summation over all configurations with $m$ partons in the final
state,
(iv) the Lorentz invariant phase space of $m$ final-state partons
$\rd\phi_m$, 
(v) a Bose symmetry factor $S_{\{ m \}}$ for identical partons in the
final state and
(vi) ${\cal N}_{in}$, which includes all factors that are QCD
independent.
Finally, the function $F_J^{(m)}$ in Eq.~(\ref{eq:LOppfin}) defines the
jet observable we want to compute. It has to fulfil the formal
requirements of infrared and collinear safety.
Furthermore, in order for our cross section to have a smooth small-mass
limit, $F_J$ should also be {\em quasi-collinear safe}.

The general definition of quasi-collinear safety is analogous to that of
collinear safety: if a pair of partons $i$ and $j$ whose masses and
relative transverse momenta are all O$(\lambda Q)$, where $Q$ is the
hard scale, then as $\lambda\to0$, $F_J^{(m+1)}(\ldots,p_i,p_j,\ldots)$
must tend to the value of $F_J^{(m)}(\ldots,p_i+p_j,\ldots)$.  Any
collinear-safe observable for massless partons can be extended to be
quasi-collinear-safe for massive partons.  However, there is a subtlety
related to the flavour-dependence of $F_J$.  To respect infrared and
collinear safety, an observable must not depend on the identities of
massless partons in the final state.  It can however depend on the
identities of massive partons, for example the thrust distribution in
$e^+e^-$ annihilation events that contain at least one bottom quark is
an infrared- and collinear-safe quantity.  However it is not
quasi-collinear safe, because events in which the $b\bar b$ pair is
quasi-collinear contribute but identical events with the $b\bar b$ pair
replaced by a gluon do not.  Careful definition of the observable can
render it quasi-collinear safe and hence finite in the high-energy
limit.  One solution in this example would be to only include events in
which the $b$ and $\bar b$ lie in opposite thrust hemispheres.  Our
method is perfectly capable of calculating cross sections for
observables that are not quasi-collinear safe, but our careful treatment
of the quasi-collinear limit only guarantees improved numerical
convergence for quasi-collinear safe observables.

According to the general notation in Eq.~(\ref{sNLO4}), the full
NLO contribution is a sum of three terms,
\beeq
\label{eq:sNLOppfin}
\sigma^{\aNLO}_{ab}(p_a,p_b;\mu_F^2) \aand =
\sigma^{\aNLO\,\{m+1\}}_{ab}(p_a,p_b) + \sigma^{\aNLO\,\{m\}}_{ab}(p_a,p_b)
\nn \\
\aand +
\int_0^1 \!\rd x \,
\left[ \,{\hat \sigma}^{\aNLO\,\{m\}}_{ab}(x;xp_a,p_b,\mu_F^2)
+ {\hat \sigma}^{\aNLO\,\{m\}}_{ab}(x;p_a,xp_b,\mu_F^2) \,\right] \:.
\eeeq

The first contribution has $(m+1)$-parton kinematics and is given by the
following expression
\beeq
\label{eq:NLOppmp}
&&
\sigma^{\aNLO\,\{m+1\}}_{ab}(p_a,p_b) =
\int_{m+1} \Bigg[ \left( \rd\sigma^\rR_{ab}(p_a,p_b) \right)_{\ep=0}
- \left( \sum_{{\rm dipoles}} \rd\sigma^\rB_{ab}(p_a,p_b) \otimes
\rd V_{{\rm dipole}}
\right)_{\!\ep=0} \Bigg]
\nn \\ && \qquad
= \int\!\rd\Phi^{(m+1)}(p_a,p_b)\Bigg[
\frac{1}{n_c(a) n_c(b)}\,| \cm_{m+1,ab}(\{p_i,m_i\};p_a,p_b)|^2\,
F_J^{(m+1)}(\{p_i\};p_a,p_b)
\nn \\ && \qquad\qquad\qquad\qquad\qquad\qquad
- \sum_{{\rm dipoles}}
\left( {\cal D} \cdot F^{(m)} \right)(\{p_i,m_i\};p_a,p_b) \Bigg]
\:,
\eeeq
where $\cm_{m+1,ab}$ is the tree-level matrix element with $m+1$ partons
in the final state and $\sum_{{\rm dipoles}} \left( {\cal D} \cdot
F^{(m)} \right)(\{p_i,m_i\};p_a,p_b)$ is the following sum of the 
dipole functions:
\beeq
\label{eq:dipoles}
&&
\sum_{\mathrm{pairs}\atop i,j}
\sum_{k\not=i,j} {\cal D}_{ij,k}(p_1, \dots ,p_{m+1};p_a,p_b) \,
F_J^{(m)}(p_1, \dots  {\widetilde p}_{ij}, {\widetilde p}_k,
\dots ,p_{m+1};p_a,p_b)
\nn \\ &&
+ \sum_{\mathrm{pairs}\atop i,j} 
\left[ {\cal D}_{ij}^{a}(p_1, \dots ,p_{m+1};p_a,p_b) \,
F_J^{(m)}(p_1, \dots  {\widetilde p}_{ij}, \dots ,p_{m+1};{\widetilde p}_a,p_b)
+ ( a \leftrightarrow b ) \right]
\nn \\ &&
+ \sum_i \sum_{k\not=i}
\left[ {\cal D}_{k}^{ai}(p_1, \dots ,p_{m+1};p_a,p_b) \;
F_J^{(m)}(p_1, \dots  {\widetilde p}_{k}, \dots ,p_{m+1};
          {\widetilde p}_{ai},p_b)
+ ( a \leftrightarrow b ) \right] 
\nn \\ &&
+ \sum_i
\left[ {\cal D}^{ai,b}(p_1, \dots ,p_{m+1};p_a,p_b) \;
F_J^{(m)}({\widetilde p}_1, \dots ,{\widetilde p}_{m+1};
          {\widetilde p}_{ai},p_b)
+ ( a \leftrightarrow b ) \right] \:,
\eeeq
where $( a \leftrightarrow b )$ indicates that the {\em roles\/} of $a$
and $b$ are reversed.  For example, in the second line, ${\cal
D}_{ij}^{a}$ becomes ${\cal D}_{ij}^{b}$ and ${\widetilde p}_a,p_b$
becomes $p_a,{\widetilde p}_b$.  In the last line of
Eq.~(\ref{eq:dipoles}), the function $F_J^{(m)}$ depends on the
final-state momenta ${\widetilde p}_1, \dots ,{\widetilde p}_{m+1}$,
which are obtained from the original momenta $p_1, \dots ,p_{m+1}$ by a
Lorentz tranformation (see Sect.~5.5 and the Erratum in
Ref.~\cite{Catani:1997vz}).

If an incoming particle is a lepton, then the terms with ${\cal D}$
carrying the corresponding parton-flavour index are absent from the sum,
i.e.\ for one initial-state hadron all terms corresponding to the 
$( a \leftrightarrow b )$ interchange and the full last line has to be
dropped and for zero initial-state hadrons, only the first line remains.

The NLO contribution with $m$-parton kinematics is obtained by adding the
virtual cross section and the singular part (with some accompanying
finite terms) of the integral of the subtraction term over the
factorized one-particle phase space. The former is given by the
interference term between the tree and one-loop amplitudes
$| \cm_{m,ab}|^2_{(\mathrm{1-loop})}$ and the latter in terms of an
insertion operator ${\bom I}(\ep)$:
\beeq
\label{eq:NLOppm}
&&\sigma^{\aNLO\,\{m\}}_{ab}(p_a,p_b) = \int_m \left[
\rd\sigma^{\rV}_{ab}(p_a,p_b) +
\rd\sigma^{\rB}_{ab}(p_a,p_b) \otimes {\bom I}
\right]_{\ep=0}
\nn \\ && 
=  \int \!\rd\Phi^{(m)}(p_a,p_b) \Bigg[  \frac{1}{n_c(a) n_c(b)}
\;| \cm_{m,ab}(\{p_i,m_i\};p_a,p_b)|^2_{(\mathrm{1-loop})}
\nn \\ && \qquad\qquad\qquad\quad
+ {}_{m,ab}\la {1, \ldots, m;a,b}\,| \,{\bom I}(\ep)\,
|{1, \ldots, m;a,b}\ra_{m,ab} \Bigg]_{\ep=0}
F_J^{(m)}(\{p_i\};p_a,p_b) \:,\qquad
\eeeq
where the operator ${\bom I}(\ep)$ is spelled out explicitly for the
various cases of interest: in Eq.~(\ref{eq:insop}) to be inserted into
$_m\la{1, \ldots, m}||{1, \ldots, m}\ra_m$ (case of no initial-state
hadrons), (\ref{eq:insop1}) into
$_{m,a}\la{1, \ldots, m;a}||{1, \ldots, m;a}\ra_{m,a}$ (case of one
initial-state hadron) and (\ref{eq:insop2}) into
$_{m,ab}\la{1, \ldots, m;a,b}||{1, \ldots, m;a,b}\ra_{m,ab}$
(case of two initial-state hadrons).

The third term on the right-hand side of Eq.~(\ref{eq:sNLOppfin}) is a
finite remainder left after cancellation of the $\ep$-poles of the
collinear counterterm. It is a sum of as many contributions as the
number of initial-state hadrons, each obtained by integrating a cross
section with $m$-parton kinematics with respect to the fraction $x$ of
the longitudinal momentum carried by one of the incoming partons. When
this parton is the parton $a$, we explicitly have:
\beeq
\label{eq:NLOppx}
&&
\int_0^1\!\rd x\,{\hat \sigma}^{\aNLO\,\{m\}}_{ab}(x;xp_a,p_b,\mu_F^2) =
\sum_{a'} \int_0^1\!\rd x\,\int_m\!
\left[ \rd \sigma^\rB_{a'b}(xp_a,p_b) \otimes
({\bom K} + {\bom P})^{a,a'}(x) \right]_{\ep=0}
\nn \\ && \quad
= \sum_{a'} \int_0^1\!\rd x\,\int\!
\rd \Phi^{(m)}(xp_a,p_b) \, F_J^{(m)}(\{p_i\};xp_a,p_b)
\nn \\ && \qquad
\times
{}_{m,a'b}\la{1, \dots, m;xp_a,p_b}| \,
{\bom K}^{a,a'}(x) + {\bom P}^{a,a'}(x,\mu_F^2) \, 
|{1, \dots, m;xp_a,p_b}\ra_{m,a'b} \:,\quad
\eeeq
where the colour-charge operators ${\bom P}$ and  ${\bom K}$ are
respectively defined in Eqs.~(\ref{eq:Pop1}) and (\ref{eq:Kab1})
to be inserted into
$_{m,a'}\la{1, \ldots, m;xp_a}||{1, \ldots, m;xp_a}\ra_{m,a'}$ and
Eqs.~(\ref{eq:Pop2}) and (\ref{eq:Kab2}) into
$_{m,a'b}\la{1, \ldots, m;xp_a,p_b}||{1, \ldots, m;xp_a,p_b}\ra_{m,a'b}$.
If there are two initial-state hadrons in the collision, then we also have 
to add the integral of ${\hat\sigma}^{\aNLO\,\{m\}}_{ab}(x;p_a,xp_b,\mu_F^2)$,
which has a completely analogous expression to Eq.~(\ref{eq:NLOppx}),
apart from the replacements $xp_a \to p_a, p_b \to xp_b$ and
$\sum_{a'} \to \sum_{b'}$.

Equations (\ref{eq:LOppfin}), (\ref{eq:NLOppmp}) and (\ref{eq:NLOppx})
are directly evaluated in four space-time dimensions.  As for
Eq.~(\ref{eq:NLOppm}), one should first cancel analytically the $\ep$
poles of the one-loop matrix element with those of the insertion
operator ${\bom I}$, perform the limit $\ep \to 0$ and then carry out
the phase-space integration in four space-time dimensions.

In the following subsections we derive the explicit form of the
${\bom I}$, ${\bom K}$ and ${\bom P}$ insertion operators, for the
various specific cases, relevant to lepton-lepton, lepton-hadron and
hadron-hadron scattering.

\subsection{Jet cross sections with no initial-state hadrons}
\label{subsec:xsee}

In processes with no initial-state hadrons, the QCD cross section for jet
observables at NLO accuracy is given by Eqs.~(\ref{eq:sigma}) and
(\ref{eq:sigmaNLO}), and the NLO contribution is rewritten into the form
of Eq.~(\ref{eq:sigmaNLOsub}).  In order to calculate the integral of
the subtraction term over the factorized one-parton phase space, we
start by writing $\int_{m+1} \rd\sigma^\rA$ using the notation
introduced in Eqs.~(\ref{eq:LOppfin}), (\ref{eq:dPhi}) and the
phase-space factorization of Eq.~(\ref{eq:psfact}). We find that the
integral of the subtraction term can be written in the following form: 
\beeq
\label{eq:intsigmaAfin}
\int_{m+1}\!\!\!\rd\sigma^\rA = {\cal N}_{in} \sum_{\{ m+1 \} }
\,\sum_{\mathrm{pairs}\atop i,j} \;\sum_{k\not=i,j}
\aand \int_m\!
\rd\phi_m(p_1, \ldots, \tpij, \ldots, \tpk, \ldots, p_{m+1};Q)
\nn \\ \nn && \quad
\times
\frac{1}{S_{\{ m+1 \} }}
F_J^{(m)}(p_1, \ldots, \tpij, \ldots, \tpk, \ldots, p_{m+1})
\\ && \quad
\times
\int [\rd p_i(\tpij, \tpk)]\,{\cal D}_{ij,k}(p_1, \ldots, p_{m+1})\:.
\eeeq
For the integration over $\rd p_i$ we write the dipole factor
as in Eq.~(\ref{eq:Dijk}),
\beq
\label{eq:Dijk2}
{\cal D}_{ij,k}(p_1, \ldots, p_{m+1}) =
- \left[ \frac{{\bom V}_{ij,k}}{(p_i + p_j)^2 - m_{ij}^2}\,
\frac{1}{{\bom T}_{ij}^2} \,
|{\cal{M}}_{m}^{ij,k}(p_1, \ldots, \tpij, \ldots, \tpk, \ldots, p_{m+1})|^2
\right]_h\:,
\eeq
where ${\cal{M}}_{m}^{ij,k}$ is a colour-correlated $m$-parton amplitude
(see Eq.~(\ref{eq:colam})) depending only on
$p_1, \ldots \tpij, \tpk, \ldots, p_{m+1}$
while ${\bom V}_{ij,k}/[(p_i + p_j)^2 - m_{ij}^2]$ depends only on
$p_i,p_j,p_k$ or, equivalently, $p_i, \tpij, \tpk$. The subscript $h$
on the square bracket in Eq.~(\ref{eq:Dijk2}) means that
${\bom V}_{ij,k}$ and ${\cal{M}}_{m}^{ij,k}$ are still coupled in
helicity space. According to the discussion in
Sect.~\ref{subsub:FFinteg}, this helicity-space coupling vanishes after
integration over $[\rd p_i(\tpij, \tpk)]$, and the integral of the
subtraction term remains the same if we exchange the splitting matrices
for the azimuthally averaged splitting matrices.  Then we can perform
the counting of the symmetry factors in order to change $S_{\{ m+1 \} }$
to $S_{\{ m \} }$, which follows exactly the procedure of the massless
case\cite{Catani:1997vz} leading to Eq.~(\ref{eq:intsigmaAfin})
rewritten in a completely factorized form of an $m$-parton contribution
multiplied by a singular integral: 
\beeq
\label{eq:dsafin}
\int_{m+1}\!\!\!\rd\sigma^\rA \aand = - {\cal N}_{in} \sum_{\{ m \} }
\int_m \!\rd\phi_{m}(p_1, \ldots , p_m;Q)
\, \frac{1}{S_{\{ m \} }} \;F_J^{(m)}(p_1, \ldots, p_m) 
\nn \\ && 
\times \sum_j\;\sum_{k\neq j} \,\frac{1}{{\bom T}_j^2}\,
|{\cal{M}}_{m}^{j,k}(p_1, \ldots, p_m) |^2
\left(\int [\rd p_i(\tpij, \tpk)]\,
\frac{\langle\bV_{ij,k}\rangle}{(p_i+p_j)^2-m_{ij}^2}
\right)_{{\widetilde {ij} \to j} \atop {{\tilde k} \to k}}.
\hspace*{2em}
\eeeq
The substitutions in the last factor are to be performed after the
integral is calculated using the results in Sect.~\ref{subsub:FFinteg}.
In this context we observe that the colour-connected $m$-parton
amplitude, ${\cal{M}}_{m}^{j,k}$ is symmetric in its upper indices, and
we sum over all pairs of $j$,$k$ in Eq.~(\ref{eq:dsafin}). 
Consequently, the antisymmetric part of
the eikonal integral, $I^{\eik}_-$ does not contribute to the integral
of $\rd\sigma^\rA$.

After these manipulations, the final result for
$\int_{m+1}\!\d\sigma^\rA$ in Eq.~(\ref{eq:dsafin})
can be written as follows,
\beq
\label{eq:dsaf}
\int_{m+1}\!\!\!\d\sigma^\rA =
\int_m\!\left[\d\sigma^\rB \otimes {\bom I}_m(\ep,\mu^2;\{p_i,m_i\})
\right] \:,
\eeq
where the notation $\left[\d\sigma^\rB \otimes {\bom I}_m \right]$
on the right-hand side means that one has to write down the expression
for $\rd\sigma^\rB$ and then replace the corresponding matrix element
squared at the Born level
\beq
\M{}{m} = {}_m\!\langle{1, \ldots, m}||{1,  \ldots, m}\rangle_m \:,
\eeq
by
\beeq
{}_m\!\langle{1, \ldots, m}| \;{\bom I}_m(\ep,\mu^2;\{p_i,m_i\})
\;|{1, \ldots, m}\rangle_m \:,
\eeeq
where the insertion operator ${\bom I}_m$ depends on the colour
charges, momenta and masses of the $m$ final-state partons in
$\rd\sigma^\rB$:
\beeq
\label{eq:insop}
&&
{\bom I}_m(\ep,\mu^2;\{p_i,m_i\})  =
- \frac{\as}{2\pi}\,
\frac{(4\pi)^\ep}{\Gamma(1-\ep)}\,
\sum_j \frac{1}{{\bom T}_j^2}
\sum_{k \neq j} {\bom T}_j\cdot{\bom T}_k
\nn \\ && \qquad
\times
\Bigg[
{\bom T}_j^2
\left(\frac{\mu^2}{s_{jk}} \right)^{\ep}
\left(\cV_j(s_{jk},m_j,m_k,\{m_F\};\ep,\kappa)
     - \frac{\pi^2}{3}\right)
+ \Gamma_j(\mu,m_j,\{m_F\};\ep) 
\nn \\ && \qquad\qquad
+ \gamma_j \ln\frac{\mu^2}{s_{jk}}
+ \gamma_j + K_j
+ {\rm O}(\eps)
\Bigg]\:,
\eeeq
where $s_{jk} = 2 p_j p_k$
and $\{m_F\}$ denotes the set of masses of those massive quarks that
may appear in the $g \to q\bar q$ splitting. The constants $\gamma_a$,
were defined in Eq.~(\ref{eq:gamma_g}), while $K_a$ denote the constants
introduced in Ref.~\cite{Sud}, 
\beq
K_q = \left( \frac{7}{2} - \frac{\pi^2}{6} \right) \CF\:,\qquad
K_g = \left( \frac{67}{18} - \frac{\pi^2}{6} \right) \CA
- \frac{10}{9} \TR N_f \:.
\label{eq:K_g}
\eeq

In Eq.~(\ref{eq:insop}) the $\cV_j$ kernels depend on the flavour of
parton $j$ and on the momenta and masses of both partons $j$ and $k$.
Thus colour correlations are accompanied by complicated mass
correlations.  We decompose these functions into a sum of two
contributions: the first is symmetric with respect to interchange of
the indices $j$ and $k$ and singular in four dimensions or for
vanishing masses and the second is neither symmetric nor singular,
\beq
\label{eq:cV_decomp}
\cV_j(s_{jk},m_j,m_k,\{m_F\};\ep,\kappa) =
\cV^{(\rm S)}(s_{jk},m_j,m_k;\ep)
+ \cV^{(\rm NS)}_j(s_{jk},m_j,m_k,\{m_F\};\kappa)
\:.
\eeq
The singular terms have the following form:
\beq
\label{eq:cVS_universal}
\cV^{(\rm S)}(s_{jk},m_j,m_k;\ep) =
\frac1{v_{jk}}\,\left(\frac{Q_{jk}^2}{s_{jk}}\right)^\ep\,
\left[ \frac{1}{\ep^2}
\left(1-\frac{1}{2}\rho_j^{-2\ep}-\frac{1}{2}\rho_k^{-2\ep}\right) 
-\frac{\pi^2}{12}\Big(\Theta(m_j) + \Theta(m_k) \Big) \right]\:,
\eeq
where we introduced the abbreviation
$Q_{jk}^2 \equiv s_{jk} + m_j^2 + m_k^2$ and $\Theta(x)$ is the step
function with $\Theta(0) = 0$.
Here and in the following the quantities $\rho$, $\rho_j$ and $\rho_k$ 
are obtained from Eq.~(\ref{eq:rhon}) with the substitutions
$\mu_n^2 \to m_n^2/Q_{jk}^2$ ($n = j,$ $k$) and $\tvijk \to v_{jk}$.
In practice the expanded forms of $\cV^{(\rm S)}$ are needed,
\beeq
\cV^{(\rm S)}(s_{jk},m_j>0,m_k>0;\ep) \aand =
\frac1{v_{jk}}\,
\Bigg[
\frac{1}{\ep}\ln\rho -\frac14\ln^2\!\rho_j^2 -\frac14\ln^2\!\rho_k^2 
- \frac{\pi^2}{6}
\Bigg]
\nn \\ \aand \quad
+\frac1{v_{jk}}\,\ln\rho\ln\left(\frac{Q_{jk}^2}{s_{jk}}\right)\:,
\nn \\ 
\cV^{(\rm S)}(s_{jk},m_j>0,0;\ep) \aand =
\frac{1}{2\ep^2} +\frac{1}{2\ep}\ln\frac{m_j^2}{s_{jk}}
- \frac14\ln^2\frac{m_j^2}{s_{jk}}
-\frac{\pi^2}{12}
\nn \\ \aand \quad
- \frac12\ln\frac{m_j^2}{s_{jk}}\,\ln\frac{s_{jk}}{Q_{jk}^2}
- \frac12\ln\frac{m_j^2}{Q_{jk}^2}\,\ln\frac{s_{jk}}{Q_{jk}^2}
\label{eq:cVS}
\:,
\nn\\ 
\cV^{(\rm S)}(s_{jk},0,0;\ep) \aand = \frac{1}{\ep^2}\:.
\eeeq
Of course, the decomposition defined in Eq.~(\ref{eq:cV_decomp}) is not
unique. One can still shift finite terms between the non-singular and
singular terms.
In particular, in Ref.~\cite{Catani:2001ef} we have not included those
terms in the singular parts that vanish for vanishing masses (terms in
the second lines of the equations in (\ref{eq:cVS})).

The non-singular terms depend on the flavours and masses of partons $j$
and $k$.  If $j$ is a quark, then $\cV^{(\rm NS)}_q$ does not depend on
$\{m_F\}$ and $\kappa$, so we suppress these arguments.  If both $j$
and $k$ are massive quarks, then
\beeq
\label{eq:VNS_QQ}
&&
\cV^{(\rm NS)}_q(s_{jk},m_j,m_k) =
\frac{\gamma_q}{{\bom T}_q^2}\ln\frac{s_{jk}}{Q_{jk}^2}
\nn\\ \nn && \qquad {}
+ \frac1{v_{jk}}\,
\left[
\ln\rho^2\ln(1+\rho^2)
+ 2\Li_2(\rho^2)- \Li_2(1-\rho_j^2) - \Li_2(1-\rho_k^2)
- \frac{\pi^2}{6} \right]
\\ \nn && \qquad {}
+ \ln\frac{Q_{jk} - m_k}{Q_{jk}}
- 2\ln\frac{(Q_{jk}-m_k)^2-m_j^2}{Q_{jk}^2}
- \frac{2m_j^2}{s_{jk}}\,\ln\frac{m_j}{Q_{jk} - m_k}
\\ && \qquad {}
- \frac{m_k}{Q_{jk} - m_k} + \frac{2m_k(2 m_k - Q_{jk})}{s_{jk}}
+\frac{\pi^2}{2}\:,
\eeeq
where we used the abbreviation $Q_{jk}=\sqrt{Q_{jk}^2}$.
If $j$ is a massive quark, but $k$ is a massless parton, this
reduces to
\beq
\cV^{(\rm NS)}_q(s_{jk},m_j,0) =
\frac{\gamma_q}{{\bom T}_q^2}\ln\frac{s_{jk}}{Q_{jk}^2}
+ \frac{\pi^2}{6}
- \Li_2\left(\frac{s_{jk}}{Q_{jk}^2}\right)
- 2\,\ln\frac{s_{jk}}{Q_{jk}^2}
- \frac{m_j^2}{s_{jk}}\,\ln\frac{m_j^2}{Q_{jk}^2} \:.
\eeq
If $j$ is a massless quark and $k$ is a massive parton, then
\beq
\label{eq:VNS_qQ}
\cV^{(\rm NS)}_q(s_{jk},0,m_k) =
\frac{\gamma_q}{{\bom T}_q^2}
\left[\ln\frac{s_{jk}}{Q_{jk}^2}-2\ln\frac{Q_{jk}-m_k}{Q_{jk}}
-\frac{2m_k}{Q_{jk}+m_k} \right]
+ \frac{\pi^2}{6} 
- \Li_2\left(\frac{s_{jk}}{Q_{jk}^2}\right)\:.
\eeq

If $j$ is a gluon, then 
\beeq
&&
\cV^{(\rm NS)}_j(s_{jk},0,m_k,\{m_F\};\kappa) =
\nn\\ \nn && \qquad
\frac{\gamma_g}{{\bom T}_g^2}
\left[\ln\frac{s_{jk}}{Q_{jk}^2}-2\ln\frac{Q_{jk}-m_k}{Q_{jk}}
-\frac{2m_k}{Q_{jk}+m_k} \right]
+ \frac{\pi^2}{6} 
- \Li_2\left(\frac{s_{jk}}{Q_{jk}^2}\right)
\nn\\ \nn && \qquad
+\frac43 \frac{\TR}{\CA}\sum_{F=1}^{N_F^{jk}}\Bigg[
 \ln\frac{Q_{jk}-m_k}{Q_{jk}}
+\frac{m_k\,\rho_1^3}{Q_{jk}+m_k}
+\ln\frac{1+\rho_1}{2} 
-\frac{\rho_1}{3}(3+\rho_1^2)
-\frac12\ln\frac{m_F^2}{Q_{jk}^2}
\Bigg]
\nn\\ \nn && \qquad
+\frac23 \frac{\TR}{\CA} \sum_{F=1}^{N_F}\ln\frac{m_F^2}{Q_{\rm aux}^2}
+ \left(\kappa - \frac23\right)\frac{m_k^2}{s_{jk}}
\Bigg[\left(2\frac{\TR}{\CA} N_f - 1\right)\,\ln\frac{2m_k}{Q_{jk} + m_k}
\\  && \qquad\qquad\qquad\qquad\quad {}
+2\frac{\TR}{\CA}\sum_{F=1}^{N_F^{jk}}\left(
\rho_2^3\ln\left(\frac{\rho_2-\rho_1}{\rho_2+\rho_1}\right)
-\ln\left(\frac{1-\rho_1}{1+\rho_1}\right)
-\frac{8\rho_1 m_F^2}{s_{jk}} \right)
\Bigg],
\label{eq:VNS_gQ}
\hspace{2em}
\eeeq
where $N_F$ is the number of heavy flavours and $N_F^{jk}$ is the number of
those heavy flavours for which $s_{jk} > 4 m_F (m_F + m_k)$. Note that 
for $m_F\to0$, we can set $N_F^{jk}=N_F$ because the region of vanishing
$s_{jk}$ gives a vanishing contribution to infrared-safe observables. Thus
$\cV^{(\rm NS)}_j(s_{jk},0,m_k,\{m_F\};\kappa)$ is finite in the
massless limit. The auxiliary mass scale $Q_{\rm aux}$ has been
introduced only to ensure this finiteness property. It can be chosen
arbitrarily\footnote{In Ref.~\cite{Catani:2001ef} $Q_{\rm aux}$ was
chosen to be the renormalization scale.} and will cancel against a
similar contribution in $\Gamma_g(\{m_F\};\ep)$
(cf.~Eq.~(\ref{eq:cgammag})).  In Eq.~(\ref{eq:VNS_gQ}) $\rho_1$ and
$\rho_2$ as defined in Eq.~(\ref{eq:rho_idef}) can be rewritten as
\beq
\label{eq:rho_idef2}
\rho_1 = \sqrt{1-\frac{4m_F^2}{(Q_{jk}-m_k)^2}}\:, \qquad
\rho_2 = \sqrt{1-\frac{4m_F^2}{Q_{jk}^2-m_k^2}}\:.
\eeq
We see that choosing $\kappa = 2/3$ simplifies Eq.~(\ref{eq:VNS_gQ})
considerably.

Finally, if both $j$ and $k$ are massless partons, then
$\cV^{(\rm NS)}_q(s_{jk},0,0) = 0$ and
Eq.~(\ref{eq:VNS_gQ}) reduces to
\beeq
&&
\cV^{(\rm NS)}_g(s_{jk},0,0,\{m_F\}) =
\nn \\ && \qquad\qquad
\frac43 \frac{\TR}{\CA}\sum_{F=1}^{N_F^{jk}}\Bigg[
\ln\frac{1+\rho_1}{2} 
-\frac{\rho_1}{3}(3+\rho_1^2)
-\frac12\,\ln\frac{m_F^2}{s_{jk}}
\Bigg]
+ \frac23 \frac{\TR}{\CA}\sum_{F=1}^{N_F}\ln\frac{m_F^2}{Q_{\rm aux}^2}\:,
\hspace*{2em}
\label{eq:VNS_gk}
\eeeq
where we dropped $\kappa$ from the arguments because the dependence on
$\kappa$ in Eq.~(\ref{eq:VNS_gQ}) is proportional to the mass of parton
$k$. If the heavy partons are completely absent, then 
$\cV^{(\rm NS)}_g(s_{jk},0,0,\{\}) = 0$, with $\{\}$ meaning the empty
set.

The functions $\Gamma_j$ depend on the flavour of the
parton $j$ and on the parton masses. In the case of gluons and
massless quarks (antiquarks) we have \cite{Catani:1997pk}
(with dummy arguments suppressed)
\beeq
\label{eq:cgammag}
\Gamma_g(\{m_F\};\ep) \aand =
\frac{1}{\ep} \, \gamma_g 
- \frac{2}{3}\,\TR \sum_{F=1}^{N_F} \ln\frac{m_F^2}{Q_{\rm aux}^2}\:,
\\
\label{eq:cgammaq0}
\Gamma_q(\ep) \aand = \frac{1}{\ep} \, \gamma_q\:,
\eeeq
while for massive quarks (antiquarks) we find
\beq
\label{eq:cgammaqm}
\Gamma_q(\mu,m_q;\ep)=
{\bom T}_q^2 \left( \frac{1}{\ep} - \ln\frac{m_q^2}{\mu^2} -2 \right) +
\gamma_q \,\ln\frac{m_q^2}{\mu^2} =
\CF \left[ \frac{1}{\ep} +
\frac{1}{2} \ln\frac{m_q^2}{\mu^2} -2 \right] \:.
\eeq
In Eq.~(\ref{eq:cgammag}) the summation runs over the heavy flavours.

Using formulae (\ref{eq:insop})--(\ref{eq:cgammaqm}), it is easy to
verify that the ${\bom I}_m(\ep,\mu^2;\{p_i,m_i\})$ operator tends to
the corresponding operator of the massless theory smoothly,
\beq
\label{eq:Immassless}
\lim_{\{m_i\} \to 0}{\bom I}_m(\ep,\mu^2;\{p_i,m_i\}) =
{\bom I}(\ep) + \Oe{}\:,
\eeq
where ${\bom I}(\ep)$ is defined in Eq.~(7.26) of Ref.~\cite{Catani:1997vz}.

The NLO QCD cross section in Eq.~(\ref{eq:sigmaNLOsub}) is finite for 
infrared-safe jet observables. We constructed the subtraction terms
such that the first term is not only finite independently of the parton
masses, but it also tends to the corresponding massless expressions
smoothly if $F_J$ is quasi-collinear safe (see
Eq.~(\ref{eq:smoothlimit})). Therefore, the second term, spelled out in
Eq.~(\ref{eq:NLOppm}), also has to be finite and, using
Eq.~(\ref{eq:Immassless}), tends smoothly to the corresponding massless
expressions in the
zero-mass limit. Thus all the $\ep$ poles and $\ln(m_i)$ terms that are
singular in the zero-mass limit in $\d\sigma^\rV$ must be cancelled by
corresponding terms in $\d\sigma^\rB \otimes {\bom I}_m(\ep)$, which
enables us to write down the singular terms of the one-loop amplitudes,
as spelled out explicitly in Ref.~\cite{Catani:2001ef}.

\subsection{Jet cross sections with one initial-state hadron}
\label{subsec:oneini}

As mentioned before,
in this paper we consider only massless partons in the initial state,
therefore, the implementation of the subtraction procedure for processes
with hadrons in the initial state is the same as described in
Ref.~\cite{Catani:1997vz} (see also Sect.~(\ref{subsec:final})).
The presence of the parton masses induces non-trivial changes in the
evaluation of the sum of the integral of the full subtraction term and
the collinear counterterm,
\beq
\label{eq:sigmaA+C}
\int_{m+1}\!\!\! \rd\sigma_a^\rA(p_a) +  \int_m\! \rd\sigma_a^\fact(p_a)\:,
\eeq
where the index $a$ refers to the flavour of the incoming parton
and $p_a$ to its momentum. The full subtraction term $\rd\sigma_a^\rA(p_a)$ is
a sum of three terms, corresponding to the first three terms in
Eq.~(\ref{eq:dff}),
\beq
\label{eq:sigmaAoneini}
\rd\sigma_a^\rA(p_a) =
\rd\sigma_a^{\rA'}(p_a) + \rd\sigma_a^{\rA''}(p_a) +
\rd\sigma_a^{\rA'''}(p_a)\:.
\eeq
The first term on the right-hand side is the subtraction term for
final-state singularities with final-state spectators and has the same
form as Eq.~(\ref{eq:intsigmaAfin}):
\beeq
\label{eq:intsigmaA'}
\int_{m+1}\!\!\!\rd\sigma_a^{\rA'} = \frac{ {\cal N}_{in} }{n_s(a) \cF(p_a)}
 \sum_{\{ m+1 \} }
\,\sum_{\mathrm{pairs}\atop i,j} \;\sum_{k\not=i,j}
\aand \int_m\!
\rd\phi_m(p_1, \ldots, \tpij, \ldots, \tpk, \ldots, p_{m+1};p_a + Q)
\nn \\ && \quad
\times
\frac{1}{S_{\{ m+1 \} }}
F_J^{(m)}(p_1, \ldots, \tpij, \ldots, \tpk, \ldots, p_{m+1};p_a)
\nn \\ && \quad
\times
\int [\rd p_i(\tpij, \tpk)]\,{\cal D}_{ij,k}(p_1, \ldots, p_{m+1};p_a)\:,
\eeeq
where the flux factor fulfils the following scaling property:
\beq
\label{eq:fluxscaling}
\cF(\eta p_a) = \eta \cF(p_a)\:.
\eeq
The integration of this term follows the steps of the previous section
and leads to the analogous result, which, for later purposes, we write in a
different form:
\beq
\label{eq:intsigmaA'2}
\int_{m+1}\!\!\!\d\sigma_a^{\rA'} =
\sum_{a'}\int_0^1\!\rd x 
\int_m\!\left[\d\sigma_{a'}^\rB(x p_a) \otimes 
\Big(\delta^{aa'}\,\delta(1-x)\, {\bom I}_m(\ep)\Big) \right] 
\:,
\eeq
with ${\bom I}_m(\ep)$ given in (\ref{eq:insop}).

The second term on the right-hand side of Eq.~(\ref{eq:sigmaAoneini})
is the subtraction term for final-state singularities with initial-state
spectators. We have seen in Eq.~(\ref{eq:psconv}) that the phase-space
factorization leads to a convolution over the boost parameter $x$, thus
\beeq
\label{eq:intsigmaA''}
\int_{m+1}\!\!\!\rd\sigma_a^{\rA''} = \frac{ {\cal N}_{in} }{n_s(a) \cF(p_a)}
 \sum_{\{ m+1 \} }
\,\sum_{\mathrm{pairs}\atop i,j}
\aand 
\int_0^1\!\rd x
\int\!\rd\phi_m(p_1, \ldots, \tpij, \ldots, p_{m+1};x p_a + Q)
\nn \\ && \quad
\times
\frac{1}{S_{\{ m+1 \} }}
F_J^{(m)}(p_1, \ldots, \tpij, \ldots, p_{m+1}; \tpa)
\nn \\ && \quad
\times
\int [\rd p_i(\tpij; p_a, x)]\,{\cal D}_{ij}^a(p_1, \ldots, p_{m+1}; p_a)\:.
\eeeq
Using the definition of the dipole function in Eq.~(\ref{eq:fidipole}),
following the same manipulations as for the integral of the
$\rd\sigma^{\rA'}$ term and finally making use of the scaling property
in Eq.~(\ref{eq:fluxscaling}), we find that Eq.~(\ref{eq:intsigmaA''})
can be recast in the following form:
\beeq
\label{eq:intsigmaA''2}
\int_{m+1}\!\!\!\rd\sigma_a^{\rA''} = 
\sum_{a'}\int_0^1\!\rd x
\int_m\!\left[\d\sigma_{a'}^\rB(x p_a) \otimes 
\Big(\delta^{aa'}\,{\bom I}_{m,a'}(x;\ep)\Big) \right] \:.
\eeeq
The new insertion operator is
\beq
\label{eq:insop_b}
{\bom I}_{m,a'}(x;\ep,\mu^2;\{p_i,m_i\}, p_a)  =
- \frac{\as}{2\pi}\,
\frac{(4\pi)^\ep}{\Gamma(1-\ep)}
\sum_j {\bom T}_j \cdot {\bom T}_{a'}
\left(\frac{\mu^2}{s_{ja}} \right)^{\ep}
\frac{1}{{\bom T}_j^2}\,\cV_j(x;s_{ja},m_j,\{m_F\};\ep)\:,
\eeq
where $s_{ja}=2p_j p_a$ . We emphasize that $p_a$ is the original
initial-state momentum of the incoming parton, while the final-state
momentum $p_j$ belongs to the boosted phase space with incoming
momentum $xp_a$.  The flavour kernel for a massive quark (or antiquark)
$j$ is
\beeq 
&&
\cV_q(x;s_{ja},m_q,\{\};\ep) =
\CF \left[J_{gQ}^a\Big(x,m_q/\sqrt{s_{ja}}\Big)\right]_+
+ \delta(1-x)\left[ \Gamma_q(\sqrt{s_{ja}}, m_q; \ep) + K_q\right]
\nn \\ && \quad
+ \delta(1-x)\,\CF \left[\frac{1}{\ep}\ln\frac{m_q^2}{s_{ja} + m_q^2}
- \frac12 \ln^2\!\frac{m_q^2}{s_{ja}}
- \frac{\pi^2}{2}
+ J_{gQ}^{a;\rm NS}\left(\frac{m_q}{\sqrt{s_{ja}}}\right)\right]\:.
\eeeq
The function $\Gamma_q(\sqrt{s_{ja}}, m_q; \ep)$ is that defined in
Eq.~(\ref{eq:cgammaqm}) with the $\mu \to \sqrt{s_{ja}}$ substitution.
Here and in the following, $\{\}$ denotes the empty set.
For a massless quark $j=q$ the flavour kernel is given by
\beq
\cV_q(x;s_{ja},0,\{\};\ep) = 
\CF\left[J_{gQ}^a(x,0)\right]_+
+ \delta(1-x)\left[\CF\left(\frac{1}{\ep^2}
- \frac{\pi^2}{3}\right) + \Gamma_q(\ep) + K_q\right]\:.
\eeq
If $j$ is a gluon, then
\beeq
\cV_g(x;s_{ja},0,\{m_F\};\ep) \aand = \CA \left[J_{gg}^a(x)\right]_+
+ \TR N_f \left[J_{Q\bar Q}^a(x,0)\right]_+
\nn \\ &&
+ \delta(1-x)\left[\CA\left(\frac{1}{\ep^2}
- \frac{\pi^2}{3}\right) + \Gamma_g(\{m_F\}; \ep)
+ K_g\right]
\nn \\ &&
+ \TR \sum_{F=1}^{N_F^{ja}}\Bigg\{
\left[J_{Q\bar Q}^a \left(x,\frac{m_F}{\sqrt{s_{ja}}}\right)\right]_{x_+}
+ \delta(x_+-x)
\left[J_{Q\bar Q}^{a;\rm NS}\left(\frac{m_F}{\sqrt{s_{ja}}}\right)
-\frac{10}{9}\right]\Biggr\}
\nn \\ &&
+ \frac{2}{3}\,\TR\Biggl[
\delta(1-x) \, \sum_{F=1}^{N_F} \ln\frac{m_F^2}{Q_{\rm aux}^2}
- \delta(x_+-x) \, \sum_{F=1}^{N_F^{ja}} \ln\frac{m_F^2}{s_{ja}}
\Biggr],
\eeeq
where $x_+ = 1 - 4 m_F^2/s_{ja}$. Note that the last line, containing the
difference of two $\delta$ functions, gives a contribution to
Eq.~(\ref{eq:intsigmaA''2}) that can easily be calculated and
has a smooth massless limit, since $N_F^{ja}$ can be replaced by $N_F$
as $m_F\to0$.

The third term on the right-hand side of Eq.~(\ref{eq:sigmaAoneini})
is the subtraction term for initial-state singularities with final-state
spectators. The relevant phase-space factorization, given in
Eq.~(\ref{eq:psconv}), is again a convolution,
\beeq
\label{eq:intsigmaA'''}
\int_{m+1}\!\!\!\rd\sigma_a^{\rA'''} = \frac{ {\cal N}_{in} }{n_s(a) \cF(p_a)}
 \sum_{\{ m+1 \} }
\,\sum_i\sum_{j\ne i}
\aand 
\int_0^1\!\rd x
\int\!\rd\phi_m(p_1, \ldots, \tpj, \ldots, p_{m+1};x p_a + Q)
\nn \\ && \quad
\times
\frac{1}{S_{\{ m+1 \} }}
F_J^{(m)}(p_1, \ldots, \tpj, \ldots, p_{m+1}; \tpai)
\nn \\ && \quad
\times
\int [\rd p_i(\tpj; p_a, x)]\,{\cal D}_j^{ai}(p_1, \ldots, p_{m+1}; p_a)\:.
\hspace*{2em}
\eeeq
The evaluation of the integral over $[\rd p_i(\tpj; p_a, x)]$ follows
that in the massless case, using the explicit results in
Sect.~\ref{subsub:IFinteg}. Then the counting of the symmetry factors
is also straightforward and using the scaling property in
Eq.~(\ref{eq:fluxscaling}), we find that Eq.~(\ref{eq:intsigmaA'''})
can be recast in the following form:
\beeq
\label{eq:intsigmaA'''2}
\int_{m+1}\!\!\!\rd\sigma_a^{\rA'''} = 
\sum_{a'}\int_0^1\!\rd x
\int_m\!\left[\d\sigma_{a'}^\rB(x p_a) \otimes 
{\bom I}_{m,aa'}(x;\ep) \right] \:,
\eeeq
where the insertion operator is
\beq
\label{eq:insop_ab}
{\bom I}_{m,aa'}(x;\ep,\mu^2;\{p_i,m_i\}, p_a) =
- \frac{\as}{2\pi}\,
\frac{(4\pi)^\ep}{\Gamma(1-\ep)}
\sum_j {\bom T}_j \cdot {\bom T}_{a'}
\left(\frac{\mu^2}{s_{ja}} \right)^{\ep}
\frac{1}{{\bom T}_{a'}^2}\,\cV^{a,a'}(x;s_{ja},m_j;\ep)\:.
\eeq
The functions $\cV^{a,a'}(x;s_{ja},m_j;\ep)$ are related to the
$I_j^{ab}(x;\mu_j;\ep)$ functions given in Eq.~(\ref{eq:Ijab}),
\beq
\cV^{a,a'}(x;s_{ja},m_j;\ep) =
I_j^{aa'}\left(x;\frac{m_j}{\sqrt{s_{ja}}};\ep\right)\:.
\eeq
Note that $\rd\sigma_a^{\rA'''}$ is the only contribution to 
$\rd\sigma_a^{\rA}$ in (\ref{eq:sigmaAoneini}) where the flavours
of the partons $a$ and $a'$ can be different. However, both $a$ and $a'$
are massless, since we do not allow for massive partons in the
initial state.
The fact that $\cV^{g,a'}$ does not receive contributions
from the splitting $g\to Q\bar Q$ into massive quarks $Q$ is not
a restriction of the presented formalism, it simply means that
the logarithmic corrections $\alps\ln m_Q$ of this origin are not
extracted from the real correction $\rd\sigma^\rR$ in (\ref{eq:sigmaNLO}).
These terms could be extracted by a further subtraction into
suitably-defined parton distributions, as discussed at the end of
Sect.~\ref{se:subwh}.

Next we sum Eqs.~(\ref{eq:intsigmaA'2}), (\ref{eq:intsigmaA''2}),
(\ref{eq:intsigmaA'''2}) and the well-known expression for the
collinear-subtraction counterterm,
\beq
\label{eq:saC}
d\sigma_a^{C}(p_a;\mu_F^2) =
- \frac{\as}{2 \pi}\,\frac{(4 \pi)^\ep}{\Gamma(1-\ep)}
\sum_{a'}
\int_0^1\!\rd x\,
\left[ - \frac{1}{\ep}
\left(\frac{\mu^2}{\mu_F^2}\right)^{\ep} P^{aa'}(x) + \Kab(x)
\right]
\,\rd\sigma_{a'}^\rB(x p_a) \:,
\eeq
to obtain the sum of the full subtraction term and the collinear
counterterm in Eq.~(\ref{eq:sigmaA+C}). In Eq.~(\ref{eq:saC})
$\mu_F$ is the factorization scale and the functions $\Kab(x)$ define
the factorization scheme. The \msbar\ scheme is defined by
$\Kab(x) = 0$.  More on Eq.~(\ref{eq:saC}) can be found e.g.~in Sect.~6
of Ref.~\cite{Catani:1997vz}.  We see that
$\int_{m+1}\!\rd\sigma_a^\rA(p_a) +  \int_m\!\rd\sigma_a^\fact(p_a)$ is
obtained from the leading-order expression
$\int_m\!\rd\sigma_a^\rB(xp_a)$ by replacing the leading-order matrix
element squared
\beq
\frac{1}{n_s(a)} \, {}_{m,a}\la{...;p_a}||{...;p_a}\ra_{m,a} \:,
\eeq
by
\beeq
\sum_{a'} \frac{1}{n_s(a')}
\,{}_{m,a'}\la{...;xp_a}| \;{\bom I}^{a,a'}(x;\ep) \;|{...;xp_a}\ra_{m,a'}
\:,
\eeeq
and performing the $x$-integration. Here the insertion operator
${\bom I}^{a,a'}(x;\ep)$ also depends on the renormalization and factorization
scales, colour charges, momenta, masses and flavours of the QCD partons,
\beeq
&&
\label{eq:Idis1}
\nn
{\bom I}^{a,a'}(x;\ep,\mu^2,\mu_F^2;\{p_i,m_i\}, p_a) =
\delta^{aa'}\delta(1-x)\,{\bom I}_m(\ep,\mu^2;\{p_i,m_i\})
\\[.5em] \nn && \qquad {}
+ \delta^{aa'} {\bom I}_{m,a'}(x;\ep,\mu^2;\{p_i,m_i\}, p_a)
+ {\bom I}_{m,aa'}(x;\ep,\mu^2;\{p_i,m_i\}, p_a)
\\[.5em] && \qquad {}
- \frac{\as}{2\pi}\,
\frac{(4\pi)^\ep}{\Gamma(1-\ep)}
\Bigg[
-\frac{1}{\ep}\left(\frac{\mu^2}{\mu_F^2}\right)^{\ep}\,P^{aa'}(x)
+\Kab(x)
\Bigg]
+\Oe{}\:.
\eeeq
One can understand the structure of the ${\bom I}^{a,a'}$ insertion
operator, if we rewrite it as
\beeq
&&
\label{eq:Idis2}
{\bom I}^{a,a'}(x;\ep,\mu^2,\mu_F^2;\{p_i,m_i\}, p_a) =
\delta^{aa'}\delta(1-x)\,{\bom I}_{m+a}(\ep,\mu^2;\{p_i,m_i\}, p_a)
\\[.5em] \nn && \qquad {}
+ {\bom P}^{a,a'}_m(x;\mu_F^2;\{p_i\}, x p_a)
+ {\bom K}^{a,a'}_m(x;\{p_i,m_i\}, p_a)
+\Oe{}\:,
\eeeq
where the operator ${\bom I}_{m+a}(\ep,\mu^2;\{p_i,m_i\}, p_a)$ is similar
to the operator defined in Eq.~(\ref{eq:insop}), 
\beq
\label{eq:insop1}
{\bom I}_{m+a}(\ep,\mu^2;\{p_i,m_i\},p_a)  =
{\bom I}_m(\ep,\mu^2;\{p_i,m_i\}) +
{\bom I}_a(\ep,\mu^2;\{p_i,m_i\},p_a)\:,
\eeq
where
\beeq
&&
{\bom I}_a(\ep,\mu^2;\{p_i,m_i\},p_a)  =
- \frac{\as}{2\pi}\, \frac{(4\pi)^\ep}{\Gamma(1-\ep)}\,
\sum_j\Bigg\{
\nn\\ && \qquad
\frac{1}{{\bom T}_j^2}\,{\bom T}_j\cdot{\bom T}_a
\Bigg[
{\bom T}_j^2
\left(\frac{\mu^2}{s_{ja}} \right)^{\ep}
\left(\cV_j(s_{ja},m_j,0,\{m_F\};\ep,\kappa)
     - \frac{\pi^2}{3}\right)
+ \Gamma_j(\mu,m_j,\{m_F\};\ep) 
\nn\\ && \qquad\qquad\qquad\qquad
+ \gamma_j \ln\frac{\mu^2}{s_{ja}}
+ \gamma_j + K_j
\Bigg]
\nn\\ && \qquad
+ \frac{1}{{\bom T}_a^2}\,{\bom T}_a\cdot{\bom T}_j 
\Bigg[
{\bom T}_a^2
\left(\frac{\mu^2}{s_{aj}} \right)^{\ep}
\left(\cV_a(s_{aj},0,m_j,\{\};\ep,2/3)
     - \frac{\pi^2}{3}\right)
+ \frac{\gamma_a}{\eps}
\nn\\ && \qquad\qquad\qquad\qquad
+ \gamma_a \ln\frac{\mu^2}{s_{aj}}
+ \gamma_a + K_a
\Bigg]
\Bigg\}
\:.
\eeeq
The only difference between ${\bom I}_m$ and ${\bom I}_{m+a}$
is that the latter depends on the additional initial-state
parton $a$ such that in the functions $\cV_a$ and $\Gamma_a$ the set
$\{m_F\}$ is empty and $\kappa = 2/3$.  The singular terms $\cV^{\rm
(S)}$ do not depend on the value of $\{m_F\}$ and $\kappa$, and crossing
the momentum of a parton from the final to the initial state does not
change the singular terms in the one-loop QCD amplitudes, therefore,
this insertion operator cancels all the singularities in the virtual
contribution $\int_m\!\rd\sigma_a^\rV(p_a)$.  As a result, the operators
${\bom P}_m^{a,a'}$ and ${\bom K}_m^{a,a'}$ are finite in four
dimensions, as can also be seen from their explicit forms below. 

The ${\bom P}_m^{a,a'}$ operator, which is defined by
\beq
\label{eq:Pop1}
{\bom P}_m^{a,a'}(x;\mu_F^2;\{p_i\}, x p_a) = \frac{\as}{2\pi}\,
P^{aa'}(x)\,\frac{1}{{\bom T}^2_{a'}}
\sum_j {\bom T}_j \cdot {\bom T}_{a'}\,\ln\frac{\mu_F^2}{x s_{ja}} \:,
\eeq
is independent of the presence of massive partons and contains the full
dependence on the factorization scale $\mu_F$. In order to obtain this
form of the $\mu_F$ dependence, we have used colour conservation 
($\sum_j {\bom T}_j = - {\bom T}_{a'}$) and expanded in $\ep$,
\beq
\left[\sum_j {\bom T}_j \cdot {\bom T}_{a'}
\left(\frac{\mu^2}{s_{ja}}\right)^{\ep}\frac{1}{{\bom T}_{a'}^2}
+ \left(\frac{\mu^2}{\mu_F^2}\right)^{\ep}
\right]\frac{1}{\ep}P^{aa'}(x) =
P^{aa'}(x)\,\frac{1}{{\bom T}_{a'}^2}
\sum_j {\bom T}_j \cdot {\bom T}_{a'}\,\ln\frac{\mu_F^2}{s_{ja}}
+\Oe{}\:.
\eeq

The ${\bom K}_m^{a,a'}$ operator contains all remaining terms, including
the factorization-scheme dependence and further finite-mass corrections:
\beeq
\label{eq:Kab1}
\hspace*{-1em}&&
{\bom K}_m^{a,a'}(x;\{p_i,m_i\}, p_a) = \frac{\as}{2\pi}
\Bigg\{ {\Kbar}(x) - \Kab(x)
- \sum_j {\bom T}_j \cdot {\bom T}_{a'}\,
\cK^{a,a'}_j(x; s_{ja}, m_j, \{m_F\})
\nn \\ \hspace*{-1em}&& \qquad
- \frac{1}{{\bom T}^2_{a'}}
\sum_j {\bom T}_j \cdot {\bom T}_{a'}
\Bigg[ P^{aa'}_\reg(x)\,\ln\frac{(1-x) s_{ja}}{(1-x) s_{ja} + m_j^2}
+ \gamma_a\,\delta^{aa'}\,\delta(1-x)
\nn \\ \hspace*{-1em}&& \qquad\qquad\qquad\qquad\quad
\times
 \left( \ln\frac{s_{ja} - 2 m_j \sqrt{s_{ja} + m_j^2}+2m_j^2}{s_{ja}}
      + \frac{2 m_j}{\sqrt{s_{ja} + m_j^2} + m_j}\right)
\Bigg]
\Bigg\}\:.
\hspace*{3em}
\eeeq
Comparing the expressions for the various insertion operators on the
right-hand side of Eq.(\ref{eq:Idis1}), we observe that the finiteness
of ${\bom K}_m^{a,a'}$ is highly non-trivial: terms from the integration
of all possible dipole contributions have to be combined, together with
the collinear subtraction, before taking the $\eps\to0$ limit.

In order to simplify the final formulae, in Eq.~(\ref{eq:Kab1}) we
used the functions $P^{aa'}_\reg(x)$ and $P^{aa'}(x)$, defined in
Eqs.~(\ref{eq:Preg}) and (\ref{eq:Pab}), respectively, and also the
${\Kbar}(x)$ functions defined in Ref.~\cite{Catani:1997vz},
\beeq
&&
{\overline K}^{aa'}(x) =
P^{aa'}_\reg(x) \ln\frac{1-x}{x} + {\hat P}^{\prime\, aa'}(x)
\nn \\ && \qquad
+ \delta^{aa'}\left[
{\bom T}_a^2 \left(\frac{2}{1-x} \ln\frac{1-x}{x}\right)_+ 
-\,\delta(1-x)\left(\gamma_a + K_a - \frac{5}{6}\,\pi^2\,{\bom T}_a^2\right)
\right]
\:.
\label{eq:kbarab}
\eeeq
Recall that the functions ${\hat P}^{\prime\, ab}(x)$ come from the
$\eps$-dependence of the splitting functions and are given in
Eq.~(\ref{eq:Pprime}).

The functions $\cK^{a,a'}_q(x; s_{ja}, m_j, \{m_F\})$ depend on the
flavours of partons $a$, $a'$ and $j$. For the sake of clarity, we
consider the various cases separately. If $j$ is a quark, then
$\cK^{a,a'}_q$ does not depend on $\{m_F\}$ and we have
\beeq
\label{eq:cKgq_q}
\cK^{g,q}_q(x; s_{ja}, m_j) \aand = 0\:,
\\ 
\label{eq:cKqq_q}
\cK^{q,q}_q(x; s_{ja}, m_j) \aand = 
2 \left[\left( \frac{\ln(1-x)}{1-x} \right)_+ - \frac{\ln(2-x)}{1-x}\right]
+ \left[J_{gQ}^a\left(x,\frac{m_j}{\sqrt{s_{ja}}}\right)\right]_+
\nn \\ \aand
  + 2 \left(\frac{1}{1-x}\right)_+
  \ln\frac{(2-x) s_{ja}}{(2-x) s_{ja} + m_j^2}
  - \frac{\gamma_q}{\CF} \delta(1-x)
\nn \\ \aand
  + \delta(1-x)
  \left(\frac{m_j^2}{s_{ja}}\ln\frac{m_j^2}{s_{ja}+m_j^2}
  + \frac12 \frac{m_j^2}{s_{ja}+m_j^2}\right)
\:,
\\ 
\label{eq:cKqg_q}
\cK^{q,g}_q(x; s_{ja}, m_j) \aand = 
2\frac\CF\CA \frac{m_j^2}{x s_{ja}} \ln\frac{m_j^2}{(1-x) s_{ja} + m_j^2}\:,
\\ 
\label{eq:cKgg_q}
\cK^{g,g}_q(x; s_{ja}, m_j) \aand = \cK^{q,q}_q(x; s_{ja}, m_j)
+ \frac\CA\CF \cK^{q,g}_q(x; s_{ja}, m_j)\:,
\eeeq
If $j$ is a gluon, then
\beeq
\label{eq:cKab_g}
&&
\cK^{a,a'}_g(x; s_{ja}, 0, \{m_F\}) =
- \delta^{aa'}\,\frac{\gamma_g}{\CA}
  \left[\left(\frac{1}{1-x}\right)_+ + \delta(1-x)\right]
\nn \\ && \qquad
+ \delta^{aa'}\,\frac\TR\CA \sum_{F=1}^{N_F^{ja}}\Bigg\{
\Big(\delta(1-x) - \delta(x_+-x)\Big)
\left[
\frac{2}{3}\left(\ln\frac{m_F^2}{s_{ja}} + \frac53\right)
- J_{Q\bar Q}^{a;\rm NS}\left(\frac{m_F}{\sqrt{s_{ja}}}\right)
\right]
\nn \\ &&\qquad\qquad\qquad
+ \left[J_{Q\bar Q}^a\left(x,\frac{m_F}{\sqrt{s_{ja}}}\right)\right]_{x_+}
+ \delta(1-x)\,\frac23\left(1-\frac{4 m_F^2}{s_{ja}}\right)^{3/2}
\Bigg\}\:,
\eeeq
where $x_+ = 1 - 4 m_F^2/s_{ja}$ and $N_F^{ja}$ is the number of
flavours with $s_{ja}>4m_F^2$. Note that in this case $m_j = 0$, so 
the terms proportional to $1/{\bom T}_{a'}^2$ in Eq.~(\ref{eq:Kab1})
vanish.  In the zero-mass limit the functions $\cK^{a,a'}_j$ simplify to
\beq
\label{eq:cKj}
\cK^{a,a'}_j(x; s_{ja}, 0, \{\}) =
- \delta^{aa'}\,\frac{\gamma_j}{{\bom T}_j^2}
  \left[\left(\frac{1}{1-x}\right)_+ + \delta(1-x)\right]\:.
\eeq

Using Eq.~(\ref{eq:cKj}), we see immediately that
in the massless limit the operator ${\bom K}_m^{a,a'}$ tends to the
corresponding massless operator smoothly,
\beq
\lim_{\{m_i\} \to 0} {\bom K}_m^{a,a'}(x;\{p_i,m_i\}, p_a) =
{\bom K}_m^{a,a'}(x)\:,
\eeq
where ${\bom K}^{a,a'}_m(x)$ is the same operator as defined in
Eq.~(8.38) of Ref.~\cite{Catani:1997vz}.  

Finally, we explain the actual evaluation of the \plusdist\ 
that contains the kinematic variable $s_{ja}$ in more detail. 
As mentioned at the end of Sect.~\ref{subsub:FIinteg}, 
$s_{ja}$ has to be kept fixed during the $x$ integration if 
$s_{ja}$ appears inside the `+'-prescription, as is the 
case in $\cK^{a,b}_j$ for $a=b$. Schematically we have to 
evaluate an integral of the form 
\beq 
\int_0^1 \rd x \int\rd\Phi(x)\, 
\left[J\left(x,s_{ja}^{(x)}\right)\right]_+ \, |{\cal M}(\Phi(x))|^2 \:,
\eeq 
where the label $(x)$ of $s_{ja}^{(x)}$ indicates that in caclulating
this variable the final-state momentum $p_j$ belongs to the phase
space $\Phi(x)$ of the $x$-boosted frame, but $p_a$ is the original
initial-state momentum of the incoming parton. At first sight this
seems to be very inconvenient, but in practice the procedure can be
simplified according to 
\beeq 
\lefteqn{ 
\int_0^1 \rd x \int\rd\Phi(x) 
\int_{-\infty}^{+\infty} \rd\bar s_{ja} \, 
\delta\left(\bar s_{ja}-s_{ja}^{(x)}\right)\, 
\left[J\left(x,\bar s_{ja}\right)\right]_+ \, |{\cal M}(\Phi(x))|^2 
} && 
\nn\\ 
&=& \int_0^1 \rd x \, 
\int_{-\infty}^{+\infty} \rd\bar s_{ja}\, J\left(x,\bar s_{ja}\right)\, 
\left\{ \int\rd\Phi(x)\, |{\cal M}(\Phi(x))|^2\, 
  \delta\left(\bar s_{ja}-s_{ja}^{(x)}\right) \right. 
\nn\\ 
&& \phantom{\int_0^1 \rd x \, \int_{-\infty}^{+\infty} \rd\bar s_{ja}\, 
            J\left(x,\bar s_{ja}\right)\, \Biggl\{ } \left.{} 
-\int\rd\Phi(1)\,|{\cal M}(\Phi(1))|^2\, 
  \delta\left(\bar s_{ja}-s_{ja}^{(1)}\right) \right\} 
\nn\\ 
&=& \int_0^1 \rd x \, \left\{ 
\int\rd\Phi(x)\, J\left(x,s_{ja}^{(x)}\right) |{\cal M}(\Phi(x))|^2 
-\int\rd\Phi(1)\, J\left(x,s_{ja}^{(1)}\right) |{\cal M}(\Phi(1))|^2 
\right\}. 
\hspace{2em} 
\label{eq:plusdist_sja} 
\eeeq 
Note that the variable $s_{ja}$ that appears in the function $J(x,s_{ja})$ 
is the one in the respective phase-space integral. 
A non-trivial change in the endpoint contribution occurs if 
this variable $s_{ja}$ is eliminated in favour of an $x$-dependent 
variable $f(x)$, such as $f(x)=m_j^2-x\,s_{ja}$ as described in 
App.~\ref{app:Q2fixed}. More precisely, if $s_{ja}^{(1)}$ in 
Eq.~(\ref{eq:plusdist_sja}) is replaced by $f(1)$, the endpoint changes; 
if $s_{ja}^{(1)}$ is replaced by $f(x)$ the endpoint remains
unchanged. 

\subsection{Jet cross sections with two initial-state hadrons}
\label{subsec:twoini}

The derivation of the insertion operators ${\bom I}_{m+a+b}$,
${\bom P}_{m+b}$ and ${\bom K}_{m+b}$, relevant for calculating the NLO
correction of Eqs.~(\ref{eq:NLOppm}) and~(\ref{eq:NLOppx}), does not
contain any new features as
compared to the combination of arguments given in Sect.~10 of
Ref.~\cite{Catani:1997vz} and in the previous subsection. Therefore, we
simply list the corresponding results.

The ${\bom I}_{m+a+b}$ insertion operator is again very similar to ${\bom
I}_m$ of Eq.~(\ref{eq:insop}).  The only difference is that it depends
on the additional initial-state partons $a$ and $b$ such that in the
functions $\cV_{a(b)}$ and $\Gamma_{a(b)}$ the set $\{m_F\}$ is empty
and $\kappa = 2/3$, which does not influence the singularity structure:
\beeq
\label{eq:insop2}
&&
{\bom I}_{m+a+b}(\ep,\mu^2;\{p_i,m_i\},p_a,p_b)  =
\nn \\ &&\qquad
{\bom I}_m(\ep,\mu^2;\{p_i,m_i\}) +
{\bom I}_a(\ep,\mu^2;\{p_i,m_i\},p_a) +
{\bom I}_b(\ep,\mu^2;\{p_i,m_i\},p_b)
\nn \\ &&\qquad
- \frac{\as}{2\pi}\, \frac{(4\pi)^\ep}{\Gamma(1-\ep)}\,
\Bigg(
\frac{1}{{\bom T}_a^2}\,{\bom T}_a\cdot{\bom T}_b 
\Bigg[
\left(\frac{\mu^2}{s_{ab}} \right)^{\ep}
\left(\frac{{\bom T}_a^2}{\ep^2} + \frac{\gamma_a}{\eps} \right)
- {\bom T}_a^2\frac{\pi^2}{3} + \gamma_a + K_a
\Bigg]
+ ( a \leftrightarrow b) \Bigg)
\:.
\nn\\
\eeeq
The operator ${\bom P}_{m+b}$ is completely analogous to ${\bom P}_m$
of Eq.~(\ref{eq:Pop1}) apart from the trivial dependence on the
additional initial-state parton,
\beq
\label{eq:Pop2}
{\bom P}_{m+b}^{a,a'}(x;\mu_F^2;\{p_i\}, x p_a, p_b) = 
{\bom P}_m^{a,a'}(x;\mu_F^2;\{p_i\}, x p_a) + \frac{\as}{2\pi}\,
P^{aa'}(x)\,\frac{1}{{\bom T}^2_{a'}} 
{\bom T}_b \cdot {\bom T}_{a'}\,\ln\frac{\mu_F^2}{x s_{ab}}\:.
\eeq
Finally, the ${\bom K}_{m+b}$ operator is
\beeq
&&
{\bom K}_{m+b}^{a,a'}(x;\{p_i,m_i\}, p_a, p_b) =
{\bom K}_m^{a,a'}(x;\{p_i,m_i\}, p_a)
\nn \\ && \quad
- \frac{\as}{2\pi} {\bom T}_b \cdot {\bom T}_{a'}
\Bigg\{ \frac{1}{{\bom T}_{a'}^2} P^{aa'}_{{\rm reg}}(x) \ln(1-x)
+ \delta^{aa'} \left[ 2\left(\frac{\ln (1-x)}{1-x} \right)_+
- \frac{\pi^2}{3} \delta(1-x) \right]
\Bigg\}\:.\qquad
\label{eq:Kab2}
\eeeq
We see that ${\bom K}_{m+b}$ has the form of ${\bom K}_m$ of
Eq.~(\ref{eq:Kab1}) with the additional term in the second line
containing parton-parton correlations between the initial-state partons.

\section{Summary}
\label{se:summary}

In this paper, we have presented an extension of the dipole subtraction
method for calculating arbitrary (phenomenologically relevant) jet
cross sections at NLO accuracy in arbitrary scattering processes
involving heavy partons in the final state. In the case of lepton
collisions, we have set up the formalism such that for those cross
sections for which the mass of the heavy parton does not set the
hard-scattering scale, i.e.~the massless limit is IR safe, we can
simply set the masses of the partons to zero to recover the massless
limit of the computations, discussed in great detail in
Ref.~\cite{Catani:1997vz}. Thus the implementation of our formalism
into a general purpose NLO partonic Monte Carlo program leads to a code
that is smooth in the massless limit or, more importantly, numerically
stable for any values of the hard-scattering scale.  In hadron
collisions, similar behaviour can be achieved by matching the
partonic calculation with a suitable definition of the heavy-parton
distributions in the massless limit. This feature will be particularly
important at the future colliders (LHC, NLC), where the ratios of
parton masses to other relevant kinematical invariants can run over a
very wide range of values.  

The factorization of the QCD matrix elements on soft poles in the
presence of massive partons has been known and used for a long time. 
In those kinematical situations in which two partons become collinear
and at least one of them is massive, the collinear divergences in the
matrix elements are screened by the finite value of the parton mass and
the regularization of the real corrections in these regions is not,
strictly speaking, necessary. Nevertheless, the cross section receives
a logarithmically enhanced contribution from these phase space regions
when the parton masses become small compared to other relevant
kinematical invariants.  In order to assure the smooth massless
behaviour of the cross section, we introduced the notion of
quasi-collinear limit and presented the factorization of the QCD matrix
elements on quasi-collinear poles (factors that become real poles in
the massless limit).  The smooth massless limit also serves as a
powerful check of both the analytic calculations and any numerical
implementation.
 
Our extension is formulated very closely along the lines of
Ref.~\cite{Catani:1997vz}. We changed the formalism to the least extent
that was necessary to incorporate the mass corrections.
Technically the generalization is cumbersome, but the complications only
concern the derivation of the insertion operators (done in this paper)
and the actual application is not much more involved than it is in
the massless case. As a result, an existing general purpose NLO Monte
Carlo program, such as {\tt NLOJET++}\cite{Nagy:2001xb,Nagy:2001fj},
written for computations in the massless theory, can be changed
straightforwardly to incorporate parton masses. In particular, the
general form of the dipole subtraction terms remains unchanged and the
dipole splitting functions receive trivial mass corrections. Only the
definition of the emitter and spectator momenta, needed for writing the
matrix elements and jet functions of the subtraction term change in a
somewhat cumbersome, but straightforward, way. 

The NLO contribution containing the virtual corrections and the
insertion operators also has the same form as in the massless case.  We
have presented these operators explicitly, thus, for constructing a
numerical program to calculate the NLO corrections to arbitrary jet
quantities in a given process, the only additional ingredients that we
need are simply related to the evaluation of the original matrix
elements, as listed in Sect.~\ref{subsec:summary}.

{\bf Acknowledgements:}
This work was supported in part by the EU Fourth Framework Programme
``Training and Mobility of Researchers'', Network ``QCD and particle
structure'', contract FMRX-CT98-0194 (DG 12 - MIHT) and by the
Hungarian Scientific Research Fund grants OTKA T-025482 and T-038240.
SD and ZT are grateful for the kind hospitality of the CERN Theory
Division, where much of this work was carried out.
We are also grateful to Arnd Brandenburg for supplying {\tt maple} files
of the results of Ref.~\cite{Brandenburg:1998pu}, which we used to
derive Eq.~(\ref{Ffinitelogs}).

\pagebreak[3]
\appendix
\section*{Appendix}

\section{The eikonal integral}
\label{app:auxint}

In Sect.~\ref{subsub:FFinteg} we have defined the eikonal integral 
$I^{\eik}(\mu_j,\mu_k)$ in Eq.~(\ref{eq:Ieik}), but did not give the
general expression, since only the part symmetric in 
$\mu_j\leftrightarrow\mu_k$ is needed in the final result.
An intermediate result of the general integral, which is given in the
following, is
\beeq
I^{\eik}(\mu_j,\mu_k) &=& \frac{1}{\tvijk} \left[
\left(1-(\mu_j+\mu_k)^2\right)^{-2\eps}\frac{1}{2\eps^2}
(1-\rho_j^{-2\eps})
+\frac{\pi^2}{12}(1-\rho_j^{-2\eps}) 
\right.
\nn\\
&& \quad \left. \vphantom{\frac{\pi^2}{12}} {}
+\Li_2\left(1-(\mu_j+\mu_k)^2\right)
-\Li_2\left(1-\rho_j^2\right)
-2f(\mu_j,\mu_k)
\right]
\nn\\
&& \quad {}
+ \Oe{}
\eeeq
with $\rho_j$, $\rho_k$, $\rho$ defined in Eq.~(\ref{eq:rhon}) 
and $f(\mu_j, \mu_k)$ is an auxiliary integral,
\beq
f(\mu_j, \mu_k) = \int_{t_-}^1\rd t\,(1-t)^{-1} \,
\ln\frac{1+\mu_j^2-\mu_k^2 +\sqrt{\lambda(1,\mu_j^2,\mu_k^2)}}
        {t+\mu_j^2-\mu_k^2 +\sqrt{\lambda(t,\mu_j^2,\mu_k^2)}}\:,
\qquad t_- = (\mu_j+\mu_k)^2\:.
\label{eq:fjk}
\eeq
The explicit result for $f(\mu_j, \mu_k)$  reads
\beeq
f(\mu_j, \mu_k) \aand = 
\Li_2\left(\frac{1-\rho_j^2}{1+\rho}\right)
+\Li_2\left(\frac{\rho(\rho_j^2-1)}{(1+\rho)\rho_j^2}\right)
+\Li_2\left(1-\rho_j^2\right)
-\Li_2\left(-\frac{\mu_k}{\mu_j}\right)
+\Li_2\left(-\frac{\mu_k}{\mu_j\rho}\right)
\nn\\* && {}
-2\ln\rho_j\ln\left(\frac{\rho_k}{1+\rho}\right)
-\ln\rho\ln\left(1+\frac{\mu_k}{\mu_j\rho}\right),
\eeeq
where the massless limit $\mu_j\to 0$ yields $f(0,\mu_k)=\pi^2/6$.

The symmetric part of the integral defined in Eq.~(\ref{eq:fjk}) is
much simpler than the full result and is given by
\beq
\frac12\left[f(\mu_j, \mu_k) + f(\mu_k, \mu_j)\right] =
\frac12 \Li_2\Big(1 - (\mu_j + \mu_k)^2\Big)
- \Li_2(-\rho) + \Li_2(1 - \rho) - \frac{\pi^2}{12}\:,
\eeq
with $\rho$ as defined in Eq.~(\ref{eq:rhon}).  We used the fact that
$\tilde{v}_{ij,k} = \tilde{v}_{ik,j}$ if the emitted parton $i$ is a
massless parton, as in the case of the eikonal integral. 

\pagebreak[3]
\section{Alternative integrated dipole functions}
\label{app:Q2fixed} 

At the end of Sect.~\ref{subsub:FIinteg} we have emphasized that 
the forms of the $J^a_{ij}$ endpoint parts of the $x$-distributions
depend on the convention for which kinematical invariant $P^2$ is kept
fixed during the $x$-integration which is performed over a plus
distribution $[f(x,P^2)]_+$. This becomes obvious if we substitute
the invariant $P^2 = 2 p_a\tpij$ used in Sect.~\ref{subsub:FIinteg}
with another invariant such as, for instance, 
\beq
P^2 = Q^2=(\tpij-x p_a)^2=m_{ij}^2-2x p_a\tpij\:, 
\eeq
which was used in Ref.~\cite{Dittmaier:2000mb}.
In this appendix we provide the necessary formulae for this
particular variant of treating the integrated dipole functions.

The transition to the parametrization in terms of the new $P^2=Q^2$
requires some changes in the separation of the one-particle
phase space described in Sect.~\ref{se:fikin}. The rescaled
parton masses have to be normalized to $P^2$, so we define
\beq
\bar\mu_n^2 = \frac{m_n^2}{-Q^2},
\qquad n=i,j,ij.
\eeq
The previously defined rescaled masses $\mu_n$ are, thus, replaced by
$\bar\mu_n$ using
\beq
\mu_n^2 = \frac{x\bar\mu_n^2}{1+\bar\mu_{ij}^2}
\eeq
in Eqs.~(\ref{eq:psfi}) and (\ref{eq:zlimits}), leading to
\beeq
\int \Big[\rd p_i\Big(Q;p_a,x\Big)\Big] \aand =
\frac{1}{4} (2\pi)^{-3+2\eps} (-Q^2)^{1-\eps}
(1+\bar\mu_{ij}^2) \int_0^{\bar x_+} \rd x\, \delta(x-\xija)
x^{-1+\eps} (1-x+\bar\mu_{ij}^2)^{-\eps}
\nn\\[.5em]
&& {}\times 
\int\rd^{d-3}\Omega
\int_{\bar z_-(x)}^{\bar z_+(x)}\rd\zi\, 
\Big[\bar z_+(x)-\zi\Big]^{-\eps}\Big[\zi-\bar z_-(x)\Big]^{-\eps},
\nn\\[.5em]
&& \hspace*{-5em}
\bar z_\pm(x) = 
\frac{1+\bar\mu_{ij}^2-x+x\bar\mu_i^2-x\bar\mu_j^2 \pm
\sqrt{(1+\bar\mu_{ij}^2-x-x\bar\mu_i^2-x\bar\mu_j^2)^2
-4\bar\mu_i^2 \bar\mu_j^2 x^2}}
{2(1+\bar\mu_{ij}^2-x)}.
\eeeq
{}From these expressions the upper limit $\bar x_+$ for $\xija$ is derived
in terms of $\bar\mu_n$,
\beq
\bar x_+ = \frac{1+\bar\mu_{ij}^2}{1+(\bar\mu_i+\bar\mu_j)^2}.
\eeq

The dipole functions $\bV_{ij}^a$, defined in Sect.~\ref{se:fidipoles},
remain unchanged. The integrated dipole functions 
\beq
\int [\rd p_i(Q;p_a,x)] \,
\frac{1}{(p_i+p_j)^2-m_{ij}^2} \, \langle\bV_{ij}^a\rangle
\equiv \frac{\alps}{2\pi}\frac{1}{\Gamma(1-\eps)}
\biggl(\frac{4\pi\mu^2}{m_{ij}^2-Q^2}\biggr)^\eps 
\bar I_{ij}^a(x;\ep)
\eeq
are given in terms of new auxiliary functions $\bar I_{ij}^a(x;\ep)$
and their calculation proceeds along the same lines as described in
Sect.~\ref{subsub:FIinteg}. The endpoints are split off according to
\beeq
\bar I_{gQ}^a(x;\ep) \aand = \CF \left(
[\bar J_{gQ}^a(x, \bar\mu_Q)]_{+} + \delta(1-x)\,
\Big[\bar J_{gQ}^{a;\rm S}(\bar\mu_Q;\ep) + 
\bar J_{gQ}^{a;\rm NS}(\bar\mu_Q)\Big]
\right)
+\Oe{}\:,
\\
\bar I_{Q\bar Q}^a(x;\ep) \aand = \TR \left(
[\bar J_{Q\bar Q}^a(x, \bar\mu_Q)]_{\bar x_+} + \delta(\bar x_+-x)\,
\Big[\bar J_{Q\bar Q}^{a;\rm S}(\bar\mu_Q;\ep) + 
\bar J_{Q\bar Q}^{a;\rm NS}(\bar\mu_Q)\Big]
\right)
+\Oe{}\:,
\eeeq
where $\bar x_+ = 1/(1 + 4\bar\mu_Q^2)$. 
The continuum part for the $Q\to gQ$ splitting,
\beeq
[\bar J_{gQ}^a(x, \bar\mu_Q)]_{+} &=&  \left(
\frac{2}{1-x} \left[ 
\ln\left(\frac{1+\bar\mu_Q^2}{1+\bar\mu_Q^2-x}\right) -1 \right]
+\frac{(1+\bar\mu_Q^2)^2(1-x)}{2(1+\bar\mu_Q^2-x)^2} \right)_+
\nn\\ && {}
+\left(\frac{2}{1-x}\right)_+
\ln\left(\frac{2+2\bar\mu_Q^2-x}{1+\bar\mu_Q^2}\right),
\eeeq
and the singular part of the corresponding endpoint contribution,
\beeq
\bar J_{gQ}^{a;\rm S}(\bar\mu_Q;\ep) \aand= 
\frac{1}{\eps^2}
-\frac{\pi^2}{3}
-\left(\frac{\bar\mu_Q^2}{1+\bar\mu_Q^2}\right)^{-\eps}
\left(\frac{1}{\eps^2}+\frac{1}{2\eps}+\frac{\pi^2}{6}+2\right)
\nn\\[.5em]
&&  {}
-\frac{1}{\eps}\ln\left(\frac{1+2\bar\mu_Q^2}{1+\bar\mu_Q^2}\right)
+\frac{1}{\CF}\left[\frac{1}{\eps}\gamma_q+K_q
\right],
\eeeq
result from their counterparts $[J_{gQ}^a(x,\mu_Q)]_{+}$
and $J_{gQ}^{a;\rm S}(\mu_Q;\ep)$, given in Eqs.~(\ref{eq:JfigQacont})
and (\ref{eq:JfigQaS}), upon the simple substitutions
$\mu_Q^2\to x\bar\mu_Q^2/(1+\bar\mu_Q^2)$ and
$\mu_Q^2\to  \bar\mu_Q^2/(1+\bar\mu_Q^2)$, respectively.
The non-singular endpoint contribution is given by
\beeq
\bar J_{gQ}^{a;\rm NS}(\bar\mu_Q) \aand= 
\frac{2\pi^2}{3}
-2\Li_2\left(\frac{1}{1+\bar\mu_Q^2}\right)
-2\Li_2\left(\frac{1+\bar\mu_Q^2}{1+2\bar\mu_Q^2}\right)
-\frac{1}{2}\ln^2\left(\frac{1+\bar\mu_Q^2}{1+2\bar\mu_Q^2}\right)
\nn\\[.5em]
&&  {}
+\frac{1}{2}\bar\mu_Q^2(2+\bar\mu_Q^2)\ln\left(\frac{1+\bar\mu_Q^2}{\bar\mu_Q^2}\right)
-\frac{1}{2}\bar\mu_Q^2.
\eeeq
Note that this does not follow from $J_{gQ}^{a;\rm NS}(\mu_Q)$, given in
Eq.~(\ref{eq:fi_intdip_gQ}), upon substituting $\mu_Q$, but a finite
difference remains
\beeq
\Delta\bar J_{gQ}^{a;\rm NS}(\bar\mu_Q) \aand= 
\bar J_{gQ}^{a;\rm NS}(\bar\mu_Q) -
J_{gQ}^{a;\rm NS}\left(\frac{\bar\mu_Q}{\sqrt{1+\bar\mu_Q^2}}\right)
\nn\\[.5em]
\aand= \frac{\pi^2}{3}
-2\Li_2\left(\frac{1}{1+\bar\mu_Q^2}\right)
+2\Li_2\left(\frac{-\bar\mu_Q^2}{1+\bar\mu_Q^2}\right)
+\frac{1}{2}\ln\left(\frac{1+\bar\mu_Q^2}{1+2\bar\mu_Q^2}\right)
\nn\\[.5em]
&&  {}
+\frac{1}{2}\bar\mu_Q^2(2+\bar\mu_Q^2)
\ln\left(\frac{1+\bar\mu_Q^2}{\bar\mu_Q^2}\right)
-\frac{\bar\mu_Q^2(1+\bar\mu_Q^2)}{1+2\bar\mu_Q^2}.
\eeeq

For the $g\to Q\bar Q$ splitting the same features are observed.
The continuum part,
\beq
[\bar J_{Q\bar Q}^a(x, \bar\mu_Q)]_{\bar x_+} =
\frac{2}{3} 
\left(
\frac{1-x+2x\bar\mu_Q^2}{(1-x)^2} 
\sqrt{1-\frac{4x\bar\mu_Q^2}{1-x}}
\right)_{\bar x_+},
\eeq
results from $[J_{Q\bar Q}^a(x, \mu_Q)]_{x_+}$, given in 
Eq.~(\ref{eq:JfiQQacont}), upon substituting $\mu_Q^2\to x\bar\mu_Q^2$ and
$x_+\to\bar x_+$. 
The singular endpoint part,
\beq
\bar J_{Q\bar Q}^{a;\rm S}(\bar\mu_Q;\ep) = 
-\frac{2}{3\eps}\left[1 - 
\left(\frac{\bar\mu_Q^2}{1+4\bar\mu_Q^2}\right)^{-\eps}\right] -\frac{10}{9},
\eeq
is obtained from
$J_{Q\bar Q}^{a;\rm S}(\mu_Q;\ep)$, given in Eq.~(\ref{eq:JfiQQaS}),
by the replacement $\mu_Q^2\to\bar x_+\bar\mu_Q^2$.
However, the non-singular
endpoint part differs from $J_{Q\bar Q}^{a;\rm NS}(\bar\mu_Q)$,
given in Eq.~(\ref{eq:fi_intdip_QQ}), in a non-trivial way:
\beq
\bar J_{Q\bar Q}^{a;\rm NS}(\bar\mu_Q) =
\frac{8}{3}\bar\mu_Q^2
+\frac{2}{3}\ln\left(\frac{\bar\mu_Q^2}{1+4\bar\mu_Q^2}\right)
+\frac{4}{3}(1-2\bar\mu_Q^2)\sqrt{1+4\bar\mu_Q^2}
\ln\left(\frac{\sqrt{1+4\bar\mu_Q^2}+1}{2\bar\mu_Q}\right).
\eeq
The difference between $\bar J_{Q\bar Q}^{a;\rm NS}(\bar\mu_Q)$
and the reparametrized $J_{Q\bar Q}^{a;\rm NS}(\bar\mu_Q)$ reads
\beeq
\Delta\bar J_{Q\bar Q}^{a;\rm NS}(\bar\mu_Q) \aand=
\bar J_{Q\bar Q}^{a;\rm NS}(\bar\mu_Q)
- J_{Q\bar Q}^{a;\rm NS}\left(\frac{\bar\mu_Q}{\sqrt{1+4\bar\mu_Q^2}}\right)
\nn\\
\aand= \frac{8}{3}\bar\mu_Q^2 - \frac{10}{9}
+\left(\frac{10}{9}+\frac{16}{3}\bar\mu_Q^2\right)(1+4\bar\mu_Q^2)^{-3/2}
\nn\\
&& {}
+\frac{4}{3}\left[ (1-2\bar\mu_Q^2)\sqrt{1+4\bar\mu_Q^2} - 1\right]
\ln\left(\frac{\sqrt{1+4\bar\mu_Q^2}+1}{2\bar\mu_Q}\right).
\eeeq                                                                            
Since the non-trivial differences between the two endpoint
parametrizations appear only in the integrated dipole functions
for final-state emitter and initial-state spectator, the necessary
changes in Sects.~\ref{subsec:oneini} and \ref{subsec:twoini}
can be easily inferred. 
The new functions $\bar\cK^{ab}_j$ follow from their previously 
defined counterparts $\cK^{ab}_j$ by simple substitutions and adding
the extra terms $\Delta\bar J_{gQ}^{a;\rm NS}$ and 
$\Delta\bar J_{Q\bar Q}^{a;\rm NS}$,
\beeq
\bar\cK^{ab}_q(x; Q_{ja}^2, m_j) \aand=
\cK^{ab}_q(x; s_{ja}, m_j)
\Big|_{s_{ja}\to(m_j^2-Q_{ja}^2)/x}
\nn\\*
&& {}
+\delta^{ab}\delta(1-x)
\Delta\bar J_{gQ}^{a;\rm NS}\left(\frac{m_j}{\sqrt{-Q_{ja}^2}}\right),
\\[.5em]
\label{eq:barcKabg}
\bar\cK^{ab}_g(x; Q_{ja}^2, 0, \{m_F\}) &=&
\cK^{ab}_g(x; s_{ja}, 0, \{m_F\})
\Big|_{s_{ja}\to -Q_{ja}^2/x,\; x_+\to\bar x_+}
\nn \\ && 
+\delta^{ab}\delta(\bar x_+ -x) \frac{\TR}{\CA} \sum_{F=1}^{N_F} 
\Delta\bar J_{Q\bar Q}^{a;\rm NS}\left(\frac{m_F}{\sqrt{-Q_{ja}^2}}\right)\:,
\eeeq
where $Q_{ja}^2=m_j^2-xs_{ja}$, not to be confused with
$Q_{ja}^2=s_{ja}+m_j^2$ used in Sect.~\ref{subsec:oneini} of the main text.
Note that in the scheme considered in this appendix, all masses of quark
are accessible at all values of $Q_{ja}^2$, so $N_F$ appears in
Eq.~(\ref{eq:barcKabg}) rather than $N_F^{ja}$.  In fact {\em all\/}
occurrences of $N_F^{ja}$ in ${\bom I}$ and ${\bom K}$ should be replaced
by $N_F$ in this scheme.

\pagebreak[3]
\section{Dipole functions in SUSY QCD}
\label{app:SUSYdipoles}

In this appendix we define the dipole functions $\cD_{ij,k}$ and
$\cD_{ij}^a$ needed to construct the dipole subtraction function
specified in Eq.~(\ref{eq:dff}) for SUSY QCD calculations (see for
example Refs.~\cite{Beenakker:1996ch, Beenakker:1997ut, Berger:2000iu,
Kramer:1997hh}). The new dipoles fall
into two classes: one that involves gluinos and gluons and the other that
involves squarks and gluons. We consider only those dipoles that lead to
$1/\ep$ poles after integration in $d = 4 -2\ep$ dimensions. Owing to the
large masses of the SUSY particles in typical supersymmetric extensions
of the SM, the massless limit is not too relevant even at LHC or NLC
energies. Thus we do not define dipole functions corresponding to
either $g \to \tg\tg,\tq\tq$ or splittings involving a squark, a gluino
and a quark.

In principle, to calculate one-loop amplitudes in supersymmetric
theories one has to use a regularization prescription that respects the
supersymmetric Ward identities, for instance, dimensional reduction
(DR). Within the dipole formalism, the use of different regularization
prescriptions only affects the actual calculation of the cross section
contribution $\sigma^{\aNLO\,\{m\}}$ in Eq.~(\ref{eq:NLOppm}).  In this
paper we derived our formulae using conventional dimensional
regularization (CDR), therefore, the one-loop amplitudes
$|\cm_{m,ab}(\{p_i,m_i\};p_a,p_b)|^2_{(\mathrm{1-loop})}$
are also needed in CDR. To derive the insertion operator ${\bom I}(\ep)$
for SUSY QCD one has two options, either to use DR throughout and the
transition rules between CDR and DR elucidated in
Refs.~\cite{Kunszt:1994sd, Catani:1997vz, Catani:2001ef, Catani:1997pk},
or to use CDR throughout and correct for the violation of the SUSY Ward
identities in the calculation of the one-loop contribution
$|\cm_{m,ab}(\{p_i,m_i\};p_a,p_b)|^2_{(\mathrm{1-loop})}$ as explained in
Refs.~\cite{Martin:1993yx,Beenakker:1996ch}.  We choose to do the
latter.

As discussed in Sect.~\ref{se:sub}, we do not consider massive partons in
the initial state, therefore, new dipoles of the $\cD^{ai}_j$ type do
not appear in the supersymmetric theory.  The dipole functions that
involve gluinos are obtained immediately by changing the colour factor
$\CF$ in Eqs.~(\ref{eq:V_gQk}) and (\ref{eq:V_gQa}) to $\CA$. As a
result the integrated dipole functions are also obtained by this simple
change in Eqs.~(\ref{eq:I_gQk}) and (\ref{eq:I_gQa}).

The flavour and the spin of the spectators do not influence the dipole
functions, therefore, we have to consider only final-state emitter
squarks as new cases. Thus, we define the splitting functions
$\bV_{ij,k}$ and $\bV_{ij}^a$ for the $\tq\to g\tq$ splitting.
The case $\bar \tq\to g\bar \tq$ is formally identical to $\tq\to g\tq$.

When the spectator is also in the final state, then we define
\beeq
\langle s|\bV_{g \tq,k}|s'\rangle \aand =
8\pi\mu^{2\eps}\alps\CF \left\{
\frac{2}{1-\zj(1-\yijk)}
-\frac{\tvijk}{\vijk}\left(2 + \frac{m_\tq^2}{p_i p_j}\right)
\right\} \delta_{ss'}
\nn\\[.5em]
\aand = \langle\bV_{g\tq,k}\rangle \delta_{ss'}\:.
\eeeq
We write the integral of $\langle\bV_{g\tq,k}\rangle/(2p_g p_\tq)$ over
the one-parton subspace in Eq.~(\ref{eq:psff}) in the form of
Eqs.~(\ref{eq:Iijk_def}) and (\ref{eq:I_gQk}). The eikonal integral
depends only on the mass of the emitter, but not on its flavour or
spin. The collinear integral is obtained as
\beq
I^{\coll}_{g\tq,k}(\mu_\tq,\mu_k) =
\frac{2}{\eps}-\frac{\mu_\tq^{-2\eps}}{\eps}
-2\mu_\tq^{-2\eps} + 6
-2\ln\left[(1-\mu_k)^2-\mu_\tq^2\right]
+\frac{4\mu_k(\mu_k-1)}{1-\mu_\tq^2-\mu_k^2}
+\Oe{}\:.
\eeq
If the spectator is in the initial state, we define
\beeq
\label{eq:V_gtqa}
\langle s|\bV_{g\tq}^a|s'\rangle \aand =
8\pi\mu^{2\eps}\alps\CF \left\{
\frac{2}{2-\xija-\zj}-2-\frac{m_{\tq}^2}{p_i p_j} \right\} \delta_{ss'}
\nn\\[.5em]
\aand = \langle\bV_{g\tq}^a\rangle \delta_{ss'}\:.
\eeeq
We write the integral of $\langle\bV_{g\tq}^a\rangle/(2p_g p_\tq)$ over
the one-parton subspace in Eq.~(\ref{eq:psfi}) in the form of
Eqs.~(\ref{eq:Iija_def}) and (\ref{eq:I_gQa}). The continuum part is
\beq
[J_{g\tq}^a(x, \mu_\tq)]_+ = 
\left(-\frac{2}{1-x}\left[1+\ln(1-x+\mu_\tq^2)\right] \right)_+
+ \left(\frac{2}{1-x}\right)_+ \ln(2+\mu_\tq^2-x)\:.
\eeq
The endpoint parts are
\beq
J_{g\tq}^{a;\rm S}(\mu_\tq;\ep) =
\frac{1}{\eps^2} - \frac{\pi^2}{3}
-\mu_\tq^{-2\eps}
\left(\frac{1}{\eps^2}+\frac{1}{\eps}+\frac{\pi^2}{6}+2\right)
-\frac{1}{\eps}\ln(1+\mu_\tq^2)
+  \frac1\CF \Bigg(\frac{1}{\eps}\gamma_\tq + K_\tq \Bigg) \:,
\eeq
\beq
J_{g\tq}^{a;\rm NS}(\mu_\tq) = 
\frac{\pi^2}{3}-2\Li_2\biggl(\frac{1}{1+\mu_\tq^2}\biggr)
-2\Li_2(-\mu_\tq^2)
-\frac{1}{2}\ln^2(1+\mu_\tq^2)\:,
\eeq
where 
\beq
\label{eq:tqconstants}
\gamma_\tq = 2\CF \:, \qquad\qquad
K_\tq = \left( 4 - \frac{\pi^2}{6} \right) \CF\:.
\eeq

Using these results, we obtain the functions needed for the construction
of the insertion operators ${\bom I}_m$ and ${\bom K}^{a,b}$. The
singular functions $\cV^{(\rm S)}$ are independent of the flavour and
spin. The non-singular function $\cV^{(\rm NS)}_j$ for the gluino is 
\beeq
\label{eq:VNS_tg}
&&
\cV^{(\rm NS)}_\tg(s_{jk},m_j,m_k) = \cV^{(\rm NS)}_q(s_{jk},m_j,m_k)
\:,
\eeeq
and for the squark it is
\beeq
\label{eq:VNS_tq}
&&
\cV^{(\rm NS)}_\tq(s_{jk},m_j,m_k) =
\frac{\gamma_\tq}{{\bom T}_\tq^2}\ln\frac{s_{jk}}{Q_{jk}^2}
- 2\ln\frac{(Q_{jk}-m_k)^2-m_j^2}{Q_{jk}^2}
+ \frac{4m_k(m_k-Q_{jk})}{s_{jk}}
+\frac{\pi^2}{2}
\nn\\ && \qquad {}
+ \frac1{v_{jk}}\,
\left[
\ln\rho^2\ln(1+\rho^2)
+ 2\Li_2(\rho^2)- \Li_2(1-\rho_j^2) - \Li_2(1-\rho_k^2)
- \frac{\pi^2}{6} \right]
\eeeq
if $k$ is massive, and
\beq
\cV^{(\rm NS)}_\tq(s_{jk},m_j,0) =
\frac{\gamma_\tq}{{\bom T}_\tq^2}\ln\frac{s_{jk}}{Q_{jk}^2}
+ \frac{\pi^2}{6}
- \Li_2\left(\frac{s_{jk}}{Q_{jk}^2}\right)
- 2\,\ln\frac{s_{jk}}{Q_{jk}^2}
\eeq
if $k$ is massless.
The singular function $\Gamma_j$ for the gluino is
\beq
\Gamma_\tg(\mu,m_\tg;\ep)=
{\bom T}_\tg^2 \left( \frac{1}{\ep} - \ln\frac{m_\tg^2}{\mu^2} -2 \right) +
\gamma_\tg \,\ln\frac{m_\tg^2}{\mu^2} =
\CA \left[ \frac{1}{\ep} +
\frac{1}{2} \ln\frac{m_\tg^2}{\mu^2} -2 \right] \:,
\eeq
and the flavour constants are
\beq
\label{eq:tgconstants}
\gamma_\tg = \frac32\CA \:, \qquad\qquad
K_\tg = \left( \frac72 - \frac{\pi^2}{6} \right) \CA\:.
\eeq
The singular function $\Gamma_j$ for the squark is
\beq
\Gamma_\tq(\mu,m_\tq;\ep)=
{\bom T}_\tq^2 \left( \frac{1}{\ep} - \ln\frac{m_\tq^2}{\mu^2} -2 \right) +
\gamma_\tq \,\ln\frac{m_\tq^2}{\mu^2} =
\CF \left[ \frac{1}{\ep} + \ln\frac{m_\tq^2}{\mu^2} -2 \right] \:.
\eeq

Finally, the functions $\cK^{a,b}_j$ for the gluino are 
\beq
\label{eq:cKab_tg}
\cK^{a,b}_\tg(x; s_{ja}, m_j) = 
\cK^{a,b}_q(x; s_{ja}, m_j) 
\:,
\eeq
and those for the squark are
\beeq
\label{eq:cKab_tq}
&&
\cK^{a,b}_\tq(x; s_{ja}, m_j) = 
\cK^{a,b}_q(x; s_{ja}, m_j) 
-\delta^{ab}\left(\frac{s_{ja}^2(1-x)}{2[s_{ja}(1-x)+m_j^2]^2}\right)_+
\nn \\ && \qquad\qquad
  + \delta^{ab}\delta(1-x)\Bigg[-
  \left(\frac{m_j^2}{s_{ja}}\ln\frac{m_j^2}{Q_{ja}^2}
  + \frac{m_j^2}{2Q_{ja}^2}\right)
+\frac{\gamma_q - \gamma_\tq}{\CF}\Bigg]
\:.
\eeeq

\pagebreak[3]
\section{Example Applications}
\label{app:examples}

We illustrate our method with the three simplest examples, $e^+e^-$
annihilation to heavy quarks, $e^+e^-\to Q\bar Q$, to heavy quarks and a
jet, $e^+e^-\to Q\bar Qg$ and heavy quark production in deep inelastic
scattering, $ep \to eQ\bar Q+X$.

\subsection{\boldmath$e^+e^-\to Q\bar Q$}

We begin by recalling the tree-level result to set the notation.  Since
our aim is to illustrate our method, we try to keep the analytical
formulae simple by taking the average over event plane orientations and
neglecting electron polarization.  The extension to oriented and
polarized observables is straightforward.

In order to match consistently with the one-loop corrections
that are available in the literature, it is
necessary to evaluate the tree-level matrix element in $d$ dimensions.
In fact, since all singularities can be cancelled before averaging over
event orientation, it is sufficient to consistently evaluate the
hadronic tensor in $d$ dimensions.  We are then free to take the rest of
the process, as well as the remaining angular integrations, in 4
dimensions.  Since after averaging over event orientation and
polarization we only encounter Dirac traces with no or two $\gamma^5$
matrices, we can easily eliminate $\gamma^5$ by taking it to be totally
anticommuting in $d$ dimensions
and using $\gamma_5^2 = 1$.

With this prescription, the $d$-dimensional matrix element is identical
to the 4-dimen\-sional one which, labelling the momenta by $e^++e^- \to
\gamma^*/Z(q) \to Q(p_1)+\bar Q(p_2)$, is given by
\cite{Nilles:1980ic}
\beq
  |\cm_2|^2 =
  \left(g^{VV}+g^{AA}\right)
  \left(1+2\mu_Q^2\right)-g^{AA}
  \left(6\mu_Q^2\right),
  \label{QQM2}
\eeq
where $\mu_Q\equiv m_Q/\sqrt{s}$, with $s=q^2$, and the normalization is
such that the cross section is given by
\beeq
  \sigma^{\aLO} &=& \sigma_0 \; v \, |\cm_2|^2 \, F_J^{(2)}(p_1,p_2),
  \label{QQsigLO}
\eeeq
where $v=\sqrt{1-4\mu_Q^2}$ is the velocity of the heavy quark in the
centre-of-mass frame and $F_J$ is our (infrared-safe) observable.  The
coupling constants and point-like cross section appearing in
Eqs.~(\ref{QQM2}) and~(\ref{QQsigLO}) are given by
\beeq
  g^{VV} &=& Q_Q^2-2g_v^eg_v^QQ_Q\Real\left\{\chi(s)\right\}+
  ({g_v^e}^2+{g_a^e}^2){g_v^Q}^2|\chi(s)|^2, \\
  g^{AA} &=& ({g_v^e}^2+{g_a^e}^2){g_a^Q}^2|\chi(s)|^2, \\
  \sigma_0 &=& \Nc\frac{4\pi\alpha^2}{3s},
\eeeq
where $g_v^f=T_3^f-2Q_f\sin^2\theta_w$, $g_a^f=T_3^f$, $Q_f$ is the
electric charge of fermion type $f$ and $\theta_w$ is the weak mixing
angle.  The function $\chi(s)$ parametrizes the $\mathrm{Z}^0$
propagator and coupling factors,
\beq
  \chi(s) = \frac1{4\sin^2\theta_w\cos^2\theta_w}
  \;\frac{s}{s-m_Z^2+im_Z\Gamma_Z},
\eeq
with $m_Z$ and $\Gamma_Z$ the mass and width of the $\mathrm{Z}^0$.

The NLO real emission process is $e^+e^-\to\gamma^*/Z(q)\to Q(p_1)+\bar
Q(p_2)+g(p_3)$, with matrix element $\cm_3(p_1,p_2,p_3)$.  In
discussing the NLO corrections to the $Q\bar Qg$ process, we will need
to discuss the $d$-dimensional behaviour of $\cm_3$ in more detail,
so we explicitly indicate the number of dimensions in which it is
evaluated.
In four dimensions we can rewrite the known result\cite{Nilles:1980ic}
as
\beeq
&&
|\cm_3^{(4)}(p_1,p_2,p_3)|^2 =
\CF\frac{8\pi\alps}s
\nonumber\\*&&\qquad\qquad
\times
\Biggl\{
|\cm_2|^2\,
\Biggl[\frac{1}{(1-x_1)}
\left(\frac{2 (1-2\mu_Q^2)}{2-x_1-x_2} - 2 - \frac{2\mu_Q^2}{1-x_1}\right)
+ (x_1 \leftrightarrow x_2)
\Biggr]
\nonumber\\&&\qquad\qquad\quad
+ \left[g^{VV} + g^{AA}(1 + 2 \mu_Q^2)\right]
  \left(\frac{1-x_2}{1-x_1} + \frac{1-x_1}{1-x_2}\right) + g^{AA}\,4\mu_Q^2
  \Biggr\},
  \label{MQQg4}
\eeeq
where $x_i\equiv 2p_i\ldot q/q^2$.  In terms of these variables, the
phase space is given by
\beq
  \label{dPhi3}
  \rd\Phi^{(3)} = \frac{s}{16\pi^2}\rd x_1\,\rd x_2\;
  \Theta(x_+-x_2)\Theta(x_2-x_-)\,\Theta(1-x_1)\Theta(x_1-2\mu_Q),
\eeq
with
\beq
  x_\pm = \frac{(2-x_1)(1-x_1+2\mu_Q^2) \pm
    (1-x_1)\sqrt{x_1^2-4\mu_Q^2}}{2(1-x_1+\mu_Q^2)}.
\eeq
The three-parton cross section is then
\beq
  \sigma^{(3)} = \sigma_0 \int \rd\Phi^{(3)} \; |\cm_3|^2 \;
  F_J^{(3)}(p_1,p_2,p_3),
\eeq
where we used a different normalization from that in Eq.~(\ref{eq:psf}).

According to the dipole subtraction method, this integral is rendered
finite by subtracting from it an auxiliary cross section constructed
from the two dipole contributions, $\cD_{31,2}$ and
$\cD_{32,1}$, defined by Eq.~(\ref{eq:Dijk}).  The colour and spin
algebra are trivial in this case, and we obtain
\beq
  \cD_{31,2}(p_1,p_2,p_3) = \frac1{2p_3\ldot p_1}
  \;\langle\bV_{g_3Q_1,2}\rangle\;
  |\cm_2|^2,
  \label{QQD312}
\eeq
with
\beq
  \langle\bV_{g_3Q_1,2}\rangle = 8\pi\alps \CF\left\{
    \frac2{1-\tilde z_1(1-y_{31,2})}
    -\frac{\tilde v_{31,2}}{v_{31,2}}
    \left[1+\tilde z_1+\frac{m_Q^2}{p_3\ldot p_1}\right]\right\}.
\eeq
Inserting the definitions of $y_{31,2}$, $\tilde z_1$, $v_{31,2}$ and
$\tilde v_{31,2}$ from Eqs.~(\ref{eq:ziyijk}), (\ref{eq:vijk})
and~(\ref{eq:tvijk}), we obtain
\beeq
\lefteqn{
  \cD_{31,2}(p_1,p_2,p_3) = \CF\frac{8\pi\alps}s
  \;|\cm_2|^2\;
}\nonumber\\&&
  \times
  \frac1{1-x_2}
  \left\{
    \frac{2(1-2\mu_Q^2)}{2-x_1-x_2}
    -\sqrt{\frac{1-4\mu_Q^2}{x_2^2-4\mu_Q^2}}
    \frac{x_2-2\mu_Q^2}{1-2\mu_Q^2}
    \left[2 + \frac{x_1 - 1}{x_2-2\mu_Q^2}
      +\frac{2\mu_Q^2}{1-x_2}
    \right]\right\}.
\phantom{(A.99)}
\eeeq
The associated dipole kinematics are given by
\beq
  \widetilde{p}_2^\mu = \frac12q^\mu+
  \frac{\sqrt{1-4\mu_Q^2}}{\sqrt{x_2^2-4\mu_Q^2}}
  \left(p_2^\mu-\frac12x_2q^\mu\right),
\qquad
  \widetilde{p}_{31}^\mu = \frac12q^\mu-
  \frac{\sqrt{1-4\mu_Q^2}}{\sqrt{x_2^2-4\mu_Q^2}}
  \left(p_2^\mu-\frac12x_2q^\mu\right).
\eeq
The dipole $\cD_{32,1}$ is obtained from $\cD_{31,2}$ by the
replacement $p_1\leftrightarrow p_2$.

Combining the expressions for the real and auxiliary cross sections, we
obtain the three-parton integral,
\beeq
&&
\sigma^{\aNLO\{3\}} = \sigma_0\;\CF\frac{\alps}{2\pi}
\int\,\rd x_1\,\rd x_2
\nonumber\\*&&\qquad
\times
\Biggl\{
|\cm_2|^2
\Biggl[
\frac{1}{1-x_1}
\left(\frac{2 (1-2\mu_Q^2)}{2-x_1-x_2} - 2 - \frac{2\mu_Q^2}{1-x_1}\right)
F_J^{(3)}(p_1,p_2,p_3)
\nonumber\\&&\qquad\qquad
- \frac1{1-x_1}
  \left(
    \frac{2(1-2\mu_Q^2)}{2-x_1-x_2}
    -\sqrt{\frac{1-4\mu_Q^2}{x_1^2-4\mu_Q^2}}
    \frac{x_1-2\mu_Q^2}{1-2\mu_Q^2}
    \left(2 + \frac{x_2 - 1}{x_1-2\mu_Q^2}
      +\frac{2\mu_Q^2}{1-x_1}
    \right)\right)
\nonumber\\&&\qquad\qquad\qquad
 \times F_J^{(2)}(\widetilde p_1,\widetilde p_{32})
+ (x_1 \leftrightarrow x_2, \widetilde p_1 \to \widetilde p_{31},
\widetilde p_{32} \to \widetilde p_2)\Biggr]
\nonumber\\&&\qquad
+ \left[\left(g^{VV} + g^{AA}(1 + 2 \mu_Q^2)\right)
  \left(\frac{1-x_2}{1-x_1} + \frac{1-x_1}{1-x_2}\right)
+ g^{AA}\,4\mu_Q^2\right]
  F_J^{(3)}(p_1,p_2,p_3)
\Biggr\}.
\label{QQNLO3}
\eeeq
It is straightforward to check that the soft singularity cancels between
the different terms in Eq.~(\ref{QQNLO3}), provided the observable is
infrared safe (implying that $F_J^{(3)}\to F_J^{(2)}$ in the soft
limit).  Furthermore, for any quasi-collinear-safe observable (implying
also that $F_J^{(3)}\to F_J^{(2)}$ in the quasi-collinear limit),
Eq.~(\ref{QQNLO3}) is finite in the small-mass limit.  In fact, taking
the quark mass to zero, one exactly recovers the massless result in
Eq.~(D.7) of Ref.~\cite{Catani:1997vz}.

Next we have to evaluate the insertion operator ${\bom I}(\eps)$, which
gives the integral of the auxiliary cross section and combine it with
the virtual cross section.  The one-loop matrix element was calculated
in Ref.~\cite{Jersak:1982sp} and is given by\footnote{Note that we have
inserted a factor of $(4\pi\mu^2/m_Q^2)^\eps/\Gamma(1-\eps)$, where $\mu$
is the dimensional-regularization scale, which is necessary for
consistency with our notation.},
\beeq
  |\cm_2|^2_{(\mathrm{1-loop})} &=& 
  \;\left(\frac{4\pi\mu^2}{m_Q^2}\right)^\eps
  \!\!\frac1{\Gamma(1-\eps)}\;\;
  2\Real\{f_1\}\;|\cm_2|^2
\nonumber\\&&
  +2\Real\{f_2\}\left(\frac32\left(g^{VV}+g^{AA}\right)
  -\left(\frac52-4\mu_Q^2\right)g^{AA}\right).
  \label{QQ1loop}
\eeeq
The form factors appearing in Eq.~(\ref{QQ1loop}) are given
by
\beeq
  \Real\{f_1\} &=& \CF\frac{\alps}{2\pi}
    \Biggl\{
    \frac{-1}{\eps}\left(1+\frac{1+v^2}{2v}\ln\frac{1-v}{1+v}\right)
    -2-\frac{1+2v^2}{2v}\ln\frac{1-v}{1+v}
\nonumber\\&&
    +\frac{1+v^2}v\left[\Li_2\left(\frac{1-v}{1+v}\right)
      +\frac{\pi^2}3-\frac14\ln^2\frac{1-v}{1+v}
      +\ln\frac{1-v}{1+v}\ln\frac{2v}{1+v}\right]\Biggr\},
  \phantom{(A.99)}\\
  \Real\{f_2\} &=& \CF\frac{\alps}{2\pi}\frac{1-v^2}{2v}\ln\frac{1-v}{1+v}.
\eeeq
For later convenience, we define a function $f_f$,
\beq
  \Real\{f_1\} = \CF\frac{\alps}{2\pi}\Biggl\{
    \frac{-1}{\eps}\left(1+\frac{1+v^2}{2v}\ln\frac{1-v}{1+v}\right)
    -2-\frac32\ln\frac{1-v}{1+v}
    +\frac{2\pi^2}3-\frac12\ln^2\frac{1-v}{1+v}
    +f_f(v)\Biggr\},
\eeq
such that $f_f(v)$ vanishes in the small-mass (i.e.~$v\to1$) limit.

The general expression for ${\bom I}(\eps)$ is given in
Eq.~(\ref{eq:insop}).  In our case, the colour structure is trivial and
we obtain
\beeq
\lefteqn{
  {}_2\langle{1,2|{\bom I}_2(\eps,\mu^2;\left\{p_i,m_i\right\})|1,2}\rangle_2
  = |\cm_2|^2\times
  2\CF\frac{\alps}{2\pi}\,\frac{(4\pi)^\eps}{\Gamma(1-\eps)}
}\nonumber\\&&
  \times\left[\left(\frac{\mu^2}{s_{12}}\right)^\eps
    \left(\cV_q(s_{12},m_Q,m_Q;\eps)-\frac{\pi^2}3\right)
    +\frac1{\CF}\Gamma_q(\mu,m_Q;\eps)
    +\frac32\ln\frac{\mu^2}{s_{12}}
    +5-\frac{\pi^2}6\right],
\label{QQIfin}
\phantom{(A.99)}
\eeeq
where $s_{12}=2p_1\ldot p_2=s-2m_Q^2$ and we have suppressed the
dependence on $\{m_F\}$ and $\kappa$, which do not enter our cross
section.  The function $\cV_q$ is decomposed into singular and
non-singular parts according to Eq.~(\ref{eq:cV_decomp}).
The $\eps$-expansion of the singular term is given in Eq.~(\ref{eq:cVS}).
In our case it is
\beq
  \cV^{(\rm S)}(s_{12},m_Q,m_Q;\eps)=\frac{1+v^2}{2v}
  \left[\frac1\eps\ln\frac{1-v}{1+v}-\frac12\ln^2\frac{1-v}{1+v}
    -\frac{\pi^2}6
  +\ln\frac{1-v}{1+v}\ln\frac2{1+v^2}\right].
\eeq
The non-singular term is given in Eq.~(\ref{eq:VNS_QQ}), and is given by
\beeq
  \lefteqn{
  \cV_q^{(\rm NS)}(s_{12},m_Q,m_Q) =
  \frac32\ln\frac{1+v^2}2
}\nonumber\\&&
  +\frac{1+v^2}{2v}\left[2\ln\frac{1-v}{1+v}\ln\frac{2(1+v^2)}{(1+v)^2}
    +2\Li_2\left(\left(\frac{1-v}{1+v}\right)^2\right)
    -2\Li_2\left(\frac{2v}{1+v}\right)-\frac{\pi^2}6\right]
\nonumber\\&&
  +\ln(1-\smfrac12\sqrt{1-v^2})-2\ln(1-\sqrt{1-v^2})
  -\frac{1-v^2}{1+v^2}\ln\frac{\sqrt{1-v^2}}{2-\sqrt{1-v^2}}
\nonumber\\&&
  -\frac{\sqrt{1-v^2}}{2-\sqrt{1-v^2}}+2\frac{1-v^2-\sqrt{1-v^2}}{1+v^2}
  +\frac{\pi^2}2.
\eeeq
Note that in the massless limit, $\cV_q^{(\rm NS)}$ vanishes.  The function
$\Gamma_q$ appearing in Eq.~(\ref{QQIfin}) is simply given by
\beq
  \Gamma_q(\mu,m_Q;\eps) =
  \CF\left[\frac1\eps+\frac12\ln\frac{m_Q^2}{\mu^2}-2\right].
\eeq

Combining the virtual and auxiliary cross sections, we obtain a
two-parton cross section that is finite as $\eps\to0$.  Setting $\eps=0$,
we obtain (recall that the leading order cross section, $\sigma^{\aLO}$,
is defined by the observable $F_J^{(2)}(p_1,p_2)$),
\beeq
  \sigma^{\aNLO\{2\}} &=&
  \sigma^{\aLO}\,2\CF\frac{\alps}{2\pi}\Biggl\{
  1
  -\frac{(1-v)^2}v\,\frac{\pi^2}{12}
  -\frac32\ln\frac{2(1+v^2)}{(1+v)^2}
\nonumber\\&&
  +\ln\frac{1-v}{1+v}
  \left(\frac{(1-v)^2}{4v}\ln\frac{1-v}{1+v}
    +\frac{1+v^2}v\ln\frac{1+v}{1+v^2}
  \right)
  +f_f(v)+\cV_q^{(\rm NS)}
  \Biggr\}
\nonumber\\&&
  +2\Real\{f_2\}\;v\,\sigma_0
  \left(\left(g^{VV}+g^{AA}\right)\left(\frac32\right)
  -g^{AA}\left(\frac52-4\mu_Q^2\right)\right)
  F_J^{(2)}(p_1,p_2).
\phantom{(A.99)}
  \label{QQNLO2}
\eeeq
Note that not only have all singularities cancelled, but also all terms
that are singular in the small-mass limit.  Furthermore, in this
small-mass limit $\sigma^{\aNLO\{2\}}$ agrees with the exactly massless
prediction given in Eq.~(D.11) of Ref.~\cite{Catani:1997vz},
\beq
\lim_{v\to1}\sigma^{\aNLO\{2\}}=\sigma^{\aLO} \CF\frac{\alps}{\pi}\:.
\eeq

As a check of our results, we can use them to obtain the total cross
section, by setting $F_J^{(3)}=F_J^{(2)}=1$.  The integral in
Eq.~(\ref{QQNLO3}) can be performed analytically, but the result does
not have a compact form, so we give its expansion in powers of $\mu_Q$,
\beeq
  \sigma^{\aNLO\{3\}} &=& \sigma_0\;\CF\frac{\alps}{2\pi} \Biggl\{
    \left(g^{VV}+g^{AA}\right)
    \left[
      -\frac12+\mu_Q^2\,\left(-2\ln\mu_Q^2+3\right)
      +4\mu_Q^3
      +{\rm O}(\mu_Q^4\ln\mu_Q^2)
    \right]
\nonumber\\&&
    +g^{AA}
    \left[
      \mu_Q^2\,\left(-8\ln\mu_Q^2-15\right)
      +{\rm O}(\mu_Q^4\ln\mu_Q^2)
    \right]
  \Biggr\}.
\eeeq
It is worth noting that no terms linear in $\mu_Q$ arise.  Although
there are no such terms in the real or virtual cross sections, they
could in principle arise in the auxiliary cross section, cancelling
between the two- and three-parton cross sections.  If this did happen it
would worsen the numerical convergence in the limit of small but finite
masses.  Such terms do appear for all higher odd-integer powers of
$\mu_Q$, but are not problematic.

Combining with $\sigma^{\aNLO\{2\}}$ from Eq.~(\ref{QQNLO2}), we obtain
\beeq
  \sigma^{\aNLO} &=& \sigma_0\;\CF\frac{\alps}{2\pi} \Biggl\{
    \left(g^{VV}+g^{AA}\right)
    \left[
      \frac32+18\mu_Q^2
      +{\rm O}(\mu_Q^4\ln\mu_Q^2)
    \right]
\nonumber\\&&
    +g^{AA}
    \left[
      \mu_Q^2\,\left(-18\ln\mu_Q^2-27\right)
      +{\rm O}(\mu_Q^4\ln\mu_Q^2)
    \right]
  \Biggr\},
\eeeq
in agreement with the result given in, for example,
Ref.~\cite{Kniehl:1990qu}.  In fact, expanding the two results to
arbitrary order in $\mu_Q$ we obtain perfect agreement.

The example application discussed in this appendix has also been
considered in Ref.~\cite{Eynck:2001en} within the context of the
methods of Refs.~\cite{Keller:1999tf} and \cite{Phaf:2001gc}. As
mentioned in Sect.~\ref{intro}, the formalisms
of Refs.~\cite{Keller:1999tf, Phaf:2001gc} are not aimed at
smoothly controlling the small-mass limit. This is explicitly shown, for
instance, by comparing our Eq.~(\ref{QQNLO2}) with the analogous
equations (the sum of Eqs.~(15) and (16) for the phase-space
slicing method of Ref.~\cite{Keller:1999tf}; the sum of Eqs.~(16) and
(24) for the dipole method of Ref.~\cite{Phaf:2001gc}) in
Ref.~\cite{Eynck:2001en}, which are singular in the small-mass limit.

\subsection{\boldmath$e^+e^-\to Q\bar Qg$}

We continue to average over event orientation and polarization.  As in
the $e^+e^-\to Q\bar Q$ case, we require the tree level matrix element
evaluated in $d$ dimensions, which is given by
\beeq
  |\cm_3^{(d)}(p_1,p_2,p_3)|^2 &=& |\cm_3^{(4)}(p_1,p_2,p_3)|^2
  -\eps|\cm_3^{(\prime)}(p_1,p_2,p_3)|^2, \\
  |\cm_3^{(\prime)}(p_1,p_2,p_3)|^2 &=&
  \CF\frac{8\pi\alps}s\left\{
    \left(g^{VV}+g^{AA}\right)+g^{AA}\left(2\mu_Q^2\right)
  \right\}
  \frac{(2-x_1-x_2)^2}{(1-x_1)(1-x_2)}.
\phantom{(A.99)}
\eeeq

At NLO, three different real-emission processes contribute: (a)
$e^+e^-\to Q(p_1)+\bar Q(p_2)+g(p_3)+g(p_4)$; (b) $e^+e^-\to Q(p_1)+\bar
Q(p_2)+q(p_3)+\bar q(p_4)$; and (c) $e^+e^-\to Q(p_1)+\bar
Q(p_2)+Q(p_3)+\bar Q(p_4)$.  The quark $q$ could be any flavour other
than $Q$, massless or massive.  Since the matrix elements, $\cm_4$,
for these processes are rather lengthy, it is not feasible to explicitly
show the cancellation of the soft and collinear poles between the
real-emission matrix elements and the auxiliary cross sections
constructed from various dipole contributions.  Therefore we only give
the parts of the auxiliary cross sections, but do not reproduce $\cm_4$,
which can be be found in Ref.~\cite{Ballestrero:1994dv}.

For process (a), we have to evaluate ten dipole contributions,
$\cD_{31,2}$, $\cD_{31,4}$, $\cD_{41,2}$,
$\cD_{41,3}$, $\cD_{32,1}$, $\cD_{32,4}$,
$\cD_{42,1}$, $\cD_{42,3}$, $\cD_{34,1}$ and
$\cD_{34,2}$.  The associated colour algebra is again
straightforward because the different colour projections of the
three-parton matrix element fully factorize (see Appendix~A
of Ref.~\cite{Catani:1997vz}).  Thus we do not need to calculate any
colour-correlated tree amplitudes and we obtain
\beeq
  \label{QQgD312a}
  \cD_{31,2}^{\mathrm{(a)}}(p_1,p_2,p_3,p_4) &=&
  \frac1{2p_3\ldot p_1}\left(1-\frac{\CA}{2\CF}\right)
  \;\langle\bV_{g_3Q_1,2}\rangle\;
  |\cm_3(\widetilde p_{31},\widetilde p_2,p_4)|^2, \\
  \label{QQgD314a}
  \cD_{31,4}^{\mathrm{(a)}}(p_1,p_2,p_3,p_4) &=&
  \frac1{2p_3\ldot p_1}\;\frac{\CA}{2\CF}
  \;\langle\bV_{g_3Q_1,4}\rangle\;
  |\cm_3(\widetilde p_{31},p_2,\widetilde p_4)|^2, \\
  \label{QQgD341a}
  \cD_{34,1}^{\mathrm{(a)}}(p_1,p_2,p_3,p_4) &=&
  \frac1{2p_3\ldot p_4}\;\frac12
  \;\langle\mu|\bV_{g_3g_4,1}|\nu\rangle\;
  \cT_{\mu\nu}(\widetilde p_1,p_2,\widetilde p_{34}).
\eeeq
The dipole contributions $\cD_{32,1}$, $\cD_{32,4}$ and
$\cD_{34,2}$ can be obtained from $\cD_{31,2}$,
$\cD_{31,4}$ and $\cD_{34,1}$ respectively by the replacement
$p_1\leftrightarrow p_2$.  Likewise $\cD_{41,2}$ and
$\cD_{41,3}$ can be obtained from $\cD_{31,2}$ and
$\cD_{31,4}$ respectively by the replacement
$p_3\leftrightarrow p_4$.  Finally, $\cD_{42,1}$ and
$\cD_{42,3}$ can be obtained from $\cD_{31,2}$ and
$\cD_{31,4}$ respectively by the replacement
$p_1\leftrightarrow p_2$ and $p_3\leftrightarrow p_4$.

The splitting functions $\langle\bV_{g_3Q_1,2}\rangle$ and
$\langle\bV_{g_3Q_1,4}\rangle$ are given by
Eq.~(\ref{eq:V_gQk}) with $m_k=m_Q$ and $m_k=0$ respectively and
$\langle\mu|\bV_{g_3g_4,1}|\nu\rangle$ by Eq.~(\ref{eq:V_ggk}) with
$m_k=m_Q$.  The tensor $\cT_{\mu\nu}$ is given below.

For process (b) we have to calculate at most four dipole contributions,
$\cD_{34,1}$ and $\cD_{34,2}$ and, perhaps, $\cD_{12,3}$
and $\cD_{12,4}$.  We obtain
\beq
  \label{QQgD341b}
  \cD_{34,1}^{\mathrm{(b)}}(p_1,p_2,p_3,p_4) =
  \frac1{2p_3\ldot p_4+2m_q^2}\;\frac12
  \;\langle\mu|\bV_{q_3\bar q_4,1}|\nu\rangle\;
  \cT_{\mu\nu}(\widetilde p_1,p_2,\widetilde p_{34}).
\eeq
The dipole contribution $\cD_{34,2}$ can be obtained from
$\cD_{34,1}$ by the replacement $p_1\leftrightarrow p_2$.  The
splitting function $\langle\mu|\bV_{q_3\bar q_4,1}|\nu\rangle$ is given by
Eq.~(\ref{eq:V_QQk}) with $m_k=m_Q$.  Note that we have explicitly
kept $m_q$ non-zero in Eq.~(\ref{QQgD341b}).  Since all our results tend
smoothly to the massless results in the small-mass limit, we can easily
replace $m_q$ by zero if necessary.

The kinematics for the dipole contributions $\cD_{12,3}$ and
$\cD_{12,4}$ replace the heavy quark and antiquark momenta $p_1$
and $p_2$ by a massless gluon with momentum $\widetilde p_{12}$.  If our
observable, $F_J$, requires the presence of heavy quarks in the final
state then it will get zero contribution from these dipole
contributions, leaving uncancelled logarithms of $\mu_Q^2$ in the
4-parton integral.  Even in this case, it may be helpful to replace the
observable by a pseudo-observable in which the subtraction $q\bar qg$
configurations do contribute, improving the numerical convergence of the
three-parton integral and allowing the logarithms to be isolated in the
pseudo-collinear region.  This may be more amenable to analytical
treatment, allowing the logarithms to be summed to all
orders\cite{Seymour:1995bz}.  If the observable sums over
all final states, then $\cD_{12,3}$ and $\cD_{12,4}$ are
needed.  If necessary, they can be obtained from $\cD_{34,1}$ and
$\cD_{34,2}$ by the replacements $p_1\leftrightarrow p_3$,
$p_2\leftrightarrow p_4$ and $m_Q\leftrightarrow m_q$.

For process (c) we have to calculate eight dipole contributions,
$\cD_{12,3}$, $\cD_{12,4}$, $\cD_{14,2}$,
$\cD_{14,3}$, $\cD_{23,1}$, $\cD_{23,4}$,
$\cD_{34,1}$ and $\cD_{34,2}$.  They are identical to those
for process (b), with $m_q$ replaced by $m_Q$.  Note however that the
external factors in Eq.~(\ref{eq:dPhi}) introduce an extra factor
of~$\frac14$ coming from the counting of two pairs of identical
particles in the final state.  The other dipoles can be obtained by
appropriate permutations of momenta.

The tensor $\cT_{\mu\nu}$ appearing in
Eqs.~(\ref{QQgD341a},\ref{QQgD341b}) is the squared amplitude for the LO
process $e^+e^-\to Q\bar Qg$ not summed over the gluon polarization.  It
is normalized so that $-g^{\mu\nu}\cT_{\mu\nu}=|\cm_3|^2$.
Again averaging over event orientation and polarization and neglecting
terms that are antisymmetric in $\mu,\nu$, which cannot contribute to
the cross section, we obtain
\beq
  \cT_{\mu\nu}(p_1,p_2,p_3) = -\CF\frac{8\pi\alps}s
  \;\frac{
    \left(g^{VV}+g^{AA}\right)\cT_{\mu\nu}^{VV}
    +g^{AA}\mu_Q^2\cT_{\mu\nu}^{AA-VV}
  }{(1-x_1)(1-x_2)},
\eeq
with
\beeq
  \cT_{\mu\nu}^{VV} &=&
  2(1+2\mu_Q^2)
    \left[\frac{p_1^\mu p_2^\nu}s+\frac{p_2^\mu p_1^\nu}s\right]
  -2\frac{1-x_1}{1-x_2}(1+2\mu_Q^2)\frac{p_1^\mu p_1^\nu}s
  -2\frac{1-x_2}{1-x_1}(1+2\mu_Q^2)\frac{p_2^\mu p_2^\nu}s
\hspace*{-1em}
\nonumber\\&&
  -\frac{1-x_1-x_2+x_2^2-2\mu_Q^2(x_1-x_2)}{1-x_2}
    \left[\frac{p_1^\mu p_3^\nu}s+\frac{p_3^\mu p_1^\nu}s\right]
\nonumber\\&&
  -\frac{1-x_2-x_1+x_1^2-2\mu_Q^2(x_2-x_1)}{1-x_1}
    \left[\frac{p_2^\mu p_3^\nu}s+\frac{p_3^\mu p_2^\nu}s\right]
\nonumber\\&&
  +\left(1+\smfrac12x_1^2+\smfrac12x_2^2-x_1-x_2\right)g^{\mu\nu}
  +4\mu_Q^2\frac{p_3^\mu p_3^\nu}s,
\\
\hspace*{-1em}
  \cT_{\mu\nu}^{AA-VV} &=&
  -12
    \left[\frac{p_1^\mu p_2^\nu}s+\frac{p_2^\mu p_1^\nu}s\right]
  +12\frac{1-x_1}{1-x_2}\frac{p_1^\mu p_1^\nu}s
  +12\frac{1-x_2}{1-x_1}\frac{p_2^\mu p_2^\nu}s
\nonumber\\&&
  -2\frac{2+2x_1-6x_2+x_1x_2+x_2^2}{1-x_2}
    \left[\frac{p_1^\mu p_3^\nu}s+\frac{p_3^\mu p_1^\nu}s\right]
\nonumber\\&&
  -2\frac{2+2x_2-6x_1+x_1x_2+x_1^2}{1-x_1}
    \left[\frac{p_2^\mu p_3^\nu}s+\frac{p_3^\mu p_2^\nu}s\right]
\nonumber\\&&
  +(2-x_1-x_2)^2g^{\mu\nu}
  -4(3-x_1-x_2)\frac{p_3^\mu p_3^\nu}s.
\eeeq

Next we have to evaluate the insertion operator ${\bom I}(\eps)$, which
gives the integral of the auxiliary cross section, and combine it with
the virtual cross section, which was calculated in
Refs.~\cite{HQeejets, Brandenburg:1998pu}. We use the notation
of Ref.~\cite{Brandenburg:1998pu}. They obtain
\beeq
  \label{M3-1loop}
  |\cm_3(p_1,p_2,p_3)|^2_{(\mathrm{1-loop})} &=&
  -\frac{\alps}{4\pi}\Nc\Biggl\{
  \frac2{\eps^2}
  +\frac1{\eps}\Biggl[\frac{17}3+2\left(\ln\frac{4\pi\mu^2}s
    +\ln\frac{\mu_Q^2}{(1-x_1)(1-x_2)}-\gamma_E\right)
\nonumber\\&&
  -\frac{2N_f}{3\Nc}
  -\frac1{\Nc^2}\,
  \frac1\beta\left(2\beta-(1+\beta^2)\ln\frac{1+\beta}{1-\beta}\right)
  \Biggr]\Biggr\}|\cm_3^{(4)}(p_1,p_2,p_3)|^2
\nonumber\\&&
  +\frac{\alps}{2\pi}\Nc\,\frac1\eps|\cm_3^{(\prime)}(p_1,p_2,p_3)|^2
  +F_{\mathrm{finite}}(p_1,p_2,p_3),
\eeeq
where $N_f$ is the number of massless flavours and
$\beta=\sqrt{1-4\mu_Q^2/(x_1+x_2-1)}$ is the velocity of $p_1$ in the
$p_1\!+\!p_2$ frame.  The function $F_{\mathrm{finite}}$ contains all of
the finite (as $\eps\to0$) terms,
\beq
  \label{Ffinite}
  F_{\mathrm{finite}} =
  \CF\frac{8\pi\alps}s\times2\left(
    g^{VV}F_{\mathrm{finite}}^{VV}+g^{AA}F_{\mathrm{finite}}^{AA}
  \right),
\eeq
where
\beq
  F_{\mathrm{finite}}^{VV\!/\!AA} =
  F_1^{\mathrm{counter,finite,}VV\!/\!AA}+
  F_1^{\mathrm{ext.,finite,}VV\!/\!AA}+
  \frac{\alps}{4\pi}\Nc\left\{F_1^{\mathrm{lc,}VV\!/\!AA}+
    \frac1{\Nc^2}F_1^{\mathrm{sc,}VV\!/\!AA}\right\},
\label{FfinVVAA}
\eeq
with $F_1^{\mathrm{counter,finite,}VV\!/\!AA}$,
$F_1^{\mathrm{ext.,finite,}VV\!/\!AA}$, $F_1^{\mathrm{lc,}VV\!/\!AA}$
and $F_1^{\mathrm{sc,}VV\!/\!AA}$ defined in Eqs.~(4.11), (A.2)
and~(A.3) of Ref.~\cite{Brandenburg:1998pu}.
In Eqs.~(\ref{M3-1loop}--\ref{FfinVVAA}) the expansion parameter is
$\alps^\msbar$, i.e.~charge renormalization is carried out in the
\msbar\ subtraction scheme.

In checking the quasi-collinear limit of our formulae, it is useful to
note the logarithmic terms in $F_{\mathrm{finite}}$.
We obtain\footnote{Note that the result in Ref.~\cite{Brandenburg:1998pu}
  is for a single heavy flavour.  We have trivially modified it to
  incorporate $N_F$ heavy flavours, which simply involves changing the
  logarithmic term in $F_1^{\mathrm{ext.,finite,}VV\!/\!AA}$ from
  $\ln(zs/\mu^2)$ to $\sum_{F=1}^{N_F}\ln(m_F^2/\mu^2)$.}
\beeq
\label{Ffinitelogs}
\lefteqn{
  F_{\mathrm{finite}}(p_1,p_2,p_3)
  =
  \frac{\alps}{2\pi}|\cm_3^{(4)}|^2\Biggl\{
  \ln\frac{\mu^2}s\left(\frac{11}6\CA-\frac23\TR(N_f+N_F)\right)
}\nonumber\\&&
\qquad
    -\frac12\CA\left(\gamma_E-\ln\frac{4\pi\mu^2}{s}\right)^2
    +\frac{2\CF-\CA}{v_{Q\bar Q}}\ln\sqrt{\frac{1-v_{Q\bar Q}}{1+v_{Q\bar Q}}}
    \left(\gamma_E-\ln\frac{4\pi\mu^2}{s}\right)
\nonumber\\&&
\qquad
    +\left(\gamma_E-\ln\frac{4\pi\mu^2}{s}\right)
    \left(2\CF+\frac{11}6\CA-\frac23\TR N_f+\CA
      \ln\frac{sm_Q^2}{s_{Qg}s_{\bar Qg}}\right)
\nonumber\\&&
\qquad
  +\CF\ln^2\mu_Q^2-\CF\ln\mu_Q^2
  +\frac23\TR\sum_{F=1}^{N_F}\ln\frac{m_F^2}s
\Biggr\}
\nonumber\\&&
  +\frac{\alps}{2\pi}|\cm_3^{(\prime)}|^2\Biggl\{
    -\CA\left(\gamma_E-\ln\frac{4\pi\mu^2}{s}\right)
    +2\CF\ln\mu_Q^2
  \Biggr\}
  +\ldots\;,
\phantom{(A.99)}
\eeeq
where the ellipsis represents terms that stay finite or vanish for
$\mu_Q^2\to0$ and $m_F^2/s\to0$.  The first line comes from the
renormalization of
$\alpha_s$, with $\mu$ the renormalization scale, and will remain in
the final result.  All the other occurrences of $\mu$ come from the
trivial fact that in Ref.~\cite{Brandenburg:1998pu}, the natural overall
factor of $(4\pi\mu^2/s)^\eps/\Gamma(1-\eps)$ is expanded in $\eps$ and
are cancelled by our insertion operator,~${\bom I}(\eps)$.  All the
remaining mass logarithms are related to the quasi-collinear
limit and should also be cancelled by~${\bom I}(\eps)$.

The general expression for ${\bom I}(\eps)$ is given in
Eq.~(\ref{eq:insop}).  The colour structure again factorizes and we
obtain
\beeq
  \label{I123}
\lefteqn{
  {}_3\langle{1,2,3|{\bom I}_3(\eps,\mu^2;\left\{p_i,m_i\right\})|1,2,3}
  \rangle_3 =
  |\cm_3^{(d)}|^2\times
  \frac{\alps}{2\pi}\,\frac{(4\pi)^\eps}{\Gamma(1-\eps)}
  \Biggl\{
  \CA
  \left(\frac{\mu^2}{s_{Qg}}\right)^\eps
    \cV^{(\rm S)}(s_{Qg},m_Q,0;\eps)
}\nonumber\\&&
  +\CA
  \left(\frac{\mu^2}{s_{\bar Qg}}\right)^\eps
    \cV^{(\rm S)}(s_{\bar Qg},m_Q,0;\eps)
  +(2\CF-\CA)
  \left(\frac{\mu^2}{s_{Q\bar Q}}\right)^\eps
    \cV^{(\rm S)}(s_{Q\bar Q},m_Q,m_Q;\eps)
\hspace{4em}
\nonumber\\&&
  +\frac{\CA}{2}
    \cV_Q^{(\rm NS)}(s_{Qg},m_Q,0)
  +\frac{2\CF-\CA}{2}
    \cV_Q^{(\rm NS)}(s_{Q\bar Q},m_Q,m_Q)
\nonumber\\&&
  +\frac{\CA}{2}
    \cV_Q^{(\rm NS)}(s_{\bar Qg},m_Q,0)
  +\frac{2\CF-\CA}{2}
    \cV_Q^{(\rm NS)}(s_{Q\bar Q},m_Q,m_Q)
\nonumber\\&&
  +\frac{\CA}{2}
    \cV_g^{(\rm NS)}(s_{Qg},0,m_Q,\{m_F\};\kappa)
  +\frac{\CA}{2}
    \cV_g^{(\rm NS)}(s_{\bar Qg},0,m_Q,\{m_F\};\kappa)
\nonumber\\&&
    +\frac{\CA}{2\CF}\,\gamma_q\ln\frac{\mu^2}{s_{Qg}}
    +\frac12\,\gamma_g\ln\frac{\mu^2}{s_{Qg}}
    +\left(1-\frac{\CA}{2\CF}\right)\gamma_q\ln\frac{\mu^2}{s_{Q{\bar Q}}}
\nonumber\\&&
    +\frac{\CA}{2\CF}\,\gamma_q\ln\frac{\mu^2}{s_{\bar Qg}}
    +\frac12\,\gamma_g\ln\frac{\mu^2}{s_{\bar Qg}}
    +\left(1-\frac{\CA}{2\CF}\right)\gamma_q\ln\frac{\mu^2}{s_{Q\bar Q}}
\nonumber\\&&
    -(2\CF+\CA)\frac{\pi^2}3
    +2\Gamma_Q(\mu,m_Q;\eps)+2\gamma_q+2K_q
    +\Gamma_g(\{m_F\};\eps)+\gamma_g+K_g
\Biggr\},
\phantom{(A.99)}
\eeeq
where $s_{jk}=2p_j\ldot p_k$, $\cV^{(\rm S)}$ is defined in
Eq.~(\ref{eq:cVS}), $\cV_j^{(\rm NS)}$ are defined in
Eqs.~(\ref{eq:VNS_QQ}--\ref{eq:VNS_gk}), $\Gamma_j$ are defined in
Eqs.~(\ref{eq:cgammag}--\ref{eq:cgammaqm}) and $\gamma_j$ and $K_j$ are
defined in Eqs.~(\ref{eq:gamma_g}) and~(\ref{eq:K_g}).

It is worth stressing that Eq.~(\ref{I123}) is the insertion operator
for the $Q\bar Qg$ final state.  It does not therefore account for terms
coming from the splitting process $g\to Q\bar Q$ in process~(b) above,
which contribute to the $q\bar qg$ final state.  The corresponding
insertion operator can be obtained by replacing $m_Q\to m_q$ in
Eq.~(\ref{I123}), while retaining $m_Q$ in the set $\{m_F\}$.

Explicitly, we obtain
\beeq
\label{I3final}
\lefteqn{
  {}_3\langle{1,2,3|{\bom I}_3(\eps,\mu^2;\left\{p_i,m_i\right\})|1,2,3}
  \rangle_3 =
  |\cm_3^{(d)}|^2\times
  \frac{\alps}{2\pi}\,\frac1{\Gamma(1-\eps)}\left(\frac{4\pi\mu^2}s\right)^\eps
  \Biggl\{
}\nonumber\\&&
  \CA
  \left(
    \frac{y_{Qg}^{-\eps}}{2\eps^2}
    +\frac{y_{Qg}^{-\eps}}{2\eps}\ln\frac{m_Q^2}{s_{Qg}}
  \right)
  +\CA
  \left(
    \frac{y_{\bar Qg}^{-\eps}}{2\eps^2}
    +\frac{y_{\bar Qg}^{-\eps}}{2\eps}\ln\frac{m_Q^2}{s_{\bar Qg}}
  \right)
  +\frac2\eps \CF
  +\frac1\eps\left(\frac{11}6\CA-\frac23\TR N_f\right)
\nonumber\\&&
  +\frac{2\CF-\CA}{v_{Q\bar Q}}
    \frac{y_{Q\bar Q}^{-\eps}}\eps\ln\sqrt{\frac{1-v_{Q\bar Q}}{1+v_{Q\bar Q}}}
  -\frac14\CA\ln^2\frac{m_Q^2}{s_{Qg}}
  -\frac14\CA\ln^2\frac{m_Q^2}{s_{\bar Qg}}
\nonumber\\&&
  -\frac12\,\frac{2\CF-\CA}{v_{Q\bar Q}}
    \ln^2\sqrt{\frac{1-v_{Q\bar Q}}{1+v_{Q\bar Q}}}
  -\frac23\TR\sum_{F=1}^{N_F}\ln\frac{m_F^2}{Q_{\rm aux}^2}
  +\CF\ln\frac{m_Q^2}s
\nonumber\\&&
  +G(p_1,p_2,p_3)
\Biggr\},
\eeeq
where $y_{jk}=s_{jk}/s$ and $Q_{\rm aux}$ is an arbitrary scale that
cancels against an equivalent term in $G$, as discussed after
Eq.~(\ref{eq:VNS_gQ}).  The function $G$ contains the terms that do not
diverge as either $\eps\to0$ or $m_F^2\ll s$,
\beeq
\lefteqn{
  G(p_1,p_2,p_3) =
  \frac34\CA\ln\frac{{s_{Q\bar Q}^2}}{s_{Qg}s_{\bar Qg}}
  +3\CF\ln\frac s{s_{Q\bar Q}}
  +\frac12\left(\frac{11}6\CA-\frac23\TR N_f\right)
    \ln\frac{s^2}{s_{Qg}s_{\bar Qg}}
}\nonumber\\&&
  +6\CF+\frac{50}9\CA
  -\frac{16}9\TR N_f
  -\CF\pi^2-\frac23\CA\pi^2
  -\frac{2\CF-\CA}{v_{Q\bar Q}}\frac{\pi^2}6\!
\nonumber\\&&
  -\CA
    \frac12\ln\frac{m_Q^4}{s_{Qg}(s_{Qg}+m_Q^2)}
    \ln\frac{s_{Qg}}{s_{Qg}+m_Q^2}
  -\CA
    \frac12\ln\frac{m_Q^4}{s_{\bar Qg}(s_{\bar Qg}+m_Q^2)}
    \ln\frac{s_{\bar Qg}}{s_{\bar Qg}+m_Q^2}
\nonumber\\&&
  +\frac{2\CF-\CA}{v_{Q\bar Q}}
    \ln\sqrt{\frac{1-v_{Q\bar Q}}{1+v_{Q\bar Q}}}
    \ln\frac{s_{Q\bar Q}+2m_Q^2}{s_{Q\bar Q}}
\nonumber\\&&
  +\frac{\CA}{2}\cV_Q^{(\rm NS)}(s_{Qg},m_Q,0)
  +\frac{\CA}{2}\cV_Q^{(\rm NS)}(s_{\bar Qg},m_Q,0)
  +(2\CF-\CA)\cV_Q^{(\rm NS)}(s_{Q\bar Q},m_Q,m_Q)
\nonumber\\&&
  +\frac{\CA}{2}\cV_g^{(\rm NS)}(s_{Qg},0,m_Q,\{m_F\};\kappa)
  +\frac{\CA}{2}\cV_g^{(\rm NS)}(s_{\bar Qg},0,m_Q,\{m_F\};\kappa).
\eeeq
Combining Eqs.~(\ref{M3-1loop}) and~(\ref{I3final}), we obtain
\beeq
\label{QQgNLO3}
\lefteqn{
  \left[
  {}_3\langle{1,2,3|{\bom I}_3(\eps,\mu^2;\left\{p_i,m_i\right\})|1,2,3}
  \rangle_3 + |\cm_3(p_1,p_2,p_3)|^2_{(\mathrm{1-loop})}
  \right]_{\eps=0} =
}\nonumber\\*&&
  \frac{\alps}{2\pi}|\cm_3^{(4)}|^2\Biggl\{
    \frac12\CA\left(\gamma_E-\ln\frac{4\pi\mu^2}{s}\right)^2
    -\frac{2\CF-\CA}{v_{Q\bar Q}}\ln\sqrt{\frac{1-v_{Q\bar Q}}{1+v_{Q\bar Q}}}
    \left(\gamma_E-\ln\frac{4\pi\mu^2}{s_{Q\bar Q}}\right)
\nonumber\\*&&\phantom{\frac{\alps}{2\pi}|\cm_3^{(4)}|^2}
    -\left(\gamma_E-\ln\frac{4\pi\mu^2}{s}\right)
    \left(2\CF+\frac{11}6\CA-\frac23\TR N_f+\CA
      \ln\frac{sm_Q^2}{s_{Qg}s_{\bar Qg}}\right)
\nonumber\\&&\phantom{\frac{\alps}{2\pi}|\cm_3^{(4)}|^2}
    +\CA\left(\frac14\ln^2y_{Qg}+\frac14\ln^2y_{\bar Qg}
      -\frac12\ln y_{Qg}\ln\frac{m_Q^2}{s_{Qg}}
      -\frac12\ln y_{\bar Qg}\ln\frac{m_Q^2}{s_{\bar Qg}}
      -\frac{\pi^2}{12}\right)
\nonumber\\&&\phantom{\frac{\alps}{2\pi}|\cm_3^{(4)}|^2}
  -\frac14\CA\ln^2\frac{m_Q^2}{s_{Qg}}
  -\frac14\CA\ln^2\frac{m_Q^2}{s_{\bar Qg}}
  -\frac12\,\frac{2\CF-\CA}{v_{Q\bar Q}}
    \ln^2\sqrt{\frac{1-v_{Q\bar Q}}{1+v_{Q\bar Q}}}
\nonumber\\&&\phantom{\frac{\alps}{2\pi}|\cm_3^{(4)}|^2}
  -\frac23\TR\sum_{F=1}^{N_F}\ln\frac{m_F^2}{Q_{\rm aux}^2}
  +\CF\ln\frac{m_Q^2}s
  +G(p_1,p_2,p_3)
\Biggr\}
\nonumber\\&&
  +\frac{\alps}{2\pi}|\cm_3^{(\prime)}|^2\Biggl\{
    -2\CF-\left(\frac{11}6\CA-\frac23\TR N_f\right)
    -\frac{2\CF-\CA}{v_{Q\bar Q}}
    \ln\sqrt{\frac{1-v_{Q\bar Q}}{1+v_{Q\bar Q}}}
\nonumber\\&&\phantom{\frac{\alps}{2\pi}|\cm_3^{(\prime)}|^2}
    -\CA\left(\ln\frac{4\pi\mu^2m_Q^2}{s_{Qg}s_{\bar Qg}}-\gamma_E\right)
  \Biggr\}
\nonumber\\&&
  +F_{\mathrm{finite}}(p_1,p_2,p_3)
  .
\eeeq
The integration of this expression over the phase space in
Eq.~(\ref{dPhi3}) with the observable $F_J^{(3)}$ gives the three-parton
cross section $\sigma^{\aNLO\{3\}}$.

Comparing Eqs.~(\ref{QQgNLO3}) and~(\ref{Ffinitelogs}), we see that all
logarithms of $m_Q^2$ cancel, leaving a cross section that is
well-behaved for all masses.

If the observable to be calculated receives contributions from all final
states, whether containing a heavy quark or not, it is also necessary to
calculate the three-parton cross section for light quarks.  Since our
formalism guarantees a smooth small-mass limit, it is straightforward to
set $m_Q$ to zero in Eq.~(\ref{QQgNLO3}), while keeping $m_F$ non-zero.

\subsection{\boldmath$ep\to eQ\bar QX$}

Since the analytical formulae are more lengthy in this case%
\footnote{A NLO Monte Carlo program, based on the subtraction method, to
compute infrared and collinear safe observables in this process is
presented in Ref.~\cite{Harris:1997zq}.},
we only give the formulae necessary to construct the auxiliary cross
section and its integral.

For simplicity, we start by assuming that the observable of interest
requires the presence of a heavy quark pair in the final state.  That
is, we do not consider heavy-quark corrections to light-quark
processes.  Thus, at lowest order, there is only one process,
$e+g(p_a)\to e+Q(p_1)+\bar Q(p_2)$, with colour averaged matrix element
$1/(\Nc^2-1)|\cm_3(p_1,p_2,-p_a)|^2$.  Note that this matrix element is
not identical to the one given earlier for the process $e^+e^-\to Q\bar
Qg$, only because we averaged over the orientation of the three-jet
system in the $e^+e^-$ rest frame.  If the full angular information had
been retained, then $|\cm_3|^2$ could simply be obtained by crossing the
$e^+e^-$ annihilation result.  The same comment applies to the tensor
$\cT_{\mu\nu}$, to be defined shortly.

Three NLO real emission processes contribute: (a)~$e+g(p_a)\to
e+Q(p_1)+\bar Q(p_2)+g(p_3)$; (b)~$e+q(p_a)\to e+Q(p_1)+\bar
Q(p_2)+q(p_3)$; and (c)~$e+\bar q(p_a)\to e+Q(p_1)+\bar Q(p_2)+\bar
q(p_3)$.  The last is identical to the second so we do not explicitly
consider it further.

For process (a), we have to evaluate six dipole contributions, 
$\cD_{31,2}$, $\cD_{32,1}$, $\cD_{31}^a$,
$\cD_{32}^a$, $\cD_1^{a3}$ and $\cD_2^{a3}$.  The first
two are constructed in exactly the same way as in the $e^+e^-$ processes
already considered.  The third dipole contribution, $\cD_{31}^a$,
with a final-state emitter and initial-state spectator, is given by
\beq
  \cD_{31}^{a\mathrm{(a)}} = \frac1{2p_3\ldot p_1}\;\frac{\CA}{2\CF}
  \;\frac1{x_{31,a}}
  \;\langle\bV_{g_3Q_1}^a\rangle
  \times
  \frac1{\Nc^2-1}|\cm_3(\widetilde p_{31},p_2,-\widetilde p_a)|^2,
\eeq
with
\beq
\label{x31a}
  x_{31,a} = \frac{p_a\ldot p_3 + p_a\ldot p_1 - p_3\ldot p_1}
  {p_a\ldot p_3 + p_a\ldot p_1},
\eeq
\beq
\label{pa,p31}
  \widetilde p_a^\mu = x_{31,a}p_a^\mu,
\qquad
  \widetilde p_{31}^\mu = p_3^\mu + p_1^\mu - (1-x_{31,a}) p_a^\mu,
\eeq
and $\langle\bV_{g_3Q_1}^a\rangle$ in Eq.~(\ref{eq:V_gQa}).  The dipole
contribution
$\cD_{32}^a$ can be obtained from $\cD_{31}^a$ by the
replacement $p_1\leftrightarrow p_2$.

The fifth dipole contribution for process (a), $\cD_1^{a3}$, with
an initial-state emitter and final-state spectator, is given by
\beq
  \cD_1^{a3\mathrm{(a)}} = \frac1{2p_a\ldot p_3}\;\frac12
  \;\frac1{x_{31,a}}
  \;\langle\mu|\bV_1^{g_ag_3}|\nu\rangle
  \times
  \frac1{\Nc^2-1}\cT_{\mu\nu}(\widetilde p_{31},p_2,-\widetilde p_a),
\eeq
with $x_{31,a}$, $\widetilde p_a$ and $\widetilde p_{31}$ given by
Eqs.~(\ref{x31a}) and~(\ref{pa,p31}) and
$\langle\mu|\bV_1^{g_ag_3}|\nu\rangle$ given in
Eq.~(\ref{eq:V_ggIF}).  The tensor $\cT_{\mu\nu}$ is the squared
amplitude for the LO process not summed over gluon polarization,
normalized so that $-g^{\mu\nu}\cT_{\mu\nu}=|\cm_3|^2$.  The
dipole contribution $\cD_2^{a3}$ can be obtained from $\cD_1^{a3}$ by
the replacement $p_1\leftrightarrow p_2$.

In process (b), we only have to evaluate two dipole contributions,
$\cD_1^{a3}$ and $\cD_2^{a3}$ (recall that we assume for now
that our observable requires heavy quarks in the final state.  The other
two dipoles for this process, $\cD_{12,3}$ and $\cD_{12}^a$,
in which the heavy quark-antiquark pair is replaced by a massless gluon,
do not therefore contribute).  They are given by
\beq
  \cD_1^{a3\mathrm{(b)}} = \frac1{2p_a\ldot p_3}\;\frac12
  \;\frac1{x_{31,a}}
  \;\langle\mu|\bV_1^{q_aq_3}|\nu\rangle
  \times
  \frac1{\Nc^2-1}\cT_{\mu\nu}(\widetilde p_{31},p_2,-\widetilde p_a),
\eeq
with $x_{31,a}$, $\widetilde p_a$ and $\widetilde p_{31}$ again given by
Eqs.~(\ref{x31a}) and~(\ref{pa,p31}) and
$\langle\mu|\bV_1^{q_aq_3}|\nu\rangle$ given in
Eq.~(\ref{eq:V_qqIF}).  The dipole contribution $\cD_2^{a3}$ can
be obtained from $\cD_1^{a3}$ by the replacement
$p_1\leftrightarrow p_2$.

Next we need to evaluate the insertion operator
${\bom I}_{2+a}(\eps,\mu^2,\{p_i,m_i\},p_a)$, which cancels all the
singularities of the one-loop cross section, and the finite operators
related to the factorization of initial-state singularities,
${\bom P}_2^{a,a'}(x;\mu_F^2;\{p_i\},xp_a)$ and
${\bom K}_2^{a,a'}(x;\{p_i,m_i\},xp_a)$.

The expression for ${\bom I}_{2+a}(\eps)$ is identical to that for
${\bom I}_3(\eps)$, Eq.~(\ref{I3final}), except that $\{m_F\}$ is
replaced by the empty set, $\{\}$, and $\kappa$ is replaced by 2/3.
Note that our uniform notation (see the discussion after Eq.~(C.27) of
Ref.~\cite{Catani:1997vz}) means that all dot products, $s_{ja}$, remain
positive and that $s$ should be replaced by $Q^2=-(p_a-p_1-p_2)^2$.

The operators ${\bom P}_2^{a,a'}$ are independent of the presence of
massive quarks in the final state and are given by
\beeq
\lefteqn{
  \sum_{a'}{}_{2,a'}\langle{1,2;xp_a|
  {\bom P}_2^{q,a'}(x;\mu_F^2;p_1,p_2,xp_a)
  |1,2;xp_a}\rangle_{2,a'} =
}\nonumber\\&&\qquad
  -\frac{\alps}{2\pi}P^{qg}(x)
  \ln\frac{\mu_F^2}{x\sqrt{s_{1a}s_{2a}}}
  \times
  \frac1{\Nc^2-1}|\cm_3(p_1,p_2,-xp_a)|^2, \\
\lefteqn{
  \sum_{a'}{}_{2,a'}\langle{1,2;xp_a|
  {\bom P}_2^{g,a'}(x;\mu_F^2;p_1,p_2,xp_a)
  |1,2;xp_a}\rangle_{2,a'} =
}\nonumber\\&&\qquad
  -\frac{\alps}{2\pi}P^{gg}(x)
  \ln\frac{\mu_F^2}{x\sqrt{s_{1a}s_{2a}}}
  \times
  \frac1{\Nc^2-1}|\cm_3(p_1,p_2,-xp_a)|^2,
\eeeq
for incoming quarks, $a=q$, and gluons, $a=g$, respectively.

The operators ${\bom K}_2^{a,a'}$ do depend on the quark mass and are
factorization-scheme, but not -scale, dependent.  They are given by
\beeq
\lefteqn{
  \sum_{a'}{}_{2,a'}\langle{1,2;xp_a|
  {\bom K}_2^{q,a'}(x,p_1,m_Q,p_2,m_Q,xp_a)
  |1,2;xp_a}\rangle_{2,a'} =
}\nonumber\\&&\quad
  \frac{\alps}{2\pi}
  \Biggl[
    \overline{K}^{qg}(x)-\KFS{qg}(x)
    +\smfrac12\CF\left(
      \frac{\CA}{\CF}\cK_q^{q,g}(x;s_{1a},m_Q)
      +\frac{\CA}{\CF}\cK_q^{q,g}(x;s_{2a},m_Q)
    \right)
\nonumber\\&&\qquad
    +\smfrac12P^{qg}_{\mathrm{reg}}(x)\left(
      \ln\frac{(1-x)s_{1a}}{(1-x)s_{1a}+m_Q^2}
      +\ln\frac{(1-x)s_{2a}}{(1-x)s_{2a}+m_Q^2}
    \right)
  \Biggr]
\nonumber\\&&\quad
  \times
  \frac1{\Nc^2-1}|\cm_3(p_1,p_2,-xp_a)|^2, \\
\lefteqn{
  \sum_{a'}{}_{2,a'}\langle{1,2;xp_a|
  {\bom K}_2^{g,a'}(x,p_1,m_Q,p_2,m_Q,xp_a)
  |1,2;xp_a}\rangle_{2,a'} =
}\nonumber\\&&\quad
  \frac{\alps}{2\pi}
  \Biggl[
    \overline{K}^{gg}(x)-\KFS{gg}(x)
    +\smfrac12\CA\left(
      \cK_q^{g,g}(x;s_{1a},m_Q)
      +\cK_q^{g,g}(x;s_{2a},m_Q)
    \right)
\nonumber\\&&\qquad
    +\smfrac12\gamma_g\delta(1-x)\left(
      \ln\frac{s_{1a}-2m_Q\sqrt{s_{1a}+m_Q^2}+2m_Q^2}{s_{1a}}
      +\ln\frac{s_{2a}-2m_Q\sqrt{s_{2a}+m_Q^2}+2m_Q^2}{s_{2a}}
    \right)
\hspace*{-1em}
\nonumber\\&&\qquad
    +\smfrac12P^{gg}_{\mathrm{reg}}(x)\left(
      \ln\frac{(1-x)s_{1a}}{(1-x)s_{1a}+m_Q^2}
      +\ln\frac{(1-x)s_{2a}}{(1-x)s_{2a}+m_Q^2}
    \right)
  \Biggr]
\nonumber\\&&\quad
  \times
  \frac1{\Nc^2-1}|\cm_3(p_1,p_2,-xp_a)|^2,
\eeeq
for incoming quarks and gluons respectively.  The functions
$\overline{K}^{aa'}$, $\cK_q^{q,g}$, $\cK_q^{g,g}$ and
$P^{ab}_{\mathrm{reg}}$ are defined in Eqs.~(\ref{eq:kbarab}),
(\ref{eq:cKqg_q}), (\ref{eq:cKgg_q}) and~(\ref{eq:Preg})
respectively, while $\KFS{aa'}$ define the factorization scheme
($\KFS{aa'}(x)=0$ in the $\msbar$ scheme) as discussed in
Ref.~\cite{Catani:1997vz}.

The ingredients given above are sufficient to construct a complete NLO
calculation for any observable that requires the presence of heavy
quarks in the final state.  The dipole contributions are subtracted from
the real matrix elements as in Eq.~(\ref{eq:NLOppmp}) to give a finite
3-parton integral.  The operator ${\bom I}_{2+a}(\eps)$ is inserted
into Eq.~(\ref{eq:NLOppm}), cancelling the singularities in the one-loop
matrix element to give a finite 2-parton integral.  Finally, the
operators ${\bom P}_2^{a,a'}(x;\mu_F^2)$ and ${\bom K}_2^{a,a'}(x)$
are inserted into Eq.~(\ref{eq:NLOppx}) to give the finite remainder
from initial-state factorization.

However, for a complete description of all final states in deep
inelastic scattering, one must also consider heavy-quark corrections to
light-quark processes.  Since we do not consider incoming massive
partons, the only such correction comes from the `QCD Compton process',
$e+q\to e+q+g$, followed by the gluon decay $g\to Q+\bar Q$.  Rather
than giving all the formulae relevant for a complete NLO calculation of
the QCD Compton process, we assume that such a calculation already
exists with $N_f$ massless quark flavours and give the additional terms
that must be added to it owing to the presence of the $N_F$ massive
quark flavours.

Firstly, we have to evaluate the two dipole corrections we neglected in
process~(b) earlier, $\cD_{12,3}$ and $\cD_{12}^a$, in which
the massive quark-antiquark pair is replaced by a massless gluon.  The
first is constructed in exactly the same way as in the $e^+e^-$
processes discussed earlier.  The second, with a final-state emitter and
initial-state spectator, is given by (recall that the momenta are
labelled $e+q(p_a)\to e+Q(p_1)+\bar Q(p_2)+q(p_3)$),
\beq
  \cD_{12}^a = \frac1{2p_1\ldot p_2+2m_Q^2}\;\frac12
  \;\frac1{x_{12,a}}
  \;\langle\mu|\bV_{Q_1\bar Q_2}^{a}|\nu\rangle
  \times\frac1{\Nc}\cT_{\mu\nu}(p_3,-\widetilde p_a,\widetilde p_{12}),
\eeq
with
\beq
  x_{12,a} = \frac{p_a\ldot p_1 + p_a\ldot p_2 - p_1\ldot p_2-m_Q^2}
  {p_a\ldot p_1 + p_a\ldot p_2},
\eeq
\beq
  \widetilde p_a^\mu = x_{12,a}p_a^\mu,
\qquad
  \widetilde p_{12}^\mu = p_1^\mu + p_2^\mu - (1-x_{12,a}) p_a^\mu,
\eeq
and $\langle\mu|\bV_{Q_1\bar Q_2}^{a}|\nu\rangle$ in
Eq.~(\ref{eq:V_QQbara}).  The matrix
element for the Born process, $e+q(p_a)\to e+q(p_1)+g(p_2)$, not summed
over gluon spin, is given by $\frac1{\Nc}\cT_{\mu\nu}(p_1,-p_a,p_2)$.

Next we have to evaluate the additional contribution to the operator
${\bom I}(\eps)$ from the $N_F$ flavours of massive quark,
\beeq
\lefteqn{
  {}_{2,a}\langle{1,2;a|\delta{\bom I}_{2+a}(\eps)|1,2;a}\rangle_{2,a}
}\nonumber\\
  &\equiv&
  {}_{2,a}\langle{1,2;a|{\bom I}_{2+a}(\eps,\{m_F\})|1,2;a}\rangle_{2,a} -
  {}_{2,a}\langle{1,2;a|{\bom I}_{2+a}(\eps,\{\})|1,2;a}\rangle_{2,a} \\
  &=&
  \frac{\alps}{2\pi}
    \times\frac23\TR\Biggl\{
      \sum_{F=1}^{N_F^{21}}\left[
        \ln\frac{1+\rho_{F,21}}2-\frac{\rho_{F,21}}3(3+\rho_{F,21}^2)
        -\frac12\ln\frac{m_F^2}{s_{21}}
      \right]
\nonumber\\&&
    +
      \sum_{F=1}^{N_F^{2a}}\left[
        \ln\frac{1+\rho_{F,2a}}2-\frac{\rho_{F,2a}}3(3+\rho_{F,2a}^2)
        -\frac12\ln\frac{m_F^2}{s_{2a}}
        \right]
  \Biggr\}\times\frac1{\Nc}|\cm_3(p_1,-p_a,p_2)|^2,
\phantom{(A.99)}
\eeeq
with $\rho_{F,jk}=\sqrt{1-4m_F^2/s_{jk}}$ and $N_F^{jk}$
defined to be the number of flavours for which $s_{jk}>4m_F^2$.
Note that this contribution is finite, so we have set $\eps\to0$.  It
does however diverge as $m_F\to0$, which is cancelled by a corresponding
divergence from the virtual matrix element, yielding a two-parton
integral that has a smooth small-mass limit.

The operator ${\bom P}(x)$ is unaffected by the quark mass.  Finally,
therefore, we just have to evaluate the extra contribution to the
operator ${\bom K}(x)$ from the $N_F$ flavours of massive quark,
\beeq
\lefteqn{
  \sum_{a'}
  {}_{2,a'}\langle{1,2;xp_a|\delta{\bom K}^{a,a'}(x)|1,2;xp_a}
  \rangle_{2,a'}
}\nonumber\\
  \equiv&&\!\!\!\!\!\!
  \sum_{a'}
  {}_{2,a'}\langle{1,2;xp_a|{\bom K}^{a,a'}(x,\{m_F\})|1,2;xp_a}
  \rangle_{2,a'} -
  \sum_{a'}
  {}_{2,a'}\langle{1,2;xp_a|{\bom K}^{a,a'}(x,\{\})|1,2;xp_a}
  \rangle_{2,a'}
\nonumber\\[-2ex]\\
\label{eq:deltaK}
  =&&\!\!\!\!\!\!
  \frac{\alps}{2\pi}\times\frac23\TR\sum_{F=1}^{N_F^{2a}}
  \Biggl\{\left(\delta(x_+-x)-\delta(1-x)\right)
    \left[\ln\frac{1+\rho_{F,2a}}2
      -\frac{\rho_{F,2a}}3(3+\rho_{F,2a}^2)
      -\frac12\ln\frac{m_F^2}{s_{2a}}
    \right]
\nonumber\\&&\!\!\!\!\!\!
  +\frac12\left(\frac{1-x+2m_F^2/s_{2a}}{(1-x)^2}
  \sqrt{1-\frac{4m_F^2/s_{2a}}{1-x}}\right)_{\!x+}\!\!\!
  +\delta(x_+-x)\frac12\rho_{F,2a}^3
  \Biggr\}\times\frac1{\Nc}|\cm_3(p_1,-xp_a,p_2)|^2,
\nonumber\\[-2ex]
\eeeq
for $a=q$, with $x_+=1-4m_F^2/s_{2a}=\rho_{F,2a}^2$.  For $a=g$,
$\delta{\bom K}(x)$ is zero.
Note that, as discussed at the ends of Sects.~\ref{se:finalinitial}
and~\ref{subsec:oneini} and in App.~\ref{app:Q2fixed}, $s_{2a}$ is given
by $2 p_2\ldot p_a$, i.e.~it is calculated from $p_a$, the momentum of
the initial parton, rather than $xp_a$, the momentum of the parton
entering the Born cross section and from $p_2$, the final-state
momentum belonging to the boosted phase space.

In the small-mass limit, since $x_+\to1$, the first line of
Eq.~(\ref{eq:deltaK}) does not contribute, so the whole contribution
remains finite.  It is straightforward to check that in this limit, it
gives the same results as if the $N_F$ extra flavours were massless.

These ingredients, together with the massless results in
Ref.~\cite{Catani:1997vz}, are sufficient to provide a complete NLO
calculation of all final states in deep inelastic scattering involving
either massive or massless partons.  However, as discussed in
Sect.~\ref{se:subwh},
care must be taken in taking the small-mass limit, since we do not
include incoming massive partons.  Logarithms of $m_F^2$ will therefore
remain in the final cross section, preventing a smooth small-mass
limit.  Since we have explicitly demonstrated that all the insertion
operators do have smooth limits, this behaviour is isolated in the
three-parton integral, $\int(d\sigma^R-d\sigma^A)$.
To achieve a smooth small-mass limit, the logarithmic behaviour of the
three-parton integral can be evaluated and properly matched with
a suitable definition of the heavy-quark parton
distribution (see Ref.~\cite{Demina:1999ze}).


\begin{thebibliography}{99}
\frenchspacing

\bibitem{Altarelli:2000ye}
Proceedings of the Workshop on
{\it Standard model physics (and more) at the LHC},
Eds. G.~Altarelli and M.L.~Mangano, CERN-2000-04, Geneva 2000.

\bibitem{Accomando:1998wt}
E.~Accomando {\it et al.}  [ECFA/DESY LC Physics Working Group Collaboration],
Phys.\ Rept.\  {\bf 299} (1998) 1
[hep-ph/9705442];\\
J.~A.~Aguilar-Saavedra {\it et al.},
TESLA Technical Design Report Part III: Physics at an $\mathrm{e^+e^-}$
Linear Collider,
hep-ph/0106315.


\bibitem{HQhtot}
P.~Nason, S.~Dawson and R.~K.~Ellis,
Nucl.\ Phys.\ B {\bf 303} (1988) 607;
\\
W.~Beenakker, H.~Kuijf, W.~L.~van Neerven and J.~Smith,
Phys.\ Rev.\ D {\bf 40} (1989) 54.

\bibitem{HQdist}
P.~Nason, S.~Dawson and R.~K.~Ellis,
Nucl.\ Phys.\ B {\bf 327} (1989) 49
[Erratum-ibid.\ B {\bf 335} (1989) 260];
\\
M.~L.~Mangano, P.~Nason and G.~Ridolfi,
Nucl.\ Phys.\ B {\bf 373} (1992) 295.

\bibitem{Giele:1992vf}
W.~T.~Giele and E.~W.~Glover,
Phys.\ Rev.\ D {\bf 46} (1992) 1980.

\bibitem{Giele:1993dj}
W.~T.~Giele, E.~W.~Glover and D.~A.~Kosower,
Nucl.\ Phys.\ B {\bf 403} (1993) 633
[hep-ph/9302225].

\bibitem{Keller:1999tf}
S.~Keller and E.~Laenen,
Phys.\ Rev.\ D {\bf 59} (1999) 114004
[hep-ph/9812415].

\bibitem{Harris:2001sx}
B.~W.~Harris and J.~F.~Owens,
preprint ANL-HEP-PR-00-044
[hep-ph/0102128].

\bibitem{Frixione:1996ms}
S.~Frixione, Z.~Kunszt and A.~Signer,
Nucl.\ Phys.\ B {\bf 467} (1996) 399
[hep-ph/9512328];
\\
S.~Frixione,
Nucl.\ Phys.\ B {\bf 507} (1997) 295
[hep-ph/9706545].

\bibitem{Catani:1996jh}
S.~Catani and M.~H.~Seymour,
Phys.\ Lett.\ B {\bf 378} (1996) 287
[hep-ph/9602277].

\bibitem{Catani:1997vz}
S.~Catani and M.~H.~Seymour,
Nucl.\ Phys.\ B {\bf 485} (1997) 291
[Erratum-ibid.\ B {\bf 510} (1997) 291]
[hep-ph/9605323].

\bibitem{Nagy:1996bz}
Z.~Nagy and Z.~Tr\'ocs\'anyi,
Nucl.\ Phys.\ B {\bf 486} (1997) 189
[hep-ph/9610498].

\bibitem{Fabricius:1981sx}
K.~Fabricius, I.~Schmitt, G.~Kramer and G.~Schierholz,
Z.\ Phys.\ C {\bf 11} (1981) 315;
\\
G.~Kramer and B.~Lampe,
Fortsch.\ Phys.\  {\bf 37} (1989) 161.

\bibitem{Ellis:1980wv}
R.~K.~Ellis, D.~A.~Ross and A.~E.~Terrano,
Nucl.\ Phys.\ B {\bf 178} (1981) 421.

\bibitem{Nagy:1997yn}
Z.~Nagy and Z.~Tr\'ocs\'anyi,
Phys.\ Rev.\ Lett.\  {\bf 79} (1997) 3604
[hep-ph/9707309],
Phys.\ Rev.\ D {\bf 57} (1998) 5793
[hep-ph/9712385],
Phys.\ Rev.\ D {\bf 59} (1999) 014020
[Erratum-ibid.\ D {\bf 62} (1999) 099902]
[hep-ph/9806317].

\bibitem{Weinzierl:1999yf}
S.~Weinzierl and D.~A.~Kosower,
Phys.\ Rev.\ D {\bf 60} (1999) 054028
[hep-ph/9901277].

\bibitem{Nagy:2001xb}
Z.~Nagy and Z.~Tr\'ocs\'anyi,
Phys.\ Rev.\ Lett.\  {\bf 87} (2001) 082001
[hep-ph/0104315].

\bibitem{Nagy:2001fj}
Z.~Nagy,
preprint IPPP/01/48
[hep-ph/0110315].

\bibitem{Campbell:1999ah}
J.~M.~Campbell and R.~K.~Ellis,
Phys.\ Rev.\ D {\bf 60} (1999) 113006
[hep-ph/9905386].

\bibitem{Ellis:1998fv}
R.~K.~Ellis and S.~Veseli,
Phys.\ Rev.\ D {\bf 60} (1999) 011501
[hep-ph/9810489];
\\
J.~M.~Campbell and R.~K.~Ellis,
Phys.\ Rev.\ D {\bf 62} (2000) 114012
[hep-ph/0006304].

\bibitem{Beenakker:1999xh}
W.~Beenakker, M.~Klasen, M.~Kr\"amer, T.~Plehn, M.~Spira and P.~M.~Zerwas,
Phys.\ Rev.\ Lett.\  {\bf 83} (1999) 3780
[hep-ph/9906298].

\bibitem{Maina:1996ep}
E.~Maina, R.~Pittau and M.~Pizzio,
Phys.\ Lett.\ B {\bf 393} (1997) 445
[hep-ph/9609468],
Phys.\ Lett.\ B {\bf 429} (1998) 354
[hep-ph/9710375].

\bibitem{Catani:1999nf}
S.~Catani and M.~H.~Seymour,
JHEP {\bf 9907} (1999) 023
[hep-ph/9905424].

\bibitem{Kramer:1999bf}
M.~Kr\"amer,
Phys.\ Rev.\ D {\bf 60} (1999) 111503
[hep-ph/9904416].


\bibitem{Dittmaier:2000mb}
S.~Dittmaier,
Nucl.\ Phys.\ B {\bf 565} (2000) 69
[hep-ph/9904440].

\bibitem{Roth:1999kk}
M.~Roth,
PhD thesis, ETH Z\"urich No. 13363 (1999),
hep-ph/0008033.


\bibitem{Denner:2000bj}
A.~Denner, S.~Dittmaier, M.~Roth and D.~Wackeroth,
Nucl.\ Phys.\ B {\bf 587} (2000) 67
[hep-ph/0006307].

\bibitem{Dittmaier:2001ay}
S.~Dittmaier and M.~Kr\"amer,
preprint DESY-01-121, to appear in Phys.\ Rev.\ D
[hep-ph/0109062].

\bibitem{Dittmaier:2000tc}
S.~Dittmaier, M.~Kr\"amer, Y.~Liao, M.~Spira and P.~M.~Zerwas,
Phys.\ Lett.\ B {\bf 478} (2000) 247
[hep-ph/0002035].

\bibitem{Phaf:2001gc}
L.~Phaf and S.~Weinzierl,
JHEP {\bf 0104} (2001) 006
[hep-ph/0102207].

\bibitem{Catani:2001ef}
S.~Catani, S.~Dittmaier and Z.~Tr\'ocs\'anyi,
Phys.\ Lett.\ B {\bf 500} (2001) 149
[hep-ph/0011222].

\bibitem{Beenakker:2001rj}
W.~Beenakker, S.~Dittmaier, M.~Kr\"amer, B.~Pl\"umper, M.~Spira and
P.~M.~Zerwas,
Phys.\ Rev.\ Lett.\  {\bf 87} (2001) 201805
[hep-ph/0107081].

\bibitem{Catani:1997fq}
S.~Catani and M.~H.~Seymour,
Acta Phys.\ Polon.\ B {\bf 28} (1997) 863
[hep-ph/9612236].

\bibitem{HQeejets}
G.~Rodrigo, A.~Santamaria and M.~Bilenky,
Phys.\ Rev.\ Lett.\  {\bf 79} (1997) 193
[hep-ph/9703358],
Nucl.\ Phys.\ B {\bf 554} (1999) 257
[hep-ph/9905276];
\\
W.~Bernreuther, A.~Brandenburg and P.~Uwer,
Phys.\ Rev.\ Lett.\  {\bf 79} (1997) 189
[hep-ph/9703305];
\\
P.~Nason and C.~Oleari,
Nucl.\ Phys.\ B {\bf 521} (1998) 237
[hep-ph/9709360].

\bibitem{factform}
See, for instance:\\
A.~Bassetto, M.~Ciafaloni and G.~Marchesini,
Phys.\ Rept.\  {\bf 100} (1983) 201;
\\
Yu.~L.~Dokshitzer, V.~A.~Khoze, A.~H.~Mueller and S.~I.~Troian,
{\it Basics of perturbative QCD\/} (Editions Fronti\`eres, Gif-sur-Yvette, 1991).

\bibitem{Altarelli:1977zs}
G.~Altarelli and G.~Parisi,
Nucl.\ Phys.\ B {\bf 126} (1977) 298.

\bibitem{Kunszt:1994mc}
Z.~Kunszt, A.~Signer and Z.~Tr\'ocs\'anyi,
Nucl.\ Phys.\ B {\bf 420} (1994) 550
[hep-ph/9401294].

\bibitem{Collins:1989gx}
See, J.~C.~Collins, D.~E.~Soper and G.~Sterman,
in {\it  Perturbative Quantum Chromodynamics}, Ed. A.H.~Mueller
(World Scientific, Singapore, 1989), p.~1, and references therein. 

\bibitem{Bloch:1937pw}
F.~Bloch and A.~Nordsieck,
Phys.\ Rev.\  {\bf 52} (1937) 54.

\bibitem{BNviol}
R.~Doria, J.~Frenkel and J.~C.~Taylor,
Nucl.\ Phys.\ B {\bf 168} (1980) 93;
\\
C.~Di'Lieto, S.~Gendron, I.~G.~Halliday and C.~T.~Sachrajda,
Nucl.\ Phys.\ B {\bf 183} (1981) 223;
\\
S.~Catani, M.~Ciafaloni and G.~Marchesini,
Nucl.\ Phys.\ B {\bf 264} (1986) 588;
\\
S.~Catani,
Z.\ Phys.\ C {\bf 37} (1988) 357.

\bibitem{perff}
B.~Mele and P.~Nason,
Nucl.\ Phys.\ B {\bf 361} (1991) 626;
\\
M.~Cacciari and M.~Greco,
Nucl.\ Phys.\ B {\bf 421} (1994) 530
[hep-ph/9311260].

\bibitem{Demina:1999ze}
R.~Demina {\it et al.},
hep-ph/0005112.

\bibitem{asren}
J.~Collins, F.~Wilczek and A.~Zee,
Phys.\ Rev.\ D {\bf 18} (1978) 242;
\\
W.~Bernreuther and W.~Wetzel,
Nucl.\ Phys.\ B {\bf 197} (1982) 228
[E ibid.\ B {\bf 513} (1982) 758].

\bibitem{Kunszt:1994sd}
Z.~Kunszt, A.~Signer and Z.~Tr\'ocs\'anyi,
Nucl.\ Phys.\ B {\bf 411} (1994) 397
[hep-ph/9305239].

\bibitem{Catani:1997pk}
S.~Catani, M.~H.~Seymour and Z.~Tr\'ocs\'anyi,
Phys.\ Rev.\ D {\bf 55} (1997) 6819
[hep-ph/9610553].

\bibitem{Baier:1973ms}
V.~N.~Baier, V.~S.~Fadin and V.~A.~Khoze,
Nucl.\ Phys.\ B {\bf 65} (1973) 381.

\bibitem{Sud}
J.~Kodaira and L.~Trentadue,
Phys.\ Lett.\ B {\bf 123} (1983) 335;\\
S.~Catani and L.~Trentadue,
Nucl.\ Phys.\ B {\bf 327} (1989) 323;\\
S.~Catani, E.~D'Emilio and L.~Trentadue,
Phys.\ Lett.\ B {\bf 211} (1988) 335;\\
S.~Catani, L.~Trentadue, G.~Turnock and B.~R.~Webber,
Nucl.\ Phys.\ B {\bf 407} (1993) 3;\\
S.~Catani, B.~R.~Webber and G.~Marchesini,
Nucl.\ Phys.\ B {\bf 349} (1991) 635.

\bibitem{Beenakker:1996ch} 
W.~Beenakker, R.~H\"opker, M.~Spira and P.~M.~Zerwas, 
Nucl.\ Phys.\ B {\bf 492} (1997) 51 
[hep-ph/9610490]. 

\bibitem{Beenakker:1997ut} 
W.~Beenakker, M.~Kr\"amer, T.~Plehn, M.~Spira and P.~M.~Zerwas, 
Nucl.\ Phys.\ B {\bf 515} (1998) 3 
[hep-ph/9710451]. 

\bibitem{Berger:2000iu} 
E.~L.~Berger, M.~Klasen and T.~M.~Tait, 
Phys.\ Rev.\ D {\bf 62} (2000) 095014 
[hep-ph/0005196]. 

\bibitem{Kramer:1997hh} 
M.~Kr\"amer, T.~Plehn, M.~Spira and P.~M.~Zerwas, 
Phys.\ Rev.\ Lett.\  {\bf 79} (1997) 341 
[hep-ph/9704322]. 

\bibitem{Martin:1993yx} 
S.~P.~Martin and M.~T.~Vaughn, 
Phys.\ Lett.\ B {\bf 318}, 331 (1993) 
[hep-ph/9308222]. 

\bibitem{Nilles:1980ic}
H.~P.~Nilles,
Phys.\ Rev.\ Lett.\  {\bf 45}, 319 (1980).

\bibitem{Jersak:1982sp}
J.~Jers\'ak, E.~Laermann and P.~M.~Zerwas,
Phys.\ Rev.\ D {\bf 25} (1982) 1218
[Erratum-ibid.\ D {\bf 36} (1982) 310].

\bibitem{Kniehl:1990qu}
B.~A.~Kniehl and J.~H.~K\"uhn,
Nucl.\ Phys.\ B {\bf 329} (1990) 547.

\bibitem{Eynck:2001en}
T.~O.~Eynck, E.~Laenen, L.~Phaf and S.~Weinzierl,
preprint NIKHEF 2001-013 [hep-ph/0109246].

\bibitem{Ballestrero:1994dv}
A.~Ballestrero, E.~Maina and S.~Moretti,
Nucl.\ Phys.\ B {\bf 415} (1994) 265
[hep-ph/9212246].

\bibitem{Seymour:1995bz}
M.~H.~Seymour,
Nucl.\ Phys.\ B {\bf 436} (1995) 163.
\\
D.~J.~Miller and M.~H.~Seymour,
Phys.\ Lett.\ B {\bf 435} (1998) 213
[hep-ph/9805414].

\bibitem{Brandenburg:1998pu}
A.~Brandenburg and P.~Uwer,
Nucl.\ Phys.\ B {\bf 515} (1998) 279
[hep-ph/9708350].

\bibitem{Harris:1997zq}
B.~W.~Harris and J.~Smith,
Phys.\ Rev.\ D {\bf 57} (1998) 2806
[hep-ph/9706334].

\end{thebibliography}
\end{document}